\documentclass[10pt,aps,a4paper,prd,floatfix,onecolumn,notitlepage,nofootinbib,showkeys,showpacs]{revtex4-1}

\usepackage[left=2.5cm,right=2.5cm,top=2.5cm,bottom=2.5cm]{geometry}

\usepackage[utf8]{inputenc}
\usepackage[british,UKenglish,USenglish,american]{babel}
\usepackage{amsmath,amsfonts}

\usepackage{slashed}
\usepackage{amssymb}
\usepackage{bbm}
\usepackage{nicefrac}
\usepackage{physics}
\usepackage{graphicx}
\usepackage{hhline,color}
\usepackage{epsfig}
\usepackage{enumitem}
\usepackage{multirow}
\usepackage[font=small,labelfont=bf,justification=justified,format=plain]{caption}
\usepackage{subcaption}
\newcommand\numberthis{\addtocounter{equation}{1}\tag{\theequation}}

\DeclareMathOperator{\diag}{diag}

\renewcommand{\bar}{\overline}
\usepackage{bm}\renewcommand{\vec}{\bm}

\newcommand{\Tcc}{T^\chi_c}
\newcommand{\Z}{\mathbb{Z}}

\usepackage{environ}
\NewEnviron{myequation}{%
\begin{equation}
\scalebox{0.9}{$\BODY$}
\end{equation}}

\begin{document}

\title{Phase Diagram, Scalar-Pseudoscalar Meson Behavior and Restoration of 
Symmetries in (2~+~1) Polyakov-Nambu-Jona-Lasinio Model}

\author{Pedro Costa}
\affiliation{Centro de F\'{\i}sica da Universidade de Coimbra (CFisUC), 
Department of Physics, University of Coimbra, P-3004 - 516  Coimbra, Portugal}

\author{Renan C\^{a}mara Pereira}
\affiliation{Centro de  F\'{\i}sica da Universidade de Coimbra (CFisUC), 
Department of Physics, University of Coimbra, P-3004 - 516  Coimbra, Portugal}

\begin{abstract}
We explore the phase diagram and the modification of mesonic observables in a
hot and dense medium using the (2~+~1) Polyakov-Nambu-Jona-Lasinio model.
We present the phase diagram in the ($T,\,\mu_B)$-plane, with its isentropic
trajectories, paying special attention to the chiral critical end point
(CEP).
Chiral and deconfinement transitions are examined.
The modifications of mesonic observables in the medium are explored as a tool
to analyze the effective restoration of chiral symmetry for different
regions of the phase diagram.
It is shown that the meson masses, namely that of the kaons, change abruptly
near the CEP, which can be relevant for its experimental search.
\end{abstract}
\keywords{phase transition; chiral symmetry; restoration of symmetry}

\pacs{11.10.Wx; 11.30.Rd; 14.40.Aq}

\maketitle

\section{Introduction}\label{sec:introd}

Since its creation in 1961 by Yoichiro Nambu and Giovanni Jona-Lasinio
\cite{Nambu:1961tp,Nambu:1961fr}, the Nambu- Jona-Lasinio (NJL) model has been
used in multiple applications.
The NJL model is a pre-Quantum Chromodynamics (QCD) model that was introduced
to describe the pion as a bound state of a nucleon and an antinucleon.
Indeed, in its first version, the NJL model was formulated as an effective
theory of nucleons and mesons, constructed from Dirac fermions that interact
via four-fermion interactions with chiral symmetry, analogous with Cooper pairs
in the Bardeen-Cooper-Schrieffer theory (BCS) theory of
superconductivity (simultaneously and independently, the analogy between four 
fermion masses and gaps in superconductors was also proposed by
V. G. Vaks and A. I. Larkin ~\cite{VaksLarkin:1961}).
The key idea was that the mass gap in the Dirac spectrum of the nucleon can be
generated analogously with the energy gap of a superconductor in BCS theory.
Later, the nucleon fields were replaced by quark fields ($u$, $d$ and $s$) and
it is still used to this day as an effective low-energy and model of QCD.
The introduction of quark degrees of freedom, in the chiral limit (the
bare quark mass is $m_i=0$), and the description of hadrons were made by T.
Eguchi and K. Kikkawa~\cite{Eguchi:1976iz,Kikkawa:1976fe} and in a more
realistic version with $m_i\neq 0$ by D. Ebert and M. Volkov
\cite{Volkov:1982zx,Ebert:1982pk,Volkov:1984kq}.

The central point of the NJL model becomes the fact that its Lagrangian density
contains the most important symmetries of QCD that are also observed in nature,
namely chiral symmetry, which~is fundamental to understand the physics of the
lightest hadrons.
The NJL model also embodies a spontaneous symmetry breaking mechanism.
As matter of fact, within this approach, mesons can be interpreted as
quark-antiquark excitations of the vacuum and baryons as bound states of quarks
(solitons or quark-diquarks structures)~\cite{Vogl:1991qt}.

Another relevant aspect is the fact that the interaction between quarks is
assumed to be point-like (the gluon degrees of freedom are considered to be
frozen into point-like quark-(anti-)quark vertices), attractive leading to a
quark-antiquark pair condensation in the vacuum.
However, the result of this point-like interaction is that the NJL model is not
a renormalizable field theory and a regularization scheme must be specified to
deal with the improper integrals that occur.
Another well-known shortcoming of NJL-type models is the absence of
gluons and the lack of the QCD property of color confinement, which implies
that some care must be taken when applying the theory to high~energies.

Over time, the model has undergone several improvements, a very important
one was the introduction of a term simulating the U$_A$(1) anomaly
\cite{Kunihiro:1987bb,Bernard:1987sg,Reinhardt:1988xu}, which is known as the
Kobayashi-Maskawa-'t Hooft (KMT) term
\cite{Kobayashi:1970ji,Kobayashi:1971qz,tHooft:1976snw,tHooft:1986ooh}.
This new KMT term is a six-fermion interaction that can be written in a
determinantal form. It breaks the unwanted axial U$_A$(1) symmetry of the
four-quark NJL Lagrangian.
Indeed, it is well know that in the chiral limit, i.e., $m_u=m_d=m_s= 0$,
QCD has a U(3) chiral symmetry. When this symmetry is spontaneously broken, it
implies the existence of nine massless Goldstone bosons. However, in nature,
only eight light pseudoscalar mesons exist: $\pi^0,\,\pi^+,\pi^-$,\,
$K^+,\,K^-,\,K^0,\,\bar{K}^0$ and $\eta$. The Adler-Bell-Jackiw U$_A$(1) anomaly
solves this discrepancy, and the eventual ninth Goldstone boson, the
$\eta^\prime$, gets a mass (around 1 GeV) due to the fact that the density of
topological charges in the QCD vacuum is nonzero
\cite{Adler:1969gk,Bell:1969ts}. Since the origin of the mass of the
$\eta^\prime$-meson is different from the masses of the other pseudoscalar
mesons, this meson cannot be seen as the remnant of a Goldstone boson.
Interestingly, the KMT term also induces a mixing between different quark
flavors with very important consequences in the scalar and pseudoscalar mesons.
We~emphasize the following review works on this model
\cite{Vogl:1991qt,Klevansky:1992qe,Hatsuda:1994pi,Buballa:2003qv}.

Over the past decade, two major improvements of the NJL model took place:
the introduction of eight-quark interactions with the purpose of stabilizing
the asymmetric ground state of the model with four and six-quark 't Hooft
interactions~\cite{Osipov:2005tq,Osipov:2006ns,Osipov:2007mk};
the introduction of the Polyakov loop effective field in order to consider
characteristics of both the chiral symmetry breaking and the deconfinement.
The latter resulted in the so-called Polyakov loop---Nambu-Jona-Lasinio (PNJL)
model~\cite{Fukushima:2003fw,Ratti:2005jh,Hansen:2006ee}.

In the PNJL model, the Lagrangian contains static degrees of freedom that are
introduced through an effective gluon potential in terms of the Polyakov loop
\cite{Meisinger:1995ih,Pisarski:2000eq,Pisarski:2002ji,Meisinger:2001cq,
Fukushima:2003fw,Mocsy:2003qw,Ratti:2005jh,Hansen:2006ee}.
This coupling of quarks to the Polyakov loop effectively induces the reduction
of the weight of the quark degrees of freedom at low temperature.
This is a consequence of the restoration of the $\Z_{N_c}$ symmetry related with
the color confinement.
Indeed, in these new types of models, the Polyakov loop effective field is not a
dynamical degree of freedom because the Lagrangian does not contain any
dynamical term. The gluon dynamics are simply reduced to a chiral-point coupling
between quarks, combined with a static background field representing the
Polyakov loop~\cite{Hansen:2006ee}.

One of the first big achievements of the NJL model was the study of meson
properties at finite temperature and density, first in SU(2) sector
\cite{Hatsuda:1986gu,Bernard:1987im,Bernard:1987ir} then in the
(2~+~1)-flavors, where the role of strangeness was taken in consideration
\cite{Bernard:1987sg,Bernard:1988db,Klimt:1989pm,Vogl:1989ea,Lutz:1992dv,
Ruivo:1994px,Ruivo:1996np,deSousa:1997nu,Ruivo:1999pr,Bhattacharyya:1998ps}.
The chiral symmetry restoration through the point of view of the chiral
partners has entered in the agenda. Indeed, experimentally, the manifestation of
chiral symmetry would be the existence of parity doublets, that is, a multiplet
of particles with the same mass and opposite parity for each multiplet of
isospin (the so-called chiral partners) in the hadronic spectrum, a situation
is not verified in nature. However, as the temperature increases, the~chiral
partners should have the same mass
\cite{Rehberg:1995kh,Costa:2004db,Costa:2005cz}.
The~same happens for increasing densities, even~for different environment
scenarios
\cite{Costa:2005cz,Blanquier:2011zz,Blanquier:2014kja,Dubinin:2016wvt}.

The study on the modification of the mesonic observables in the hot medium (used
as a tool for understanding the restoration of chiral and axial symmetries) was
extended to the (2~+~1)-flavor PNJL model to infer the role of the Polyakov loop.
It was concluded that the partial restoration of the chiral symmetry is faster
in the PNJL model than in the NJL model~\cite{Costa:2008dp}.
The properties of the mesons in the PNJL were also studied in
\cite{Blanquier:2011zz,Blanquier:2012du,Dubinin:2013yga,Blanquier:2014kja,
Dubinin:2016wvt}.

Interestingly, the NJL model has also been used to study mesons containing
charmed quarks like $D$ and $B$-mesons
\cite{Gottfried:1992bz,Blaschke:2003ji,Guo:2012tm,Carames:2016qhr}, in order to
describe both the light and heavy quarks in one model.
The same mesons were studied within the PNJL model, with particular focus on the
modification of the $D$-meson properties (masses and widths) in hot and dense
matter. The purpose was to find the consequences of the possible non dropping
of the $D$-meson masses in the medium for $J/\psi$ suppression scenarios
\cite{Blaschke:2011yv}.


Another aspect where the NJL model (and its extensions) proved to be very
useful was in the study of the possible phases of strongly-interacting matter.
Since the first conjecture by N.~Cabibbo and G. Parisi~\cite{Cabibbo:1975ig},
the QCD phase diagram has been widely studied by both theoretical and
experimental physicists (for a general review see~\cite{Brambilla:2014jmp}).
In fact, in the hot and dense region of the phase diagram, the~QCD phase
structure is replete of rich details~\cite{Halasz:1998qr,Gupta:2011wh},
namely the nature of the hadron matter-quark gluon plasma (QGP) transition
and the possible existence of the QCD chiral critical endpoint (CEP), have
been subject of remarkable theoretical and experimental efforts.
In~this region, QCD~is non-perturbative, meaning that the set of theoretical
tools available to study the phase transitions and meson behavior is limited.
Some options on the table are lattice gauge theory applied to QCD (lattice QCD),
Dyson-Schwinger equations and effective models.
Lattice QCD is a first principles method however, at finite density, it suffers
from the so-called sign problem, which renders the importance sampling needed in
Monte Carlo simulations not appropriate~\cite{Schmidt:2017bjt}. Different
methods are currently trying to fix, or circumvent, this issue like reweighting,
Taylor expansions, considering an imaginary chemical potential and complex
Langevin~\cite{Schmidt:2017bjt,Seiler:2017wvd}.
Dyson-Schwinger equations are a method based on the QCD effective action
\cite{Roberts:1994dr,Roberts:2012sv}. This method generates an infinite tour of
integro-differential equations for the Green's function of the theory that need
to be truncated at some order. Making a proper truncation is not a simple task
and several techniques have been developed and applied throughout the years
\cite{Fischer:2018sdj}.
The use of effective models of QCD allows access to the entire phase diagram
rooting the model to experimental or lattice QCD data. The main disadvantage of
using effective models is that they are not derived from first principles; one
should only study the model inside its range of applicability.

One interesting and very timely topic of the phase diagram of strong
interactions, is the possible existence and location of the chiral CEP. The CEP
is a conjectured second order phase transition point in the ($T-\mu_B(\rho_B)$
)-plane (belonging to the three-dimensional Ising universality class),
which~separates the crossover transition at zero density predicted by lattice
QCD calculations~\cite{Borsanyi:2010bp,Bazavov:2014pvz} and a possible
first-order phase transition in the cold and dense region of the diagram.
Old lattice results~\cite{Fodor:2004nz}, Dyson-Schwinger calculations
\cite{Fischer:2013eca,Fischer:2014ata} and several models predict its existence
but this remains a matter of debate.

Experimentally, a major goal of heavy ion collision (HIC) experiments is not
only to map the the QCD phase boundaries but also settle the question about the
existence of the elusive CEP. Indeed, the~search for the CEP is already being
carried out in several facilities such as the Relativistic Heavy Ion Collider
(RHIC) (STAR Collaboration) at Brookhaven National Laboratory
\cite{Adamczyk:2014fia,Adamczyk:2017wsl,Adamczyk:2017iwn} and in the Super
Proton Synchrotron (SPS) (NA61/SHINE Collaboration) at CERN
\cite{Aduszkiewicz:2015jna,Grebieszkow:2017gqx};
Future facilities like the J-PARC Heavy Ion Project at Japan Proton Accelerator
Research Complex (J-PARC)~\cite{Sako:J-PARC}, the Facility for Antiproton and
Ion Research (FAIR) at GSI Helmholtzzentrum für Schwerionenforschung
\cite{Ablyazimov:2017guv} and the Nuclotron-based Ion Collider fAcility (NICA)
at Joint Institute for Nuclear Research~\cite{NICAWP}, have planned experiments
to its search (a review on the experimental search of the CEP can be found
in Ref.~\cite{Akiba:2015jwa}).

One of the first suggestions for a first-order phase transition in the framework
of the standard SU(2) NJL model was made in~\cite{Asakawa:1989bq} and the study
of the phase diagram in the same version of the model at finite temperature and
density was presented in~\cite{Schwarz:1999dj}. Also in the framework of the
standard SU(2) NJL model, the calculation of the hadronization cross section in
a quark plasma at finite temperatures and densities was done in
Reference~\cite{Zhuang:2000ub}.
After that, several studies have addressed the peculiarities of the different
versions of the model, namely the SU(2)
\cite{Scavenius:2000qd,Fujii:2003bz,Biguet:2014sga} and (2~+~1)-flavors
\cite{Mishustin:2000ss,Costa:2007ie,Costa:2008yh}.
The PNJL model brings with it some changes relative to the NJL model, in both
SU(2) and (2~+~1)-flavor versions, particularly concerning to a higher temperature
for the position of the CEP and also a larger size of the critical region
\cite{Fu:2007xc,Costa:2008gr,Costa:2010zw,Friesen:2014mha,Torres-Rincon:2017zbr}.

The applications of NJL and PNJL models are indeed very wide, see for example
their usefulness in the study of neutron stars
\cite{Schertler:1999xn,Hanauske:2001nc,Buballa:2003et,Menezes:2003xa,
Lenzi:2010mz,Bonanno:2011ch,Pereira:2016dfg}
or the influence of strong magnetic fields
\cite{Menezes:2008qt,Ferreira:2013tba,Costa:2013zca,Ferreira:2014kpa,
Costa:2015bza,Ferreira:2017wtx,Ferreira:2018sun}.

In the present work, we will consider the PNJL model to explore the QCD phase
diagram and the in-medium behavior of scalar and pseudoscalar mesons, especially
the hot and dense regions near the chiral transition.

This paper is organized as follows.
In Section \ref{sec:modelo_forma}, we present the PNJL model used throughout
the work to study the phase diagram of QCD and the behavior of scalar and
pseudoscalar mesons.
In Section \ref{sec:Phase Diagram}, we present the phase diagram and analyze some
features of both the chiral and deconfinement transitions. Special attention is
given to the isentropic trajectories near the CEP.
In~Section \ref{sec:Mesons}, we study the behavior of scalar and pseudoscalar
mesons in six different paths that cross the phase diagram: zero density, zero
temperature, crossover, CEP, first-order and a path following an isentropic line.
Finally, we present the concluding remarks in Section \ref{sec:conclusions}.

\section{Model and Formalism\label{sec:modelo_forma}}

\subsection{The PNJL Model}

The SU$_f(3)$ Nambu$-$Jona-Lasinio model, including the 't Hooft
interaction (the inclusion of this term is not only important to
correctly reproduce the symmetries of QCD, but also allows to reproduce the
correct mass split between the $\eta$ and $\eta'$ mesons in SU$_f$(3)
\cite{Klevansky:1992qe,Kobayashi:1970ji}), which explicitly breaks U$_A$(1),
is~defined via the following Lagrangian density:
\begin{align*}
\mathcal{L}_\mathrm{NJL} =
\bar{\psi} \qty(i\slashed{\partial}-\hat{m})\psi
& + \frac{g_S}{2} \sum_{a=0}^{8} \qty[ \qty(\bar{\psi} \lambda^a \psi)^2 + \qty(\bar{\psi} i \gamma^5 \lambda^a \psi)^2 ]\nonumber
\\
& + g_D \qty[
\det\qty( \bar{\psi} \qty( 1 + \gamma_5 ) \psi ) +
\det\qty( \bar{\psi} \qty( 1 - \gamma_5 ) \psi )  ].
\numberthis
\label{eq:SU3_NJL_lagrangian}
\end{align*}

Here, the quark field $\psi$, is a $3$-component vector in flavor space, where
each component is a Dirac spinor, $\hat{m}=\diag\qty(m_u,m_d,m_s )$ is the quark
current mass matrix, diagonal in flavor space. The matrices $\lambda^a$ with
$a=1,...,8$, are the Gell-Mann matrices and
$\lambda^0=\sqrt{\nicefrac{2}{3}} \mathbbm{1}_{3 \times 3}$.

Following~\cite{Costa:2008dp}, the bosonization procedure can be easily carried
out after the 't Hooft six-quark interaction is reduced to a four-quark
interaction. One gets:
\begin{align}\label{eq:lagr_eff}
{\cal L}_\mathrm{eff} =
\bar{\psi} \qty( i\slashed{\partial}-\hat{m} )\psi
+ \frac{1}{2} S_{ab} \qty[ \qty( \bar{\psi} \lambda^a \psi )
\qty( \bar{\psi} \lambda^b \psi  ) ]
+ \frac{1}{2} P_{ab} \qty[ \qty( \bar{\psi} i \gamma_5 \lambda^a \psi )
\qty( \bar{\psi} i \gamma_5 \lambda^b \psi ) ],
\end{align}
where  the projectors $S_{ab}\,, P_{ab}$ are given by
\begin{align}
S_{ab} &= g_S \delta_{ab} + g_D D_{abc}\left\langle \bar{q} \lambda^c
q\right\rangle,
\label{eq:sab}
\\
P_{ab} &= g_S \delta_{ab} - g_D D_{abc}\left\langle \bar{q} \lambda^c
q\right\rangle.
\label{eq:pab}
\end{align}

The constants $D_{abc}$ coincide with the SU$_f$(3) structure constants for
$a,b,c=(1,2,\ldots ,8)$, while $D_{0ab}=-\frac{1}{\sqrt{6}}\delta_{ab}$ and
$D_{000}=\sqrt{\frac{2}{3}}$.
The quark fields can then be integrated out using the Hubbard-Stratonovich
transformation.

The main shortcoming of the NJL model as an effective model of QCD is its
inability to describe confinement physics. To remedy this situation and study
the deconfinement transition, the Polyakov-NJL model was introduced by K.
Fukushima~\cite{Fukushima:2003fw}, by coupling the NJL model to an order
parameter that describes the $Z(N_c)$ symmetry breaking: the Polyakov loop.

One important global symmetry of QCD is the center symmetry of SU$(N_c)$, the
$Z(N_c)$ symmetry. The breaking of this symmetry is associated to the
deconfinement transition: $Z(N_c)$ is respected in the confined phase and broken
in the deconfined phase. When considering pure glue theory at finite
temperature, the boundary conditions of QCD are respected by the $Z(N_c)$
symmetry. An order parameter for the possible $Z(N_c)$ symmetry breaking can be
defined using the thermal Wilson~line~$L\left( \vec{x} \right)$,
\begin{align}
L\left( \vec{x} \right) = \mathcal{P} \exp \left[ i \int_0^\beta d\tau \; A_4
\left( \tau, \vec{x} \right) \right].
\label{eq:thermal.wilson.line}
\end{align}
$\mathcal{P}$ is the path ordering operator and $A_4$ the gluon field in the
time direction,
\begin{align}
A_4 & = i g \mathcal{A}_\mu^a \frac{\lambda_a}{2} \delta_0^\mu  , \quad a=1,...,N_c^2-1
\\
& = i A_0  , \quad A_0 = g \mathcal{A}_\mu^a \frac{\lambda_a}{2} \delta_0^\mu .
\label{eq:A_4.definiton}
\end{align}

Here, $\mathcal{A}_\mu^a$ is the gluon field of color index $a$.
The Polyakov loop $\Phi$ can be defined as the trace over color of the thermal
Wilson line:
\begin{align}
\Phi = \frac{1}{N_c} \; \underset{c}{ \tr} L\left( \vec{x} \right).
\label{eq:loop.polyakov.definiton}
\end{align}

In the confined phase $\Phi \rightarrow 0$ while in the deconfined phase,
$\Phi \rightarrow 1$.

The NJL model can be minimally coupled to the gluon field in the temporal
direction, through~the introduction of the covariant derivative:
\begin{align}
\partial_\mu \rightarrow D_\mu = \partial_\mu - A_4^0 \delta_\mu^0 .
\end{align}

One has also to add to the Lagrangian density the effective Polyakov-loop
potential, $\mathcal{U}\left(\Phi,\bar{\Phi};T\right)$, which~represents the
effective glue potential at finite temperature.

The ansatz for the Polyakov loop effective potential can be written in terms of
the order parameter, using the Ginzburg-Landau theory of phase transitions.
The potential has to respect the $Z\qty(N_c)$ symmetry and to reproduce its
spontaneous breaking at some high temperature. There are several potentials in
the literature that fulfill these properties, e.g.,
\cite{Ratti:2005jh,Roessner:2006xn,Fukushima:2008wg}. We choose to adopt the one
proposed in Reference~\cite{Fukushima:2003fw,Roessner:2006xn}:
\begin{align}
\frac{\mathcal{U}\left(\Phi,\bar{\Phi};T\right)}{T^4} = -\frac{1}{2} a\left( T \right) \bar{\Phi} \Phi + b\left( T \right) \ln \left[ X(\Phi,\bar{\Phi}) \right],
\label{eq:Polyakov.loop.potential}
\end{align}
with the $T$-dependent parameters~\cite{Roessner:2006xn},
\begin{align}
a\left( T \right) & = a_0 + a_1 \left( \frac{T_0}{T} \right) + a_2 \left( \frac{T_0}{T} \right)^2 ,
\\
b\left( T \right) & =  b_3 \left( \frac{T_0}{T} \right)^3 ,
\end{align}
and where the argument in the logarithm is written as:
\begin{align}
X(\Phi,\bar{\Phi}) = 1-6\Phi\bar{\Phi}+4(\Phi^3+\bar{\Phi}^3)-3(\Phi\bar{\Phi})^2 .
\label{eq:simplification.X.effective.polyakov.potential}
\end{align}

The parameters $T_0$, $a_0$, $a_1$, $a_2$ and $a_3$ are fixed by reproducing
lattice QCD results at $\mu=0$~\cite{Ejiri:1998xd,Boyd:1996bx,Kaczmarek:2002mc}.
A commonly used set is:
\begin{align}
T_0 & = 270 \quad \text{(in the pure gauge sector)},\nonumber
\\
a_0 & = 3.51, \quad a_1 = -2.47,\nonumber
\\
a_2 & = 15.2, \quad a_3 = -1.75.\nonumber
\end{align}

To study the effects of finite density one can add to the Lagrangian density the
term $\bar{\psi} \gamma^0 \hat{\mu} \psi$, where~$\hat{\mu}=\diag\qty(\mu_u,\mu_d,\mu_s)$.
If considering the $A_4$ as a constant background mean field, it can be absorbed
in the definition of the chemical potential,
$\hat{\mu} \rightarrow \hat{\mu} - i A_4 = \tilde{\mu}$.
The main effect of $A_4$ in the effective chemical potential is to change the
distribution functions for particles and antiparticles, as~shown in
Reference~\cite{Hansen:2006ee}.

In the present work, temperature and finite density effects will be introduced
through the Matsubara formalism, which can be translated in the usual
substitution, $p_0 \rightarrow i\omega_n + \hat{\mu}$:
\begin{align}
\int \frac{ \dd[4]{p} }{ \qty(2 \pi)^4 } \rightarrow \frac{1}{-i\beta} \int \frac{ \dd[3]{p} }{ \qty(2 \pi)^3 } \sum_n .
\end{align}
where $\beta = 1/T$ and the sum is made over the Matsubara frequencies,
$\omega_n$.

\subsection{Gap Equations}

Considering the mean field approximation, the effective quark masses
(the gap equations) can be derive from Equation~(\ref{eq:lagr_eff})
(see Reference~\cite{Klevansky:1992qe}). One gets:
\begin{align}
M_{i}=m_{i}-2g_{_{S}}\left\langle\bar{q}_{i}q_{i}\right\rangle
-2g_{_{D}}\left\langle\bar{q}_{j}q_{j}\right\rangle\left\langle\bar
{q}_{k}q_{k}\right\rangle\,,
\label{eq:gap}
\end{align}
where  the quark condensates $\left\langle\bar{q}_{i}q_{i}\right\rangle$, with
$i,j,k=u,d,s$, in cyclic order.

The SU$_f$(3) PNJL grand canonical potential is:
\begin{align*}
\Omega  =
\mathcal{U}\left(\Phi,\bar{\Phi};T\right) &
+ g_{_{S}} \sum_{i=u,d,s} \left\langle\bar{q}_{i}q_{i}\right\rangle^2
+ 4 g_{_{D}}\left\langle\bar{q}_{u}q_{u}\right\rangle
\left\langle\bar{q}_{d}q_{d}\right\rangle\left\langle\bar{q}_{s}q_{s}\right\rangle
-  2N_c \sum_{i=u,d,s} \int \frac{d^3p}{(2\pi)^3} E_i
\\
& - 2 \sum_{i=u,d,s} \int \frac{d^3p}{(2\pi)^3} \left[ \mathcal{F}\left(\vec{p},T,\mu_i \right) + \mathcal{F}^*\left(\vec{p},T,\mu_i \right)  \right] .
\numberthis
\label{eq:pot.termo.PNJL}
\end{align*}

Here, $E_i$ is the quasiparticle energy for the quark $i$:
$E_{i}=\sqrt{\mathbf{p}^{2}+M_{i}^{2}}$ and the thermal functions $\mathcal{F}$
and $\mathcal{F}^*$ are defined as:
\begin{align}
\mathcal{F}\left(\vec{p},T,\mu_i \right) & = T \ln \left[ 1 + e^{-3\left(E_i-\mu_i \right)/T} + N_c \bar{\Phi} e^{-\left(E_i-\mu_i \right)/T} + N_c \Phi e^{-2\left(E_i-\mu_i \right)/T} \right] ,
\label{eq:Fthermal}
\\
\mathcal{F}^*\left(\vec{p},T,\mu_i \right) & = T \ln \left[ 1 + e^{-3\left(E_i+\mu_i \right)/T} + N_c\Phi e^{-\left(E_i+\mu_i \right)/T} + N_c\bar{\Phi} e^{-2\left(E_i+\mu_i \right)/T} \right]   ,
\label{eq:F*thermal}
\end{align}

To obtain the mean field equations, we apply the thermodynamic consistency
relations, i.e., calculate the minima of the thermodynamical potential density,
Equation~(\ref{eq:pot.termo.PNJL}), with respect to
\mbox{$\left\langle\bar{q}_iq_{i}\right\rangle$ ($i=u,d,s$), $\Phi$, and $\bar\Phi$}:
\begin{align}
\frac{\partial \Omega}{\partial
\left\langle\bar{q}_iq_{i}\right\rangle
} = \frac{\partial \Omega}{\partial \Phi} = \frac{\partial \Omega}{\partial \bar{\Phi}} = 0 ,
\quad  i=u,d,s .
\label{eq:thermodynamic.consistency.PNJL}
\end{align}

These relations define the value of the $i-$flavor quark condensate:
\begin{align}
\left\langle\bar{q}_{i}q_{i}\right\rangle & =
- i \tr \left[ S_i(q)  \right]
= - 2 \; N_c  \int \frac{d^3 p}{(2\pi)^3} \frac{M_i}{E_i} \left( 1 - \nu_i - \bar{\nu}_i \right) .
\label{eq:condensate.f.PNJL}
\end{align}

Here, $S_i(q)$ is the quark propagator of flavor $i$, $\nu_i$ and $\bar{\nu}_i$
are the particle and antiparticle occupation numbers in the PNJL model, defined
as:

\begin{align}
\nu_i  & = \frac{ \frac{3}{N_c} e^{-3\left(E_i-\mu_i \right)/T} + \bar{\Phi} e^{-\left(E_i-\mu_i \right)/T} + 2 \Phi e^{-2\left(E_i-\mu_i \right)/T} }{1 + e^{-3\left(E_i-\mu_i \right)/T} + N_c \bar{\Phi} e^{-\left(E_i-\mu_i \right)/T} + N_c \Phi e^{-2\left(E_i-\mu_i \right)/T}}  ,
\label{eq:particle.ocu.PNJL}
\\
\bar{\nu}_i & = \frac{ \frac{3}{N_c}e^{-3\left(E_i+\mu_i \right)/T} + \Phi e^{-\left(E_i+\mu_i \right)/T} + 2\bar{\Phi} e^{-2\left(E_i+\mu_i \right)/T} }{1 + e^{-3\left(E_i+\mu_i \right)/T} + N_c\Phi e^{-\left(E_i+\mu_i \right)/T} + N_c\bar{\Phi} e^{-2\left(E_i+\mu_i \right)/T} }  .
\label{eq:antiparticle.ocu.PNJL}
\end{align}

The \textit{gap} equations for the Polyakov loop fields $\Phi$ and $\bar{\Phi}$
are:
\begin{align*}
& -\frac{1}{2} a\left( T \right) \bar{\Phi} - \frac{6b\left( T \right) \left( \bar{\Phi} -2 \Phi^2 + \bar{\Phi}^2 \Phi \right) }{ X(\Phi,\bar{\Phi}) }  =
\frac{2 N_c}{T^3}  \sum_{i=u,d,s}  \int \frac{d^3p}{(2\pi)^3} \left[  \frac{ e^{-\left(E_i+\mu_i \right)/T}}{e^{\mathcal{F}^*\left(\vec{p},T,\mu_i \right)/T}}  +  \frac{  e^{-2\left(E_i-\mu_i \right)/T}}{e^{\mathcal{F}\left(\vec{p},T,\mu_i \right)/T}}  \right]  ,  \numberthis
\label{eq:gap.PHI.PNJL}
\\
&  -\frac{1}{2} a\left( T \right) \Phi - \frac{6b\left( T \right) \left( \Phi -2 \bar{\Phi}^2 + \bar{\Phi} \Phi^2  \right) }{ X(\Phi,\bar{\Phi})  }  =
\frac{2 N_c}{T^3}  \sum_{i=u,d,s}  \int \frac{d^3p}{(2\pi)^3} \left[    \frac{ e^{-\left(E_i-\mu_i \right)/T}}{e^{\mathcal{F}\left(\vec{p},T,\mu_i \right)/T}}  +  \frac{ e^{-2\left(E_i+\mu_i \right)/T}}{e^{\mathcal{F}^*\left(\vec{p},T,\mu_i \right)/T}}   \right]  .  \numberthis
\label{eq:gap.PHIbar.PNJL}
\end{align*}

In the $T=0$ limit, the PNJL grand canonical potential is reduced to the usual
NJL model. Indeed, in this limit, the Polyakov loop potential and the thermal
function $\mathcal{F}^*$ vanish while the function $\mathcal{F}$ becomes a
step-function. For more details, see the Appendix \ref{append2}. We draw
attention to the fact that this feature is a consequence of the definition of
the Polyakov loop potential in Equation~(\ref{eq:Polyakov.loop.potential}).
Actually, one can try to build a different Polyakov loop potential that does
not vanish in the $T \to 0$ limit, by~including for example, an explicit
dependence in the chemical potential. Of course, such a modified potential would
have to respect the $Z(N_c)$ of QCD, as well as, reproduce lattice observables.

\subsection{Pseudoscalar and Scalar Meson Nonets}

To study the meson mass spectrum and decays, we need to calculate the meson
propagators.
Following the same procedure outlined in detail in
Reference~\cite{Rehberg:1995kh,Costa:2008dp}, we expand the effective action to
second order in the meson fields, yielding the following meson propagator:
\begin{equation}
D^{M}_{ab} (q)=
\frac{1}{ M_{ab} ^{-1} - \Pi^M_{ab} (q)} .
\label{eq:propmeson}
\end{equation}

Here, $M_{ab}$ are the so-called  projectors which, for scalar mesons are
$S_{ab}$ and for pseudoscalar mesons $P_{ab}$, defined in Equations (\ref{eq:sab})
and (\ref{eq:pab}). The polarization operator for the meson channel $M$, is:
\begin{equation}
\Pi_{ab}^{M}(q)=
iN_{c}\int\frac{d^{4}p}{(2\pi)^{4}}\mbox{tr}\left[
S_{a}(p)\Gamma_a^M
S_{b}(p+q)\Gamma_b^M
\right].
\label{eq:polarization}
\end{equation}
The trace has to be made over flavor and Dirac spaces and
$\Gamma_a^M =\lambda^{a}\otimes\Gamma^M$, where
$\Gamma^M = \{ \mathbbm{1} , i\gamma_{5} \}$.
The~explicit expression is presented in the Appendix \ref{append}.
As already stated, the introduction of the Polyakov loop is made in the quark
propagator, $S_i(q)$,  by the use of the modified Fermi functions, defined in
Equations (\ref{eq:particle.ocu.PNJL}) and (\ref{eq:antiparticle.ocu.PNJL}).

The calculation of the masses of the scalar and pseudoscalar mesons is done from
the zeros of the inverse meson propagators in the rest frame, i.e.,
\begin{equation}
1 - M_{ab} \Pi^M_{ab}
\left(
M_M-i\frac{\Gamma_M}{2}, \vec{q} = \vec{0}
\right) = 0.
\label{pole}
\end{equation}

Here, $M_M$ and $\Gamma_M$ are the mass and decay width of the meson channel $M$.
The Mott dissociation is commonly identified by mass poles for mesons becoming
complex: the real part being the ``mass'', $M_M$, of the resonance; and the
imaginary part being related to a finite width, $\Gamma_M$, due to its decay
into the quark constituents (reflecting the fact that the NJL model does not
confine the quarks).
These two quantities are extracted from the zeroes of the complete real and
imaginary components of Equation~(\ref{pole}) that can be written in the form
of a system of two coupled equations.
Usually, different approaches can also be used to compute $M_M$ and $\Gamma_M$.
In~\cite{Torres-Rincon:2017zbr}, the meson masses were calculated by supposing
that the pole is near the real axis and the imaginary part of the solution in
the argument of Equation~(\ref{Iij2}) of the Appendix \ref{append} is neglected.
In~\cite{Ruivo:2004xf,Costa:2005cz,Costa:2008dp} only the $\Gamma_M^2$
contribution coming from $(M_M-i\Gamma_M/2)$ was~neglected.

The quark-meson coupling constants are given by the residue at the poles of
the propagators defined in Equation~(\ref{eq:propmeson}). It yields:
\begin{eqnarray}\label{mesq}
g_{M\overline{q}q}^{-2} = -\frac{1}{2 M} \frac{\partial}{\partial q_0}
\left[\Pi_{ab}^M (q_0) \right]_{ \vert_{ q_0=M_M}} \, ,
\end{eqnarray}
where $M_M$ is the mass of the bound state containing quark flavors $a,b$.

To calculate the masses of the $\eta$ and ${\eta'}$ mesons, we consider the
basis of $\pi^0 - \eta - \eta'$ system (for~more details see
\cite{Costa:2003uu}).
In the case where the $\pi^0$ is decoupled (this happens if
$\left\langle\bar{q}_{u}\,q_{u}\right\rangle=\left\langle\bar{q}_{d}\,q_{d}\right\rangle$)
from the $\eta-\eta'$, the following inverse propagators can be defined:
\begin{align}
D_\eta^{-1} (q) & = \left( { A}+{ C}\right) -  \sqrt{({ C}-{ A})^2+4{ B}^2}
\label{D_eta}
\\
D_{\eta'}^{-1}(q) & = \left( { A}+{ C}\right) +  \sqrt{({ C}-{ A})^2+4{ B}^2}
\label{D_eta_prime}
\end{align}
with ${ A} = P_{88} -\Delta \Pi_{00}^P (q),
{ C} = P_{00} -\Delta \Pi_{88}^P (q),
{ B} = - (P_{08} +\Delta \Pi_{08}^P (q)$ and
$\Delta = P_{00} P_{88}- P_{08}^2 $.
Here, $P_{ab}$ is defined in Equation~(\ref{eq:pab}).

In the rest frame,
$D_{\eta }^{-1}(q_0=M_{\eta },{\bf q }=0) =0,\,D_{\eta'}^{-1}(P_0=M_{\eta'},{\bf q }=0) =0$.
The mixing $\eta-\eta'$ pseudoscalar angle $\theta_P$, is given by:
\begin{align}
\tan 2 \theta_P = \frac{ 2B }{  A-C } .
\label{theta_P}
\end{align}

Turning now to the scalar mesons $\sigma$ and $f_0$, the approach is identical
to the one followed above. The propagators for these mesons and scalar angle
$\theta_S$, are be identical to the ones in Equations (\ref{D_eta})--(\ref{theta_P}). 
However, $A$, $B$ and $C$ are obtained by substituting $P_{ab} \rightarrow S_{ab}$ 
and $\Pi_{ab}^{P}(q) \rightarrow \Pi_{ab}^{S}(q)$.

\subsection{Thermodynamics}

Thermodynamic quantities are of great importance. Some of these quantities can
be compared with the results that have become accessible in first principles
calculations on the lattice at non-zero chemical potential (e.g., results of
fluctuations of conserved charges such as baryon number, electric charge, and
strangeness~\cite{Bazavov:2012vg}).
Also, it is interesting to study isentropic trajectories due to their importance
for studying the thermodynamics of matter created in relativistic heavy-ion
collisions.

The equations of state and other quantities of interest like the particle
density $(\rho_i)$, energy density $(\epsilon)$, and entropy density $(S)$,
can be derived from the thermodynamical potential $\Omega(T,\mu)$ 
(Equation~(\ref{eq:pot.termo.PNJL})), using the 
following relations~\cite{Kapusta:2006pm}:
\begin{align}
p & =  - \Omega,
\label{eq:def.pressao}
\\
\rho_i & =  -  \left(
\frac{\partial \Omega}{\partial \mu_i}
\right)_{T} ,
\label{eq:def.densidade}
\\
S & =  -
\left(
\frac{\partial \Omega}{\partial T} \right)_ {\mu} ,
\label{eq:def.entropia}
\\
\epsilon & = -P + TS + \sum_i \mu_i \rho_i. \label{eq:def.energia}
\end{align}

The pressure and the energy density are defined in such way that their values
are zero in the vacuum state~\cite{Buballa:2003qv}.

\subsection{Model Parameters and Regularization Procedure}

Concerning the numerical calculations, the model is fixed by the coupling
constants $g_{S}$, and~$g_{D}$ in the Lagrangian (\ref{eq:SU3_NJL_lagrangian}),
the cutoff parameter $\Lambda$ which regularizes momentum space integrals
$I_{1}^{i}$ (Equation~(\ref{Ii1})) and $I_{2}^{ij}$ (Equation~(\ref{Iij2})), and the
current quark masses $m_{i}$ ($i=l,\,s$; with $m_u=m_d=m_l$).
We employ the parameters of~\cite{Rehberg:1995kh}. The (2~+~1)-flavors version of
the NJL (PNJL) model has five parameters. These parameters are fixed in order
to fit the observables $M_\pi = 135.0$ MeV, \mbox{$f_\pi=92.4$ MeV}, $M_K=497.7$ MeV,
and $M_{\eta'}=957.8$ MeV, while $m_{l}=5.5$ MeV.
The parameters and the numerical results are given in Table \ref{table:param}.

\begin{table}[h!]
\caption{Physical quantities in the vacuum state and the parameter set used in
this work. The asterisk signalize the results of the model for such physical
quantities.}
\centering
\renewcommand{\arraystretch}{1.5}
\begin{tabular}[c]{c c}
\hline\hline
\multirow{2}{*}{\textbf{Physical Quantities}}&\textbf{Parameter set}\\
    &\textbf{and constituent quark masses}\\
\hline
$f_{\pi}=92.4$ MeV & $m_{u}=m_{d}=5.5$ MeV\\
$M_{\pi}=135.0$ MeV & $m_{s}=140.7$ MeV\\
$M_{K}=497.7$ MeV & $\Lambda=602.3$ MeV\\
${M}_{\eta^{\prime}}={957.8}$ MeV & $g_{S}\Lambda^{2}=3.67$\\
${M}_{\eta}=$ ${514.8}$ MeV~$^{\ast}$ & $g_{D}\Lambda^{5}=-12.36$\\
$f_{K}=93.1$ MeV$~^\ast$ & ${M}_{u}{=M}_{d}=367.7$ MeV~$^\ast$\\
${M}_{\sigma}={728.9}$ MeV~$^\ast$ & ${M}_{s}=549.5$ MeV~$^\ast$\\
${M}_{a_{0}}={880.2}$ MeV~$^\ast$ & \\
${M}_{\kappa}={1050.5}$ MeV~$^\ast$ & \\
${M}_{f_{0}}={1198.3}$ MeV~$^\ast$ & \\
${\theta}_{P}={-5.8}$${{}^{\circ}}~^\ast$\,; ${\theta}_{S}={16}$${{}^{\circ}}^\ast$ & \\
\hline
\end{tabular}
\label{table:param}
\end{table}
\renewcommand{\arraystretch}{1}
\vspace{-4pt}


An important aspect of both, NJL and PNJL models, is the lack of
renormalizability which comes from the point-like nature of the quark-quark
interaction. As a consequence, a procedure for regularizing divergent quantities
in both models is required.
Therefore, the regularization scheme determines the model. As pointed out in
\cite{Klevansky:1992qe}, it is needed to look to physical and not just to the
mathematical content. Therefore, the regularization process must be carried out
in such a way that physically expected properties of the model and symmetry
considerations are maintained~\cite{Klevansky:1992qe}.

Several regularization procedures are available:
three dimensional cut-off~\cite{Klevansky:1992qe}, four dimensional cut-off
\cite{Vogl:1991qt,Klevansky:1992qe,VandenBossche}
Pauli-Villars regularization
\cite{Pauli:1949zm,Schuren:1993aj,Dmitrasinovic:1995cb}, regularization in
proper time~\cite{Schwinger:1951nm,Schuren:1993aj}.
For a detailed analysis of the regularization procedures and more references to
the corresponding literature see
\cite{Klevansky:1992qe,Ripka:1997zb,VandenBossche:1997nn}.

In this work, we will use a three dimensional cut-off, $\Lambda$, in all
integrals and not just the divergent ones.
The effect of the inclusion of the Polyakov loop and the regularization
procedure in the medium was studied in~\cite{Costa:2010zw}.
A regularization that includes high momentum quark states
($\Lambda\rightarrow\infty$ in finite integrals), is necessary to get the
required increase of extensive thermodynamic quantities, allowing the
convergence to the Stefan-Boltzmann (SB) limit of QCD. However, this leads to
unphysical behavior of the quark condensates at very high temperatures (the
quark condensates change sign because the constituent quark masses
go below the respective current value)~\cite{Costa:2007fy,Costa:2009ae}.
In Reference~\cite{Bratovic:2012qs} a regularization procedure was proposed that
prevents the unphysical behavior of the quark condensates ensuring, at the same
time, that the pressure reaches the SB limit at high temperatures.
On the other hand, in~\cite{Moreira:2010bx} it was shown that the inclusion of
a temperature and chemical potential dependent term (which arises as a constant
of integration when integrate the gap equations to obtain the thermodynamic
potential) leads to the correct asymptotics for all observables considered.

The parameter $T_0$ is the temperature for the deconfinement phase transition in
pure gauge. According to lattice findings, it is usually fixed to $270$ MeV
\cite{Karsch:2001cy,Borsanyi:2012ve}. With this value of $T_0$, an almost exact
coincidence between the chiral crossover and the deconfinement transition at
zero chemical potential is obtained, as seen in lattice calculations.
However, different criteria for fixing $T_0$ are found in the literature as in
\cite{Schaefer:2007pw} where an explicit $N_f$ dependence of $T_0$
is given considering renormalization group arguments.
Since the results for the Polyakov loop coming from lattice calculations with
(2~+~1) flavors and with fairly realistic quark masses are identical to the
SU$_f$(2) case~\cite{Kaczmarek:2007pb}, it was chosen to maintain the same
parameters that were used in SU$_f$(2) PNJL~\cite{Roessner:2006xn} for the
effective potential $\mathcal{U}\left(\Phi,\bar\Phi;T\right)$.

In the present work, we rescale $T_0$ to $195$ MeV in order to get an agreement
between the deconfinement pseudocritical temperature obtained in the model, with
the results obtained on the lattice: 170 MeV~\cite{Aoki:2009sc}.
This modification of $T_0$ (which is the only free parameter of the Polyakov
loop since the effective potential is fixed) has a rescaling effect without
drastically changing the physics of the model.
With this choice of parameters, $\Phi$ and $\bar\Phi$ are always lower
than $1$.

To conclude this section, some comments regarding the the number of
parameters and the nature of the Polyakov field are needed.
It is important to mention that NJL parameters and the Polyakov potential are
not on the same footing. In fact, while the NJL parameters can be directly
related with physical quantities, the role of the Polyakov loop potential is
to insure the recovering of pure gauge lattice expectations.
This means that the potential for the Polyakov loop can be seen as an unique
but functional parameter.
The pure gauge critical temperature, $T_0$, is the only true parameter and fixes
the temperature scale of the system. However, in the Landau-Ginzburg framework,
the~characteristic temperature for a phase transition is not expected to be a
prediction.
Therefore, this parameter is changed in order to fix the correct energy scale
and obtain the result for the deconfinement pseudocritical temperature, coming
from lattice QCD calculations.

Finally it is important to point out that the Polyakov loop effective field is
not a dynamical degree of freedom. This is due to the the absence of dynamical
term in the Polyakov loop potential at the Lagrangian level:
it is a background gauge field in which quarks propagate.

\section{The Phase Diagram in the PNJL Model}\label{sec:Phase Diagram}

We will start our analysis of the PNJL model by the respective phase diagram.
We briefly discuss the chiral and deconfinement transitions as well as the Mott
dissociation of the pion and sigma mesons at finite temperature and/or baryonic
chemical potential. Isentropic trajectories will also be drawn and discussed.

\subsection{Characteristic Temperatures at Zero Density}\label{sec:FiniteT}

At zero temperature and density (baryonic chemical potential $\mu_B$), the
chiral symmetry of QCD is broken, both explicitly and spontaneously.
It is then expected that chiral symmetry gets restored at high temperatures and
an eventual phase transition will occur separating the regions of low and high~temperatures.

We start the discussion of the results by identifying the characteristic
temperatures, \mbox{at $\mu_B=0$}, that~splits the different thermodynamic phases in
the PNJL model~\cite{Hansen:2006ee}.
On the one hand, the~pseudocritical temperature associated to the
``deconfinement'' transition is $T_c^\Phi$. This temperature corresponds to the
crossover location of $\Phi$, defined by its inflexion point of, i.e.,
${\partial^2 \Phi }/{\partial T^2}=0$.
The~terminology ``deconfinement'' is used here to designate the transition
between $\Phi \simeq 0$ and $\Phi \simeq 1$ (see Reference~\cite{Hansen:2006ee} for a
detailed discussion of this subject).
On the other hand, the chiral transition characteristic temperature, $\Tcc$,
is given by the inflexion point (chiral crossover) of the light chiral
condensate $\langle\bar{q}_iq_i\rangle$ ($i=u,d$), i.e.,
${\partial^2 \left\langle \bar{q}_iq_i \right\rangle}/{\partial T^2}=0$.
However, since the presence of nonzero current quark mass terms break
the chiral symmetry, the restoration of chiral symmetry is realized through
parity doubling rather than by massless quarks.
The temperature of {\it effective} restoration of chiral symmetry (in the light
sector), $T^\chi_{eff}$, is provided by the degeneracy of the respective chiral
partners [$\pi,\,\sigma$] and [$\eta,\,{a_0}$] or, in other words, by the
merging of their spectral functions \cite {Hansen:2006ee,Costa:2008dp}.
$T^\chi_{eff}$ is then defined as the temperature where
$M_{\sigma}-M_{\pi}<1\%\,(M_{\sigma}^{vac}-M_{\pi}^{vac})$ MeV.
Indeed, one key point of this work is the effective restoration of symmetries
through the mesonic properties perspective.

\begin{figure}[h!]
\begin{center}
\begin{subfigure}{.5\textwidth}
\centering
\includegraphics[width=1.05\linewidth]{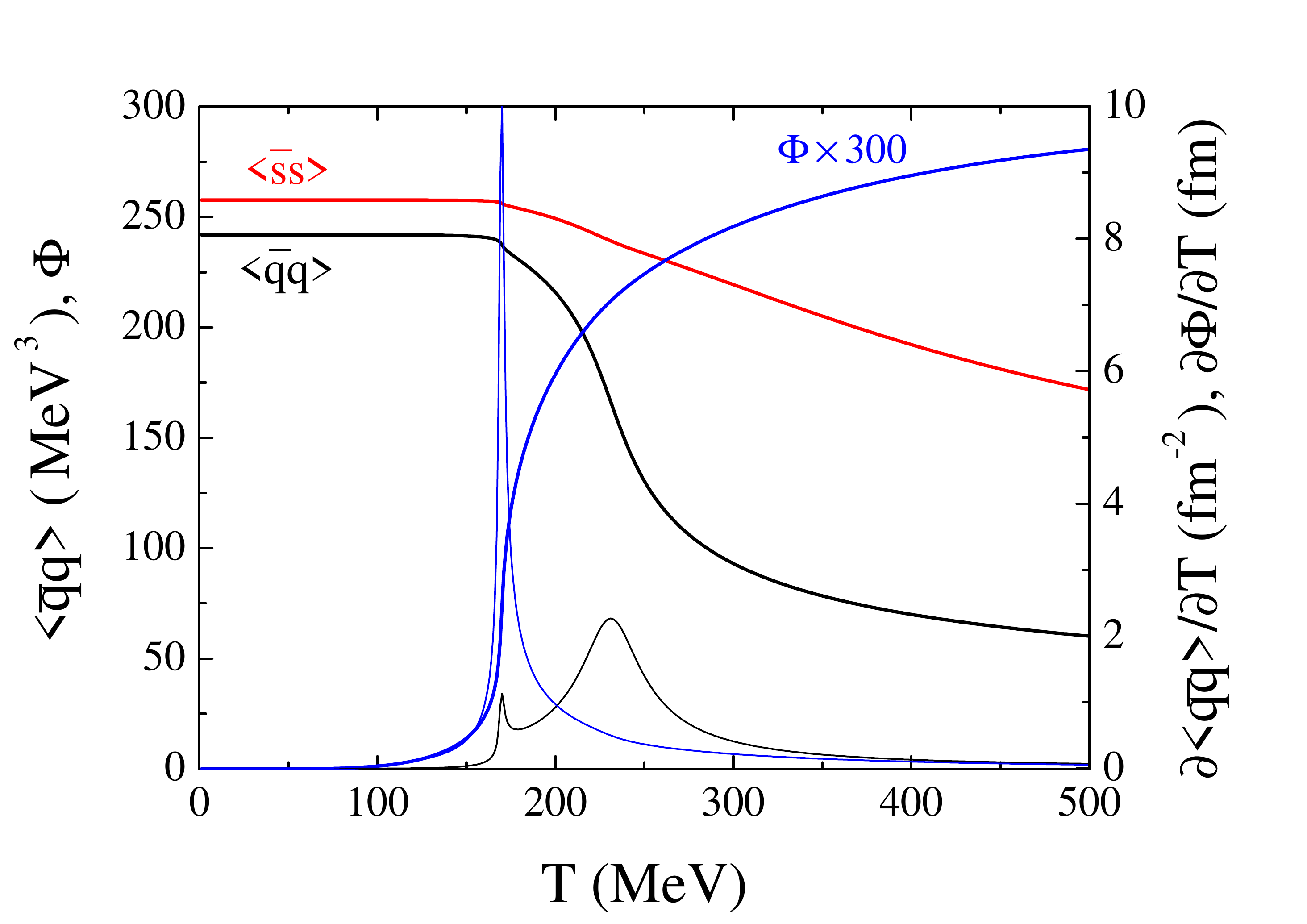}
\end{subfigure}%
\begin{subfigure}{.5\textwidth}
\centering
\includegraphics[width=1.05\linewidth]{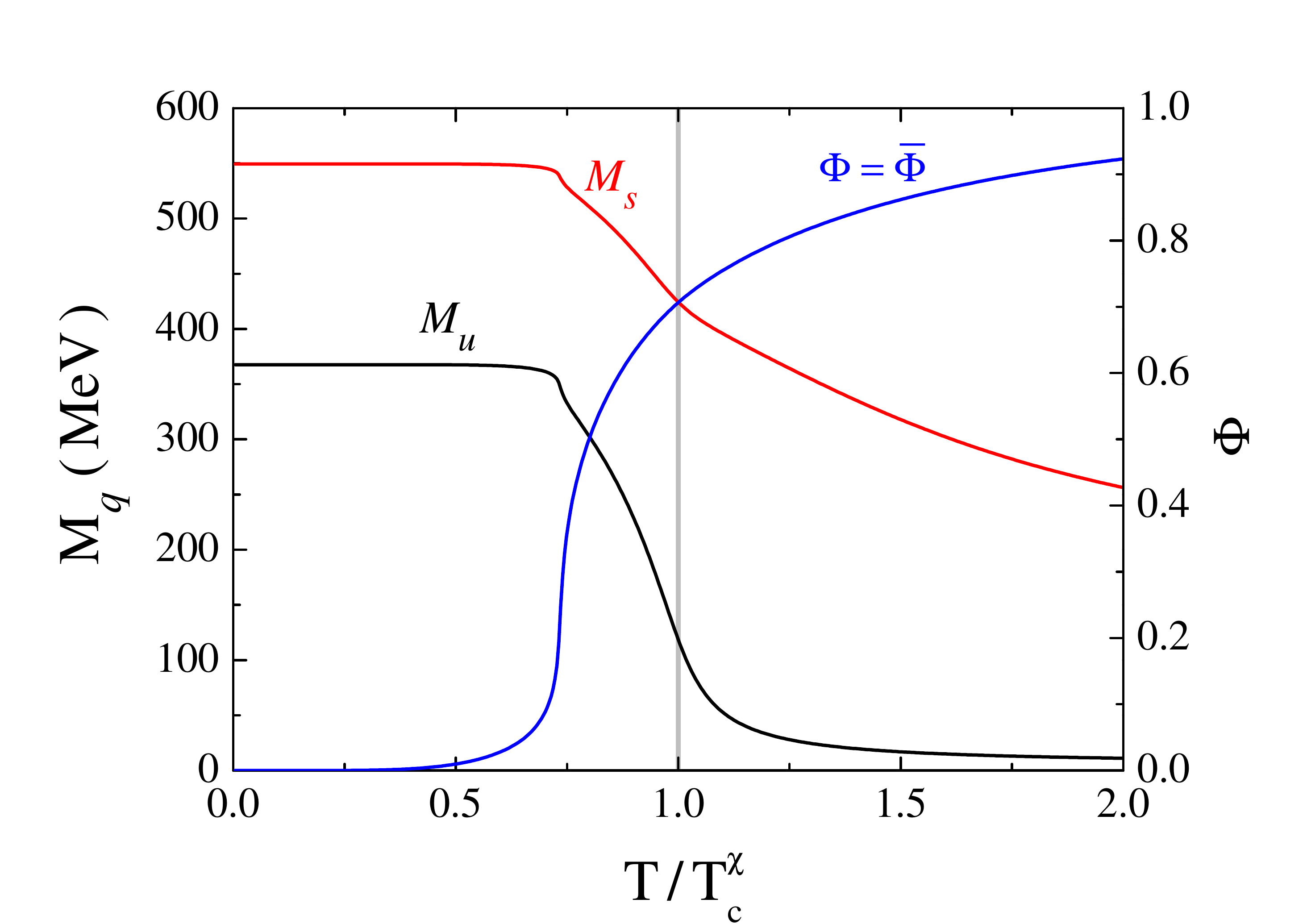}
\end{subfigure}
\end{center}
\vspace{-0.25cm}
\caption{Quark condensates and respective derivatives (\textbf{left panel}) and quark
masses (\textbf{right panel}) in the PNJL 
model as functions of the temperature; the
Polyakov loop field, $\Phi$, is also shown, together with its derivative given
by the thin blue line in the left panel.}
\label{fig:Cond_Mquark}
\end{figure}

In Figure~\ref{fig:Cond_Mquark}, the strange and nonstrange quarks
condensates are plotted (left panel), as well as the respective quark masses (right panel), and the
Polyakov loop as functions of the temperature.
In the left panel, we also plot the derivatives of the light quarks (black) and
of the Polyakov loop (blue) in order to the temperature (thin lines).
The respective peaks also indicate the pseudocritical temperatures for the
partial restoration of chiral symmetry and the deconfinement.
For temperatures near $\Tcc$, the light quark masses drop in a continuous way
to the respective current quark mass value.
This indicates the smooth crossover from the chirally broken to a partially
chirally symmetric phase, once the value of the quarks masses are still far from
the value of the corresponding current masses (partial restoration of chiral
symmetry).
The strange quark mass shows a similar behavior to that of the nonstrange
quarks, with a substantial decrease above  $\Tcc$; however, its mass is still
far away from the strange current quark mass.
As in the NJL model, regarding the strange sector~\cite{Costa:2005cz}
and since $m_u=m_d<m_s$, the (sub)group SU(2)$\otimes$SU(2) is a better
symmetry of the Lagrangian (Equation~(\ref{eq:SU3_NJL_lagrangian})).
This will have consequences in the behavior of the meson masses.

In Table \ref{table:Mott_temp}, we report the characteristic temperatures in the
PNJL model: the pseudocritical temperatures for the chiral transition
($T_c^{\chi}$) and deconfinement ($T_c^{\Phi}$); the effective chiral symmetry
restoration temperature ($T_{eff}^{\chi}$); and the Mott temperatures for pion
($T^{Mott}_{\pi}$), for eta ($T^{Mott}_{\eta}$), for kaon ($T^{Mott}_{kaon}$)
and for the sigma ($T^{Mott}_{\sigma}$) mesons.
We remind that the Mott transition is related with the composite nature of
the mesons: at the Mott temperature it becomes energetically favorable for a
meson to decay into a $\bar{q}q$ pair .

\begin{table}[h!]
\caption{\label{table:Mott_temp} Characteristic and Mott temperatures in the
PNJL model at $\mu_B$ with $T_0=195$ MeV.}
\centering
\renewcommand{\arraystretch}{1.5}
\begin{tabular}{ccccccc}
\hline\hline
\boldmath$T_c^{\chi}$ \textbf{[MeV]} & \boldmath$T_c^{\Phi}$ \textbf{[MeV]} & \boldmath$T_{eff}^{\chi}$ \textbf{[MeV] }
& \boldmath$T^{Mott}_{\pi}$ \textbf{[MeV]} & \boldmath$T^{Mott}_{\eta}$ \textbf{[MeV]}
& \boldmath$T^{Mott}_{K}$ \textbf{[MeV]}   & \boldmath$T^{Mott}_{\sigma}$ \textbf{[MeV]}\\
\hline
$231$ & $170$ & $280$ & $239$ & $211$ & $243$ & $197$\\
\hline
\end{tabular}
\renewcommand{\arraystretch}{1}
\end{table}

The difference between $T_c^\Phi$ and $T^\chi_c$ is due to the choice of $T_0$
and the regularization procedure, as~it was pointed out in
Reference~\cite{Hansen:2006ee}. In these work a three-dimensional momentum cutoff
to both the zero and the finite temperature/densities contributions is applied.
Different type of regularizations can lower $T^\chi_c$
\cite{Costa:2007fy}.
In Reference~\cite{Costa:2010zw} the meson properties were calculated without
rescaling the parameter $T_0$ to a smaller value. By adopting a higher value of
$T_0$, a smaller difference between the psudocritical temperatures of the two
transitions was obtained. However, since our goal is the general properties of
mesons, the absolute value of the pseudocritical temperatures is not the most
significant point: in fact, these properties are independent of the precise
value of $T^\chi_c$ and different values of $T_0$ do not change the conclusions.

\subsection{Finite Temperature and Chemical Potential}\label{sec:FiniteTdens}

Now, let's extend the analysis to the scenario with $\mu_B\neq 0$, i.e., the
phase diagram.
The PNJL model can mimic a region in the interior of a neutron star (where the
PNJL model is reduced to the NJL model) or a dense fireball created in a HIC.
Bearing in mind that in a relativistic HIC, the fireball evolves rapidly (it
takes about $10^{-22}$ s for hadronization), only processes mediated by the
strong interaction will attain thermal equilibration rather than the full
electroweak equilibrium.
In this work, we impose the condition $\mu_u= \mu_d = \mu_s = \mu_q$
(equal quark chemical potentials) with $\mu_B=3\mu_q$.
This~corresponds to zero charge (or isospin), $\mu_Q=0$, and zero strangeness
chemical potential, $\mu_S = 0$.
This~choice also allows for isospin symmetry, $M_u = M_d$. The net strange
quark density, $\rho_S$, will be different from zero but only at very high
values of $T$ and/or $\mu_B$.

\begin{figure}[h!]
\centering
\begin{subfigure}{.5\textwidth}
\centering
\includegraphics[width=1.1\linewidth]{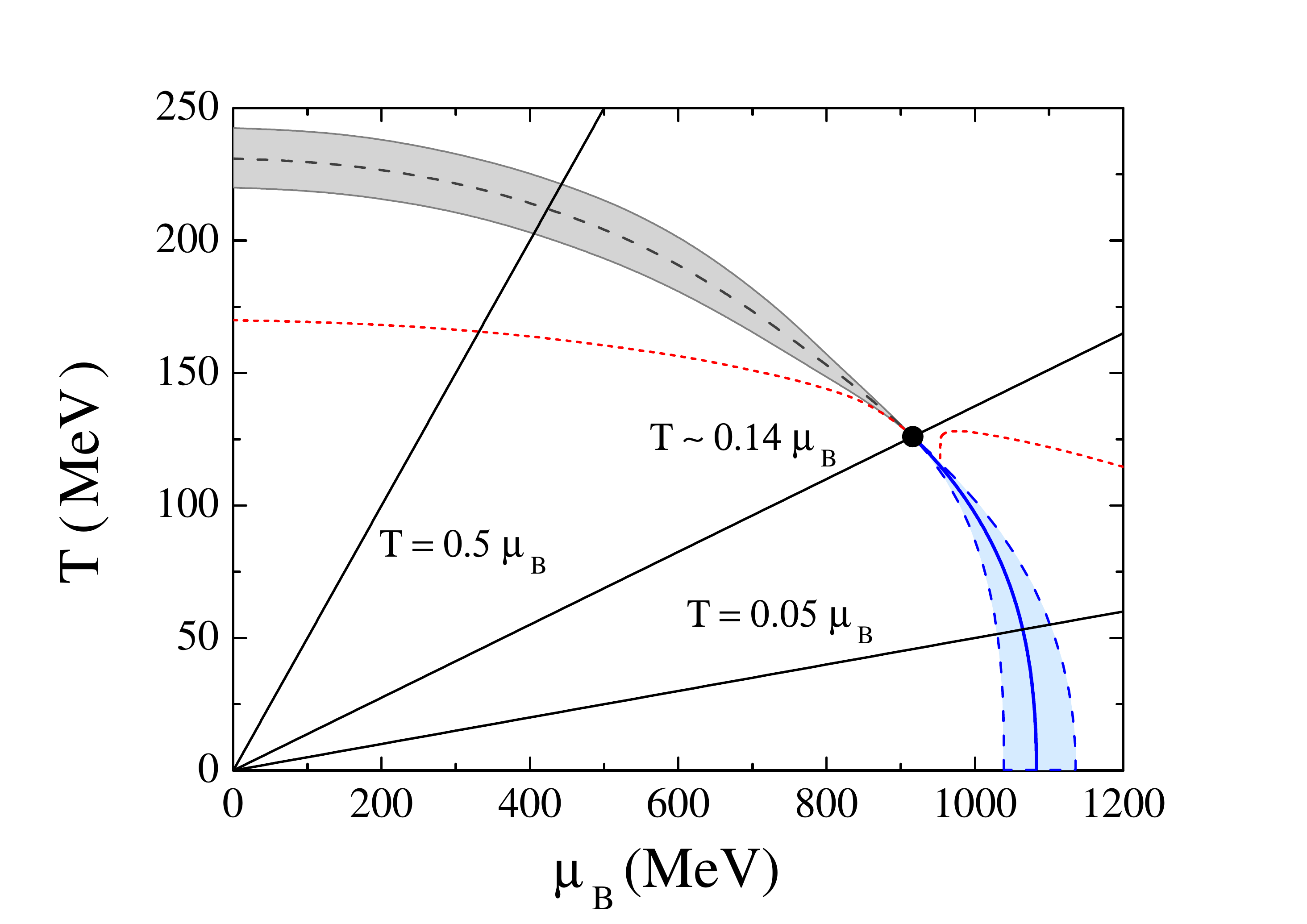}
\end{subfigure}%
\begin{subfigure}{.5\textwidth}
\centering
\includegraphics[width=1.1\linewidth]{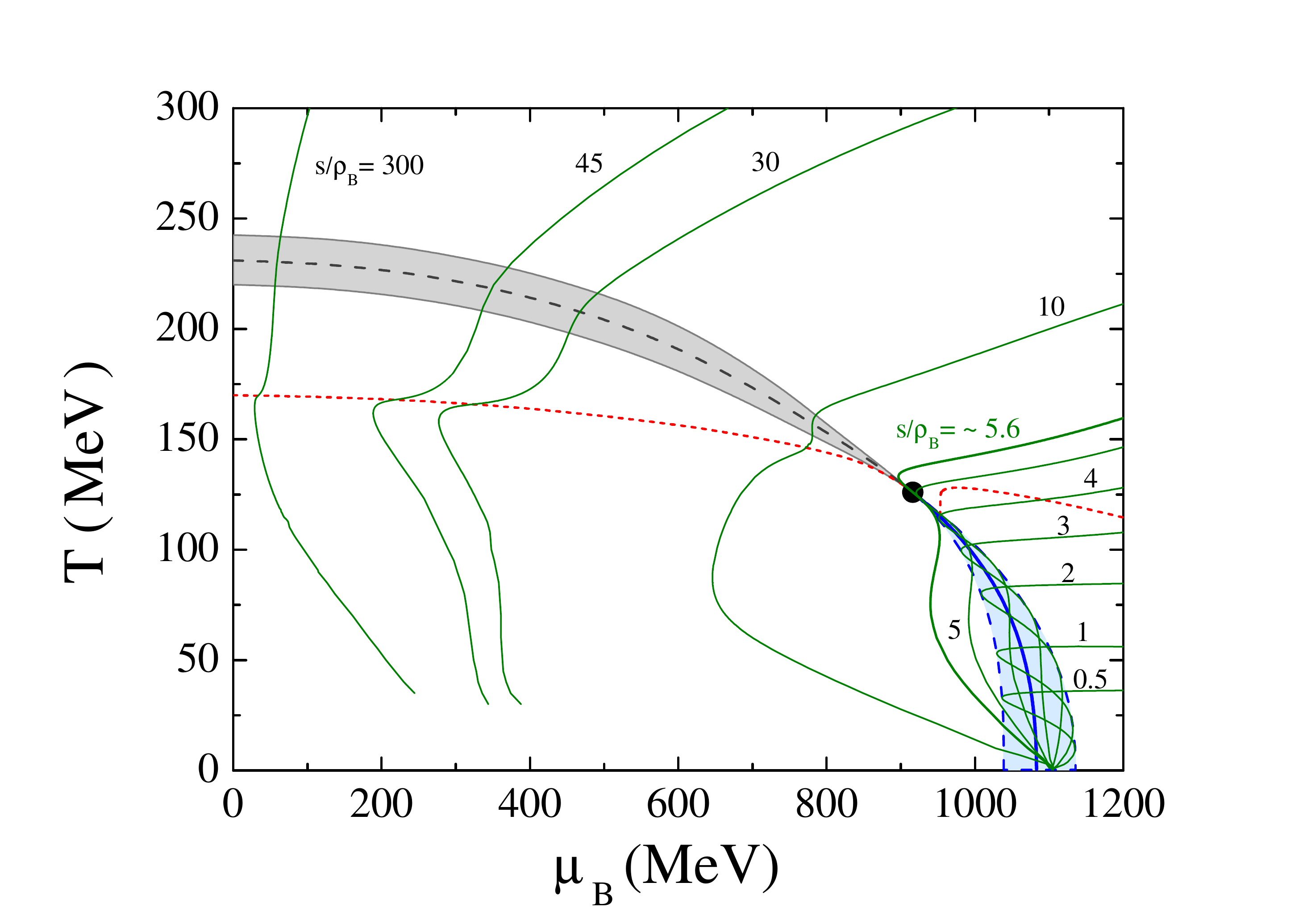}
\end{subfigure}
\vspace{-0.25cm}
\caption{
(\textbf{Left panel}) Deconfinement transition (red dots) and chiral transition with
first-order (full blue line), CEP 
(black dot) and crossover regions (dashed
black line) and spinodal lines (dashed blue line). (\textbf{Right panel}) Deconfinment and
chiral transitions alongside isentropic lines (full green lines).
}
\label{fig:PD}
\end{figure}

From both panels of Figure~\ref{fig:PD} one can see that, instead of what
happens at finite temperature, at $T=0$ and finite baryonic chemical potential a
first-order phase transition takes place with the critical chemical potential
being $\mu_B^{crit}= 1083$ MeV (or $\mu_q^{crit}= 361$ MeV). As the temperature
increases this first-order transition (blue full line) persists up to the chiral
CEP (black dot).
Along this line, the~thermodynamic potential has two degenerate minima which are
separated by a finite potential barrier. The two minima correspond to the phases
of broken and restored chiral symmetry, respectively.
As the temperature increases the height of the barrier decreases and disappears
at the CEP, which is located at $T^{CEP} = 126.3$ MeV and $\mu_B^{CEP} = 915.4$
MeV ($\rho_B^{CEP}=1.75\rho_0$). At the CEP the chiral transition becomes a
second-order phase transition.\vspace{-8pt}
\vspace{-8pt}

The blue dashed lines in Figure~\ref{fig:PD} are the borders of the coexistence
area (light blue region).
The~domain between these lines has metastable states, characterized by
large fluctuations. Although they are also solutions of the gap equations, their
thermodynamic potential is higher than for the stable solutions.
The left dashed curve is the beginning of the metastable solutions of restored
symmetry in the phase of broken symmetry, while the right dashed curve depicts
the end of the metastable solutions of broken symmetry in the restored symmetric
phase.
Above the CEP the thermodynamic potential has only one minimum, meaning that the
transition is washed out and only a smooth crossover occurs (gray dashed line).

The crossover is a fast decrease of the chiral order-like parameter (the quark
condensate) and its range is defined by the interval between the two peaks
around the zero of
${\partial^2 \left\langle \bar{q}_iq_i \right\rangle}/{\partial T^2}$.
In Figure~\ref{fig:PD} this region is presented in gray. It can be seen that, as
$\mu_B$ increases, the area where the crossover takes place is getting narrower
and narrower until it reaches the CEP.

\vspace{-0.25cm}
\subsection{Nernst Principle and Isentropic Trajectories} \label{sec:isent}

It is accepted that the expansion of the QGP in HIC is a hydrodynamic expansion
of an ideal fluid and it will nearly follow trajectories of constant entropy.
Due to the baryon number conservation, the isentropic trajectories are lines
of constant entropy per baryon, i.e., $s/\rho_B$, in the ($T,\mu_B$)-plane and
contain relevant information on the adiabatic evolution of the system.
The values of $s/\rho_B$ for AGS, SPS, and RHIC, are 30, 45, and 300,
respectively~\cite{Bluhm:2007nu}.

The numerical results for the isentropic trajectories in the $(T,\mu_B)$-plane
are given in the right panel Figure~\ref{fig:PD}. A complete study of the
behavior of isentropics trajectories in the PNJL model under the influence of a
repulsive vector interaction and by the presence of an external magnetic field
was performed in~\cite{Costa:2016vbb}.
It is important to start the discussion by analyzing the behavior of the
isentropic trajectories when $T\rightarrow 0$.
In this limit, $s \rightarrow 0$ according to the third law of thermodynamics
and, once~also $\rho_B \rightarrow 0$, it is insured the condition
$s/\rho_B\,=\,const.$.
In fact, all isentropic trajectories end at the same point ($T=0$ and
$\mu_q=\mu_B/3=367.7$ MeV) of the horizontal axes.
Since $\mu_q=M_q^{vac}> \mu_q^{crit}$, the combination
($T=0\,, \mu_q=367.7$ MeV) corresponds to the vacuum.
With the chosen set of parameters, the point where the first-order transition
occurs satisfies the condition $\mu_q^{crit}< M_q^{vac}$ for $T=0$
\cite{Buballa:2003qv}. It~also allows the existence of a strong first-order
phase transition from the vacuum solution, $M_q=M^{vac}_q$, into the partially
chiral restored phase with $M_q$ smaller the $M^{vac}_q$.
At the transition point, the total baryonic density jumps from zero to
$~2.5\rho_0$, equally carried by $u$ and $d$-quarks (the density of strange
quarks, $\rho_s$, is still zero, and the system only has $\rho_s\ne0$ when
$\mu_q>M_s$).

Close to the first-order region, for $T\neq0$, the isentropic trajectories
($s/\rho_B=1,...,5$) show the following behavior:
they come from the region of partially restored chiral symmetry reaching the
unstable region (spinodal region), delimited by the spinodal lines, going then
along with it as $T$ decreases until they reach $T = 0$.
By looking to the trajectory $s/\rho_B = 0.5$, it can be seen that the
isentropic trajectory crosses the spinodal line at
($T\approx36$ MeV, $\mu_B\approx1122$ MeV) and intersects the first-order line
twice, at ($T\approx35$ MeV, $\mu_B\approx1075$ MeV) and
($T\approx25$ MeV, $\mu_q\approx1080$ MeV), as the temperature decreases in a
``zigzag''-shaped trajectory delimited by the spinodal lines.
By analyzing the isentropic trajectory with $s/\rho_B = 5$, it is interesting
to notice that it starts by having an identical behavior to the isentropic
trajectory $s/\rho_B = 0.5$, but it then leaves the spinodal region.
Nonetheless, as~the temperature diminishes, the isentropic trajectory reaches
once again the spinodal region, now from lower values of $\mu_B$. After that,
it also goes to the horizontal axes ($T=0$).
The isentrpoic trajectory $s/\rho_B = 5.6$ nearly goes through the CEP and we
will calculate the meson masses along this path.

Concerning the behavior of the isentropic trajectories in the chiral crossover
region, ($s/\rho_B > 5.6$) it is qualitatively similar to the one obtained in
lattice calculations~\cite{Ejiri:2005uv,Borsanyi:2012cr} or in some models
\cite{Nonaka:2004pg,Fukushima:2009dx,Kahara:2008yg}: they directly go through
the the crossover region, displaying a smooth behavior. However, they suffer a
pronounced bend when they cross the deconfinement transition (red dashed curve)
reaching the spinodal region from lower values of $\mu_B$.
In conclusion, all the trajectories directly arrive in the same point of the
horizontal axes at $T=0$.

A final remark; while the critical behavior at the CEP is universal,
characteristic shape of the isentropic trajectories near the CEP can change
from model to model, even if they belong to the same universality class
\cite{Nakano:2009ps}.

\section{Scalar and Pseudoscalar Mesons in the PNJL Model}\label{sec:Mesons}

In this section, we will study the behavior of scalar and pseudoscalar mesons in
the PNJL model for different physical scenarios. We will look for signs of the
effective restoration of chiral symmetry and how the chiral transition and the
deconfinement affect their behavior. The criterion to identify an effective
restoration of the chiral symmetry will be to seek for the degeneracy of the
respective chiral partners. Indeed, we will compare the properties (e.g., the
masses) of the pseudoscalar meson nonet ($\pi$, $K$, $\eta$, and $\eta'$)
with those which can be considered as chiral partners, i.e., the scalar mesons
($\sigma$, $\kappa$, $a_0$, and $f_0$).

\subsection{Mesons Properties at Finite Temperature}\label{sec:MesonsFiniteT}

\subsubsection{Mesonic Masses and Mixing Angles }

\begin{figure}[h!]
\centering
\begin{subfigure}{.5\textwidth}
\centering
\includegraphics[width=1.1\linewidth]{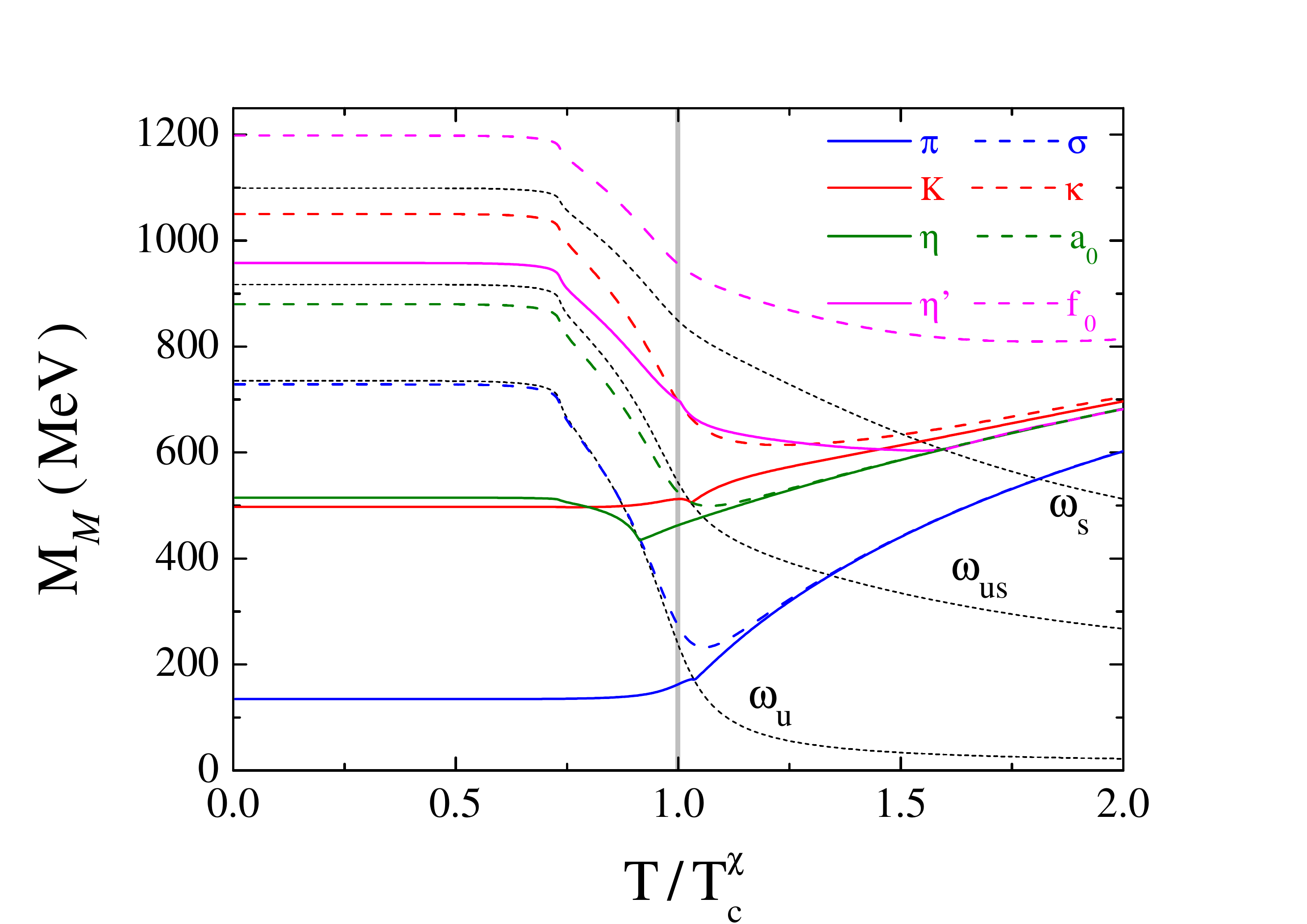}
\end{subfigure}%
\begin{subfigure}{.5\textwidth}
\centering
\includegraphics[width=1.1\linewidth]{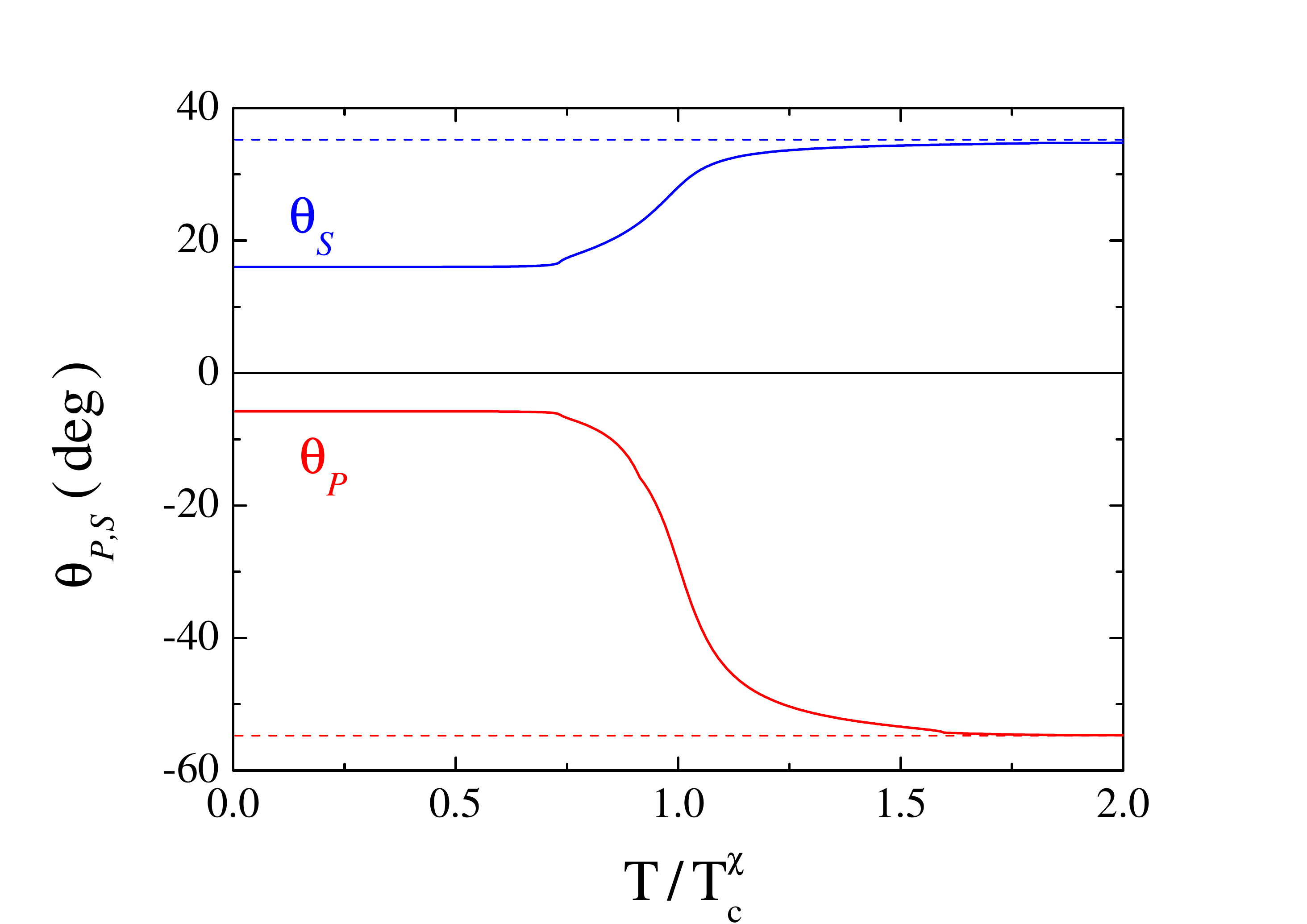}
\end{subfigure}
\vspace{-0.25cm}
\caption{
(\textbf{Left panel}) Masses of the pseudoscalar (full lines) and scalar mesons (dashed
lines) as function of the reduced temperature. (\textbf{Right panel}) Pseudoscalar and
scalar mixing angles as a function of the reduced temperature.
}
\label{fig:MesonsT}
\end{figure}

We start with the analysis of the mesons and mixing angles general behavior.
The plot of the meson masses, mixing angles and coupling constants will be
done as functions of the reduced temperature $T/\Tcc$. This will allow a better
understanding of their behavior around the chiral pseudocritical temperature.
The masses of the pseudoscalar mesons $\pi$, $K$, $\eta$, and $\eta^{\prime}$
(solid lines) and of the respective scalar chiral partners $\sigma$, $\kappa$,
$a_0$, and $f_0$ (dashed lines) as functions of the reduced temperature are
given in the left panel of Figure~\ref{fig:MesonsT}.\vspace{-8pt}
\vspace{-6pt}

Concerning the pseudoscalar mesons, they are bound states at low temperature
(except the $\eta^{\prime}$ meson that is always above the continuum
$\omega_{u}=2M_{u}$), but they become unbound at the respective Mott
temperatures (see Table \ref{table:Mott_temp}).
The Mott temperature is the temperature where the transition from a bound state
to a resonance takes place and, as already
pointed out, it comes from the fact that mesons are not elementary objects but
composed states of $\bar{q}q$ excitations.
Above the Mott temperature, the imaginary parts of the integrals $I_2^{ij}$
(see Equation~(\ref{Iij2})) must be taken into account and the finite width
approximation is used~\cite{Rehberg:1995kh}.
The lower limits of the continua belonging to each meson are also shown (black
doted lines). Indeed, the continuum starts when the $\pi$, and the $\eta$ masses
cross the quark threshold $\omega_u=2M_{u}$ and the $K$ crosses
$\omega_{us}=M_{u}+M_{s}$.
For the $\pi$ and $K$ mesons this entry into the continuum occurs at
approximately the same temperature ($T^{Mott}_{\pi}=239$ MeV,
$T^{Mott}_{K}=242$ MeV). The~$\eta'$ meson is always an unbound state and its
mass already begins to be larger than $\omega_u=2M_{u}$.

Concerning the scalar mesons, the $\sigma$-meson is the only scalar meson that
can be considered as a true (slightly) bound state for relatively small
temperatures.  At the corresponding Mott temperature, $T^{Mott}_{\sigma}=197$
MeV, it turns into a resonance. The other scalar mesons are always resonant
states.
For~the $\pi$, $\eta$ and $\sigma$ there is a second entry into the continuum
when the mass of these mesons intersect $\omega_s=2M_s$.

As shown in Ref.~\cite{Costa:2009ae}, at $T^{\chi}_{eff}$ the behavior of some
observables signalize the so-called effective restoration of chiral: the masses
of the meson chiral partners become degenerated as it can be see in Figure
\ref{fig:MesonsT} (left panel). In this case, for temperatures above $280$ MeV
the $\pi$ starts to be degenerate with the $\sigma$ meson. It also can be seen
that the partners [$\pi,\sigma$] (blue curves) and [$\eta,a_0$] (green curves)
become degenerate at almost the same temperature. In both cases, this behavior
is a clear indication of the effective restoration of chiral symmetry in the
nonstrange sector.
Differently, the masses of $\eta^{\prime}$ and $f_0$ mesons do not show a
tendency to converge in the range of temperatures considered. This pattern can
be interpreted as a sign that chiral symmetry does not show a trend to
get restored in the strange sector, a consequence of the slow decrease of $M_s$
in Figure~\ref{fig:Cond_Mquark} (right panel).

Finally, our attention will be focused in the $K$ and $\kappa$ mesons (red
curves in the left panel of Figure~\ref{fig:MesonsT}). The $\kappa$ is always
unbounded and it exhibits a tendency to get degenerate in mass with the $K$
meson for increasing temperatures.

Summarizing the above, the SU(2) chiral partners [$\pi,\sigma$] and [$\eta,a_0$]
become degenerate for temperatures higher than $T>280$ MeV (the masses of the
$\sigma$ and $\eta$ mesons become less strange, and converge, respectively,
with the non strange ones, $\pi$ and $a_0$); the [$\eta^{\prime},f_0$] do not
indicate a tendency to converge in the rage of temperatures studied;
the convergence of the partners $K$ and $\kappa$ that have a $\bar{u}s$
structure, occurs at higher values of the temperature, and is presumably slowed
down by the small decrease of the strange quark mass, $M_s$.
The degree of restoration of chiral symmetry in the different sectors
essentially drives the mesonic behavior.

It is important to refer that, as pointed out in~\cite{Costa:2008dp}, the
behavior of the mesonic masses in the PNJL model is qualitatively similar to the
corresponding one in the NJL model~\cite{Costa:2003uu,Costa:2002gk,Costa:2005cz}.

Concerning the pseudoscalar mixing angle, $\theta_{P}$, a first evidence
emerges from the right panel of Figure~\ref{fig:MesonsT}: as the temperature
increases, it goes to its ideal value $\theta_{P}=$ $-54.736^{\circ}$
asymptotically. As a result, a remarkable change in the quark content of the
mesons $\eta$ and $\eta'$ occurs, although a small percentage of mixing always
remains: the $\eta$ eventually becomes almost nonstrange, while the $\eta'$
becomes almost a purely strange meson~\cite{Costa:2002gk,Costa:2005cz}.
The scalar mixing angle $\theta_S$ shows an identical tendency: the ideal mixing
angle goes asymptotically to $\theta_{S}=$ $35.264^{\circ}$. Therefore, there is
a decrease of the strange component of the $\sigma$-meson, that never disappears,
and $f_0$ becomes almost purely strange.

The mixing angles are very sensitive to temperature (and also the medium)
effects, particularly due to its influence on the mass of strange quark.
This might be an explanation to the fact that some aspects of $\theta_S$ and
$\theta_P$  behavior are not the same in different models
\cite{Schaefer:2008hk,Horvatic:2007qs}.
It should be noticed that the mixing angles depend on the mesons masses,
namely, $\theta_S$ on the mass of the $\sigma$-meson and $\theta_P$ on the mass
of the $\eta$-meson.

The evolution of the strangeness content of both mesons, $\eta$ and
$\eta'$ determines which meson will become nonstrange, and consequently
will behave as the chiral axial partner of the $\pi$.
The behavior of $\theta_P$ at finite temperature leads to the identification
of the $\eta$ as the chiral axial partner of the $\pi$, but~the opposite is
found in References~\cite{Lenaghan:2000ey,Schaefer:2008hk} where a crossing of the
mixing angles can be seen.
A~certain degree of crossing of the pseudoscalar mixing angle and exchange of
identities of $\eta,\, \eta^\prime$ was also seen in
Reference~\cite{SchaffnerBielich:1999uj}, and an alike effect for the scalar angle
was presented in  Ref.~\cite{Schaefer:2008hk}.
Indeed, because of the mixing angles behavior, $f_0$ and $\eta'$
become essentially strange for increasing temperatures. Besides, as it was shown
in Reference~\cite{Costa:2007fy}, even when the spontaneously broken chiral symmetry
gets restored in all sectors (and unlike what is found for the nonstrange chiral
partners) a considerable difference between the masses of these mesons is still
present, a fact due to the high value of the current strange quark mass
used in this work ($m_s=140.7$ MeV).
Indeed, at high temperatures the relation
$m_{f_0}^2\simeq m_{\eta^\prime}^2 + 4 m_s^2$ is approximately valid, thus
explaining the observed behavior.

\vspace{-6pt}
\subsubsection{Pion and Kaon Coupling Constants}

\begin{figure}[h!]
\centering
\begin{subfigure}{.5\textwidth}
\centering
\includegraphics[width=1.15\linewidth]{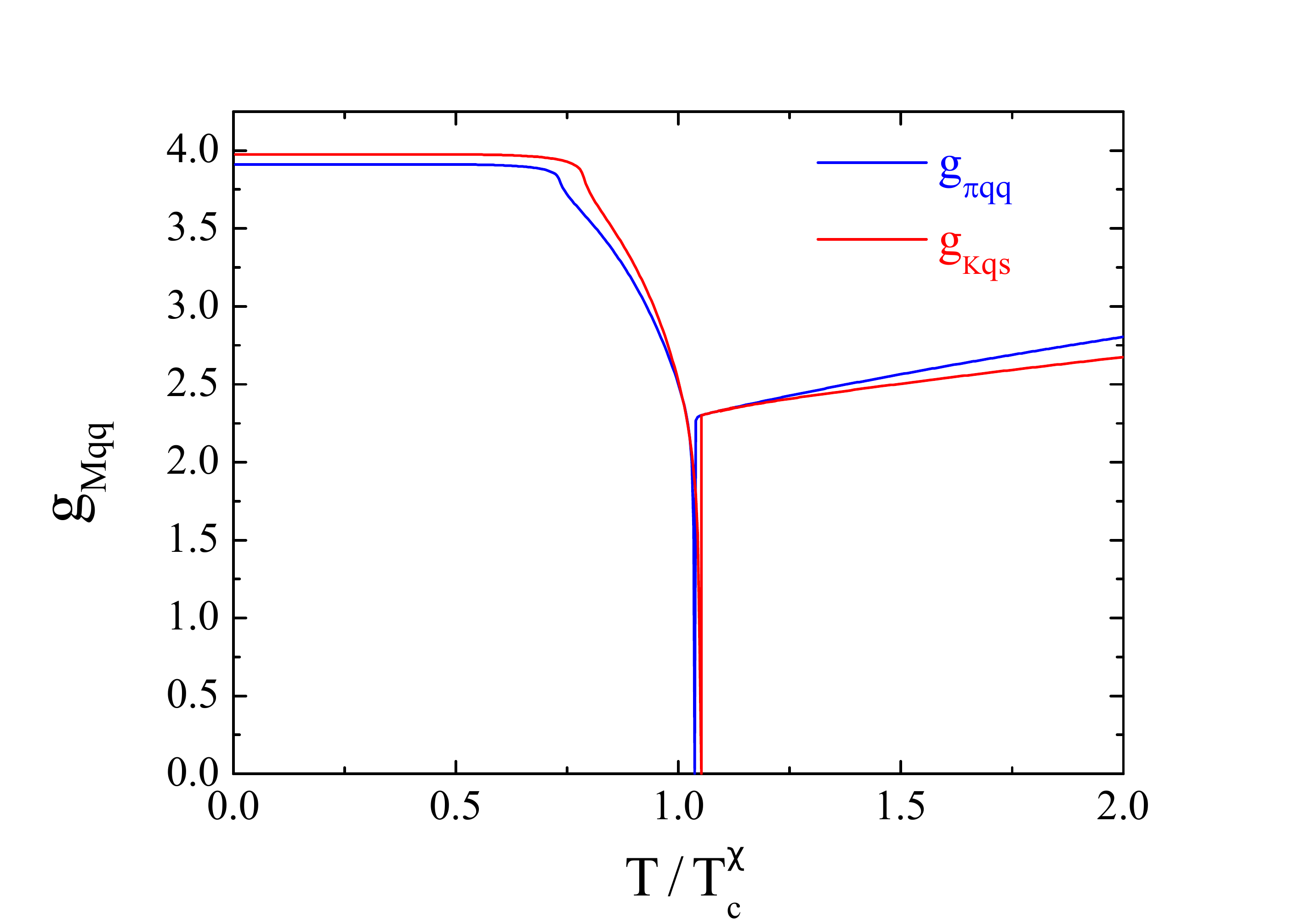}
\end{subfigure}%
\begin{subfigure}{.5\textwidth}
\centering
\includegraphics[width=1.15\linewidth]{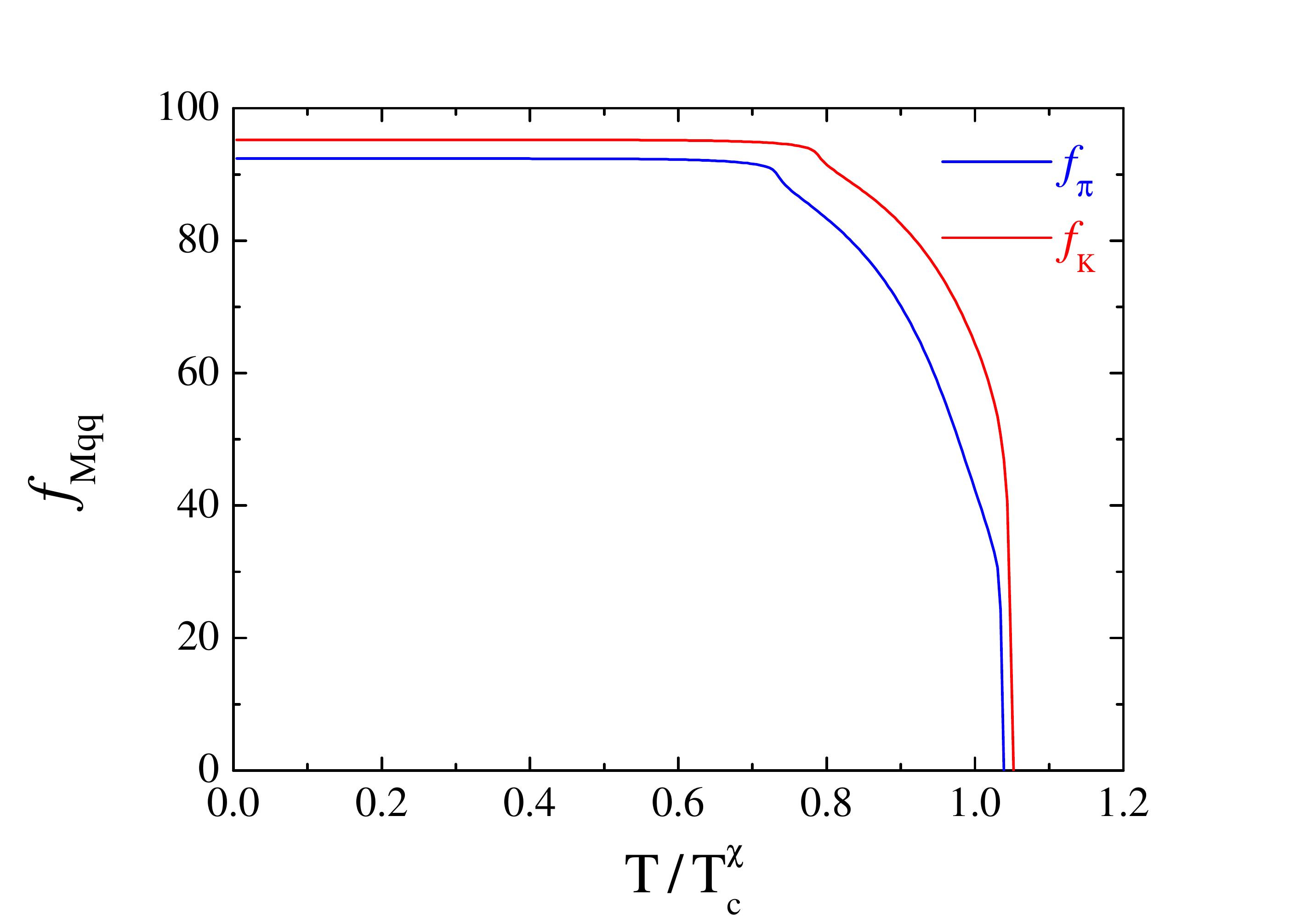}
\end{subfigure}
\vspace{-0.25cm}
\caption{
Coupling constants (\textbf{left panel}) and decay constants 
(\textbf{right panel}) as a function of the reduced temperature.}
\label{fig:Acop_DecT}
\end{figure}

The left(right) panel of Figure~\ref{fig:Acop_DecT} shows the values of the
$\pi$ and $K$ coupling(decay) constants. At~the Mott temperature, a striking
behavior for each meson can be seen: the coupling strengths approach zero for
$T\rightarrow T_{\pi(K)}^{Mott}$~\cite{Rehberg:1995kh}. This behavior occurs
because the polarization displays a kink singularity, which can also be seen
in the meson masses.
\vspace{-12pt}

It is interesting to note that the Mott temperature for each meson is above
$\Tcc$, indicating a slightly survival of these mesons as bound states in
the restored phase. This is a feature of the PNJL model, which is a
quantitative step toward confinement regarding the NJL model.
This is due to the $\Phi$ factor that suppresses the 1- and 2- quarks Boltzmann
factor at low temperatures. Indeed, the fast restoration of the $\Z_3$ symmetry
($\Phi$ goes to one with increasing temperatures) produces a quark thermal bath
with all (1-, 2- and 3-) quark contributions in a short range of temperatures,
which might help to explain the fastening of the transition~\cite{Costa:2008dp}.

\subsection{Mesons at Zero Temperature}

Next, we will study the meson behavior at finite values of $\mu_B$. We emphasize
that, as previously stated, the employed PNJL model at $T=0$ is identical to the
usual NJL model.

\begin{figure}[h!]
\centering
\begin{subfigure}{.5\textwidth}
\centering
\includegraphics[width=1\linewidth]{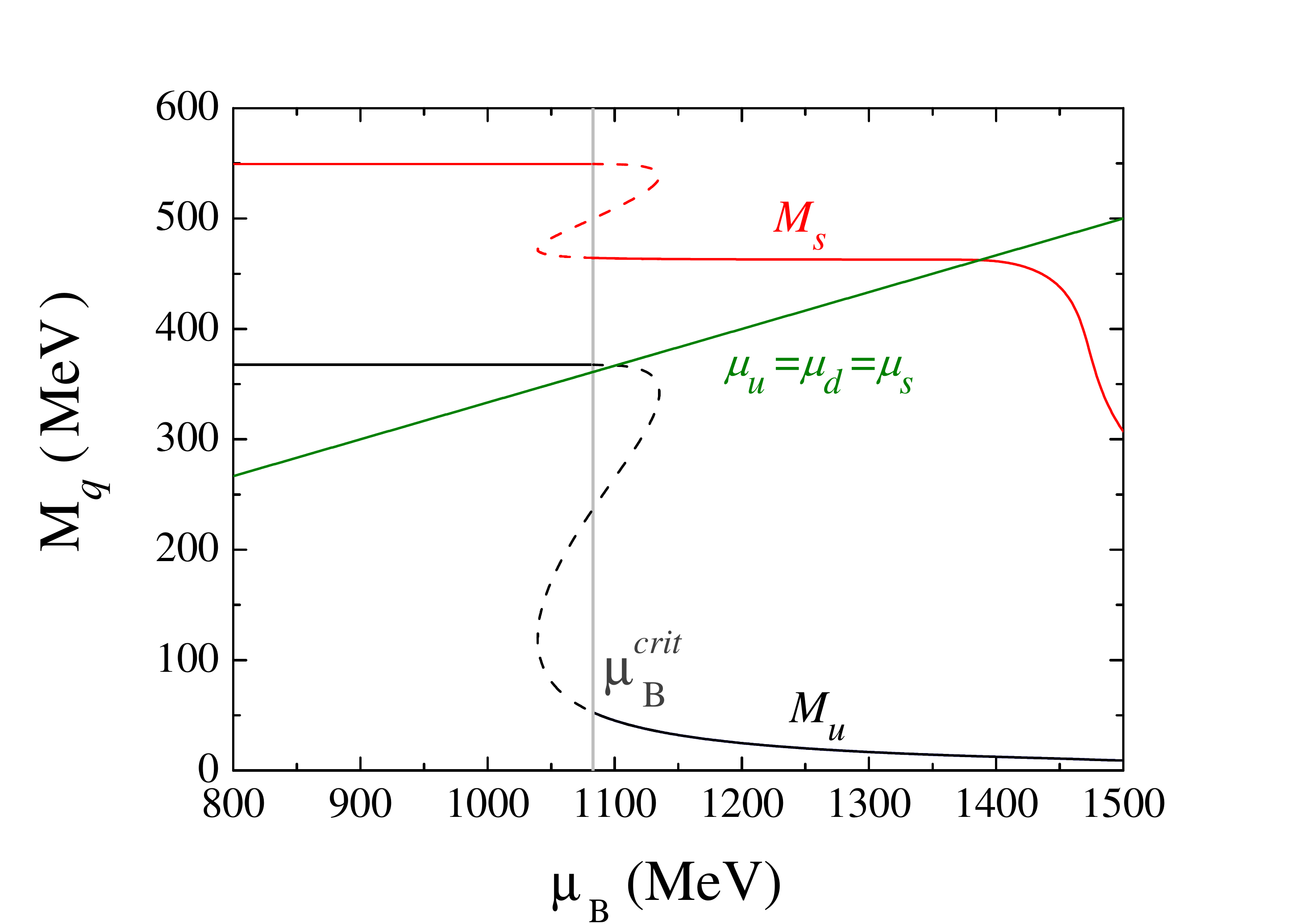}
\end{subfigure}%
\begin{subfigure}{.5\textwidth}
\centering
\includegraphics[width=1\linewidth]{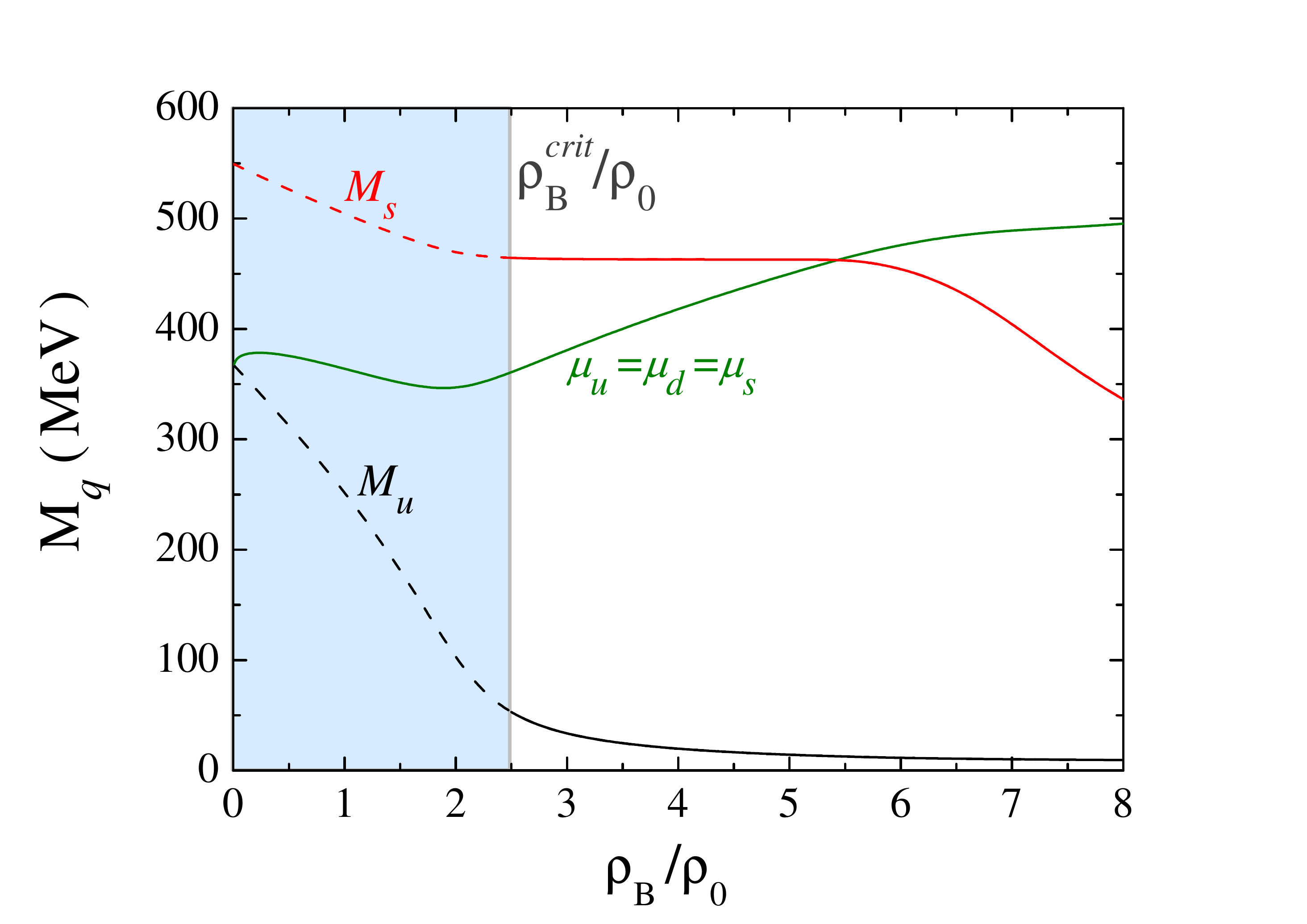}
\end{subfigure}
\vspace{-0.25cm}
\caption{
Quark masses at $T=0$ as a function of the baryonic chemical potential
(\textbf{left panel}) and reduced baryonic density (\textbf{right panel}).
The dashed lines represent the quark masses in the first-order region.
$\mu_B^{crit}$ is the baryonic critical chemical potential of the first-order
phase transition, and $\rho_B^{crit}/\rho_0$ the respective critical reduced
density ($\rho_0=0.16$ fm$^{-3}$). }
\label{fig:Quarks_dens}
\end{figure}

We consider symmetric quark matter as in Section \ref{sec:FiniteTdens}.
The first aspect that arises is that, similarly to the previous finite
temperature scenario, chiral symmetry is effectively restored only in the light
sector. This is due to some specific details of the behavior of the strange
quark mass with the density. Figure~\ref{fig:Quarks_dens} shows the quark masses
as function of $\mu_B$ around the first-order transition, left panel, and as a
function of $\rho_B/\rho_0$, right panel (the light-blue area corresponds to
the region of the phase transition). As it can be seen, in the present case
there are no strange quarks in the medium at low densities. The~mass of the
strange quark smoothly decreases, due to the  effect of the 't Hooft interaction,
and it becomes smaller than the chemical potential for strange quarks
(at $\rho_B \simeq 5.45 \rho_0$ and $\mu_B \simeq 1388$ MeV) and strange
quarks appear in the system.
Then, a pronounced decrease of the strange quark mass is~observed.

With respect to the meson spectra and the mixing angles, we will discuss new
aspects that arise, mainly in the high baryonic chemical potential/density
region.

\vspace{-8pt}

\begin{figure}[h!]
\centering
\begin{subfigure}{.5\textwidth}
\centering
\includegraphics[width=1.1\linewidth]{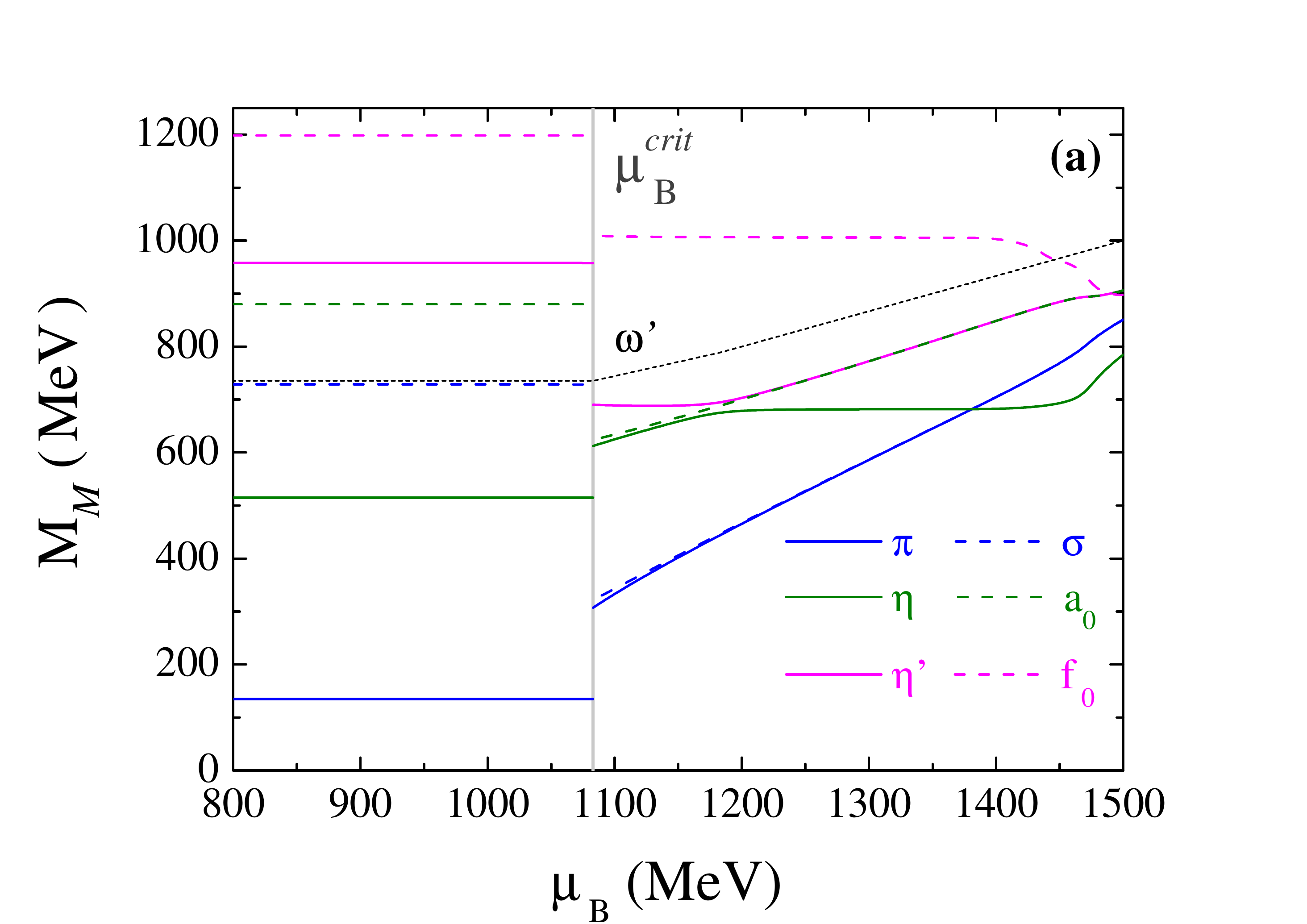}
\end{subfigure}%
\begin{subfigure}{.5\textwidth}
\centering
\includegraphics[width=1.1\linewidth]{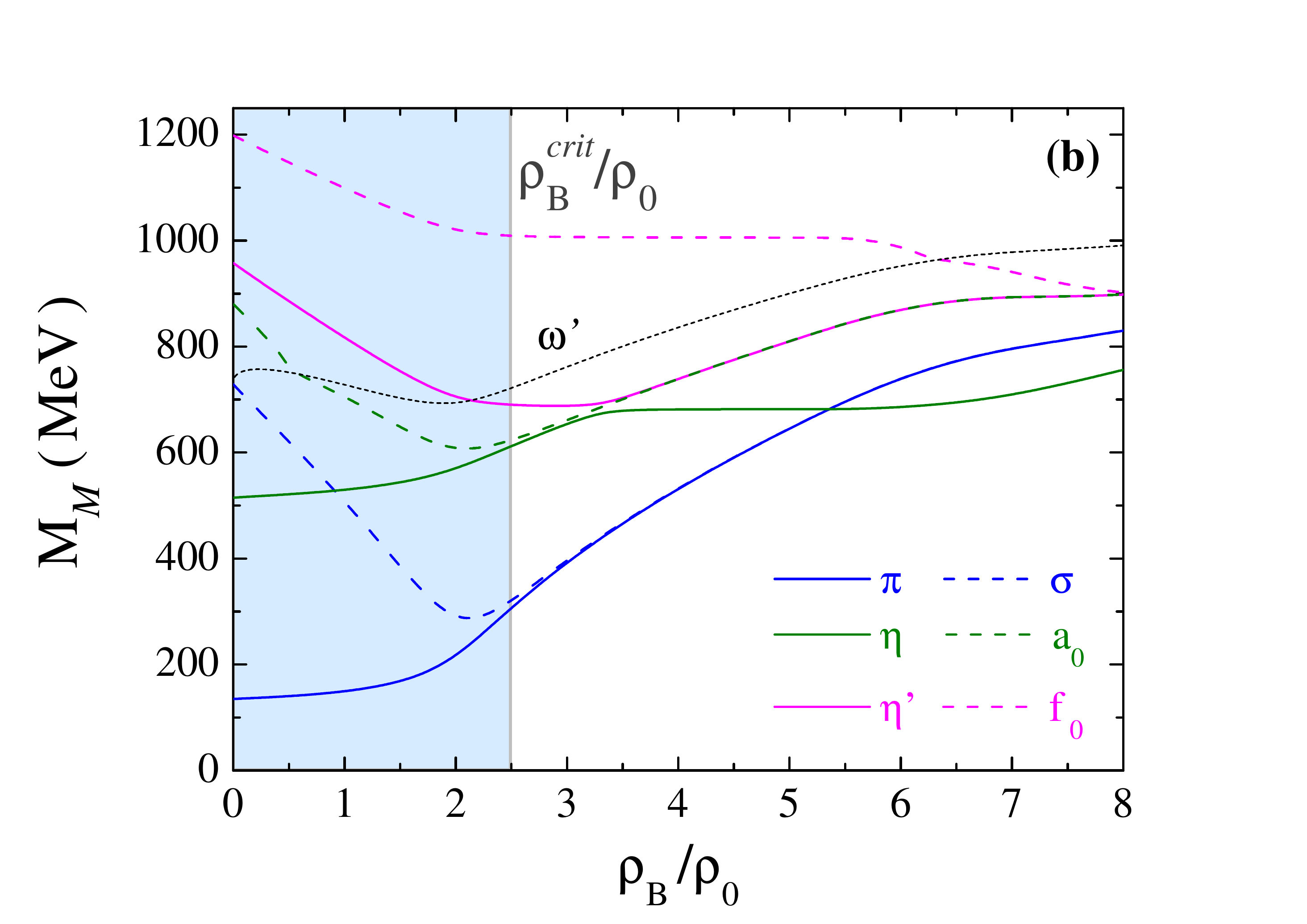}
\end{subfigure}
\begin{subfigure}{.5\textwidth}
\vspace{-0.25cm}
\centering
\includegraphics[width=1.1\linewidth]{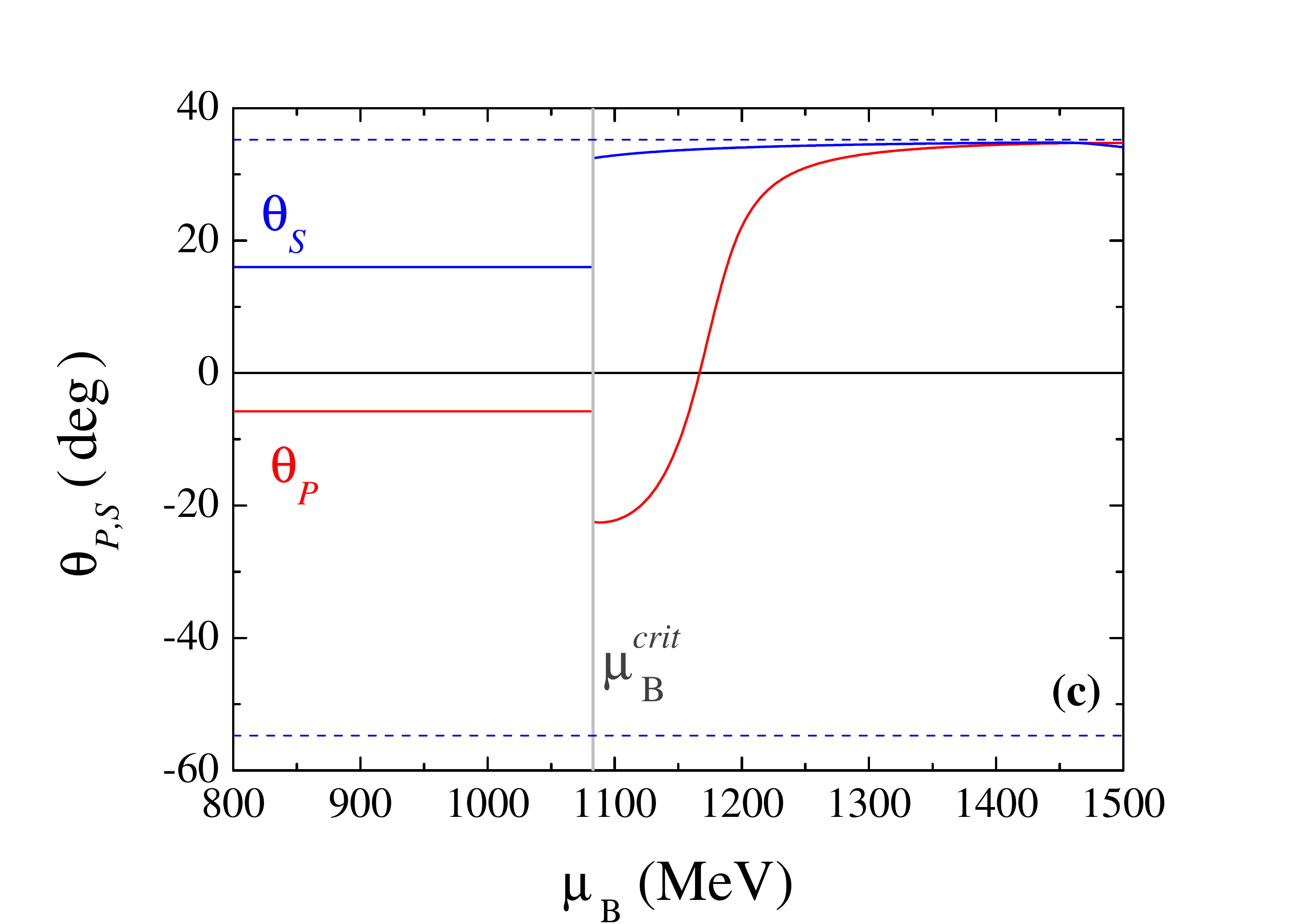}
\end{subfigure}%
\begin{subfigure}{.5\textwidth}
\centering
\vspace{-0.25cm}
\includegraphics[width=1.1\linewidth]{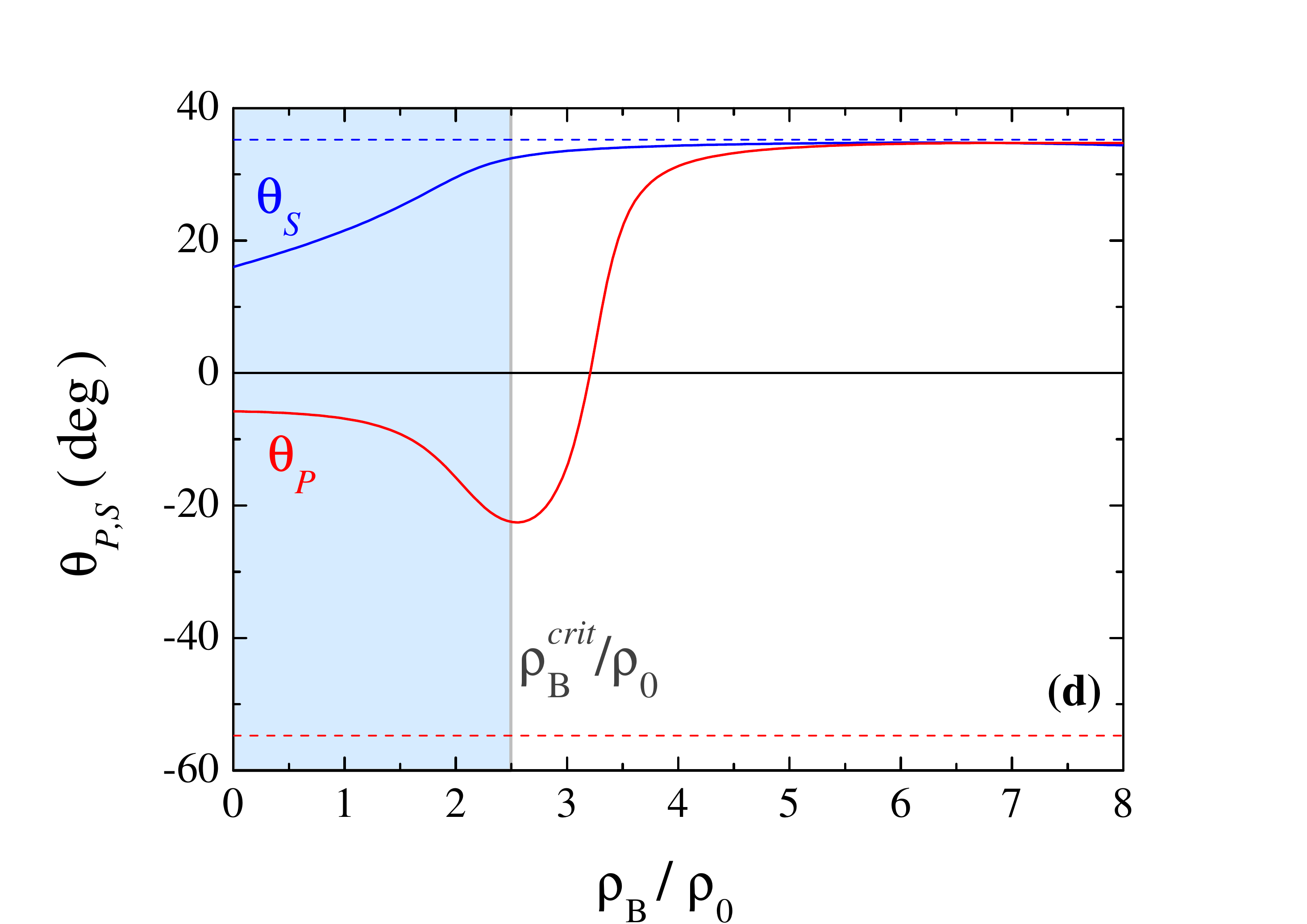}
\end{subfigure}
\begin{subfigure}{.5\textwidth}
\centering
\vspace{-0.25cm}
\includegraphics[width=1.1\linewidth]{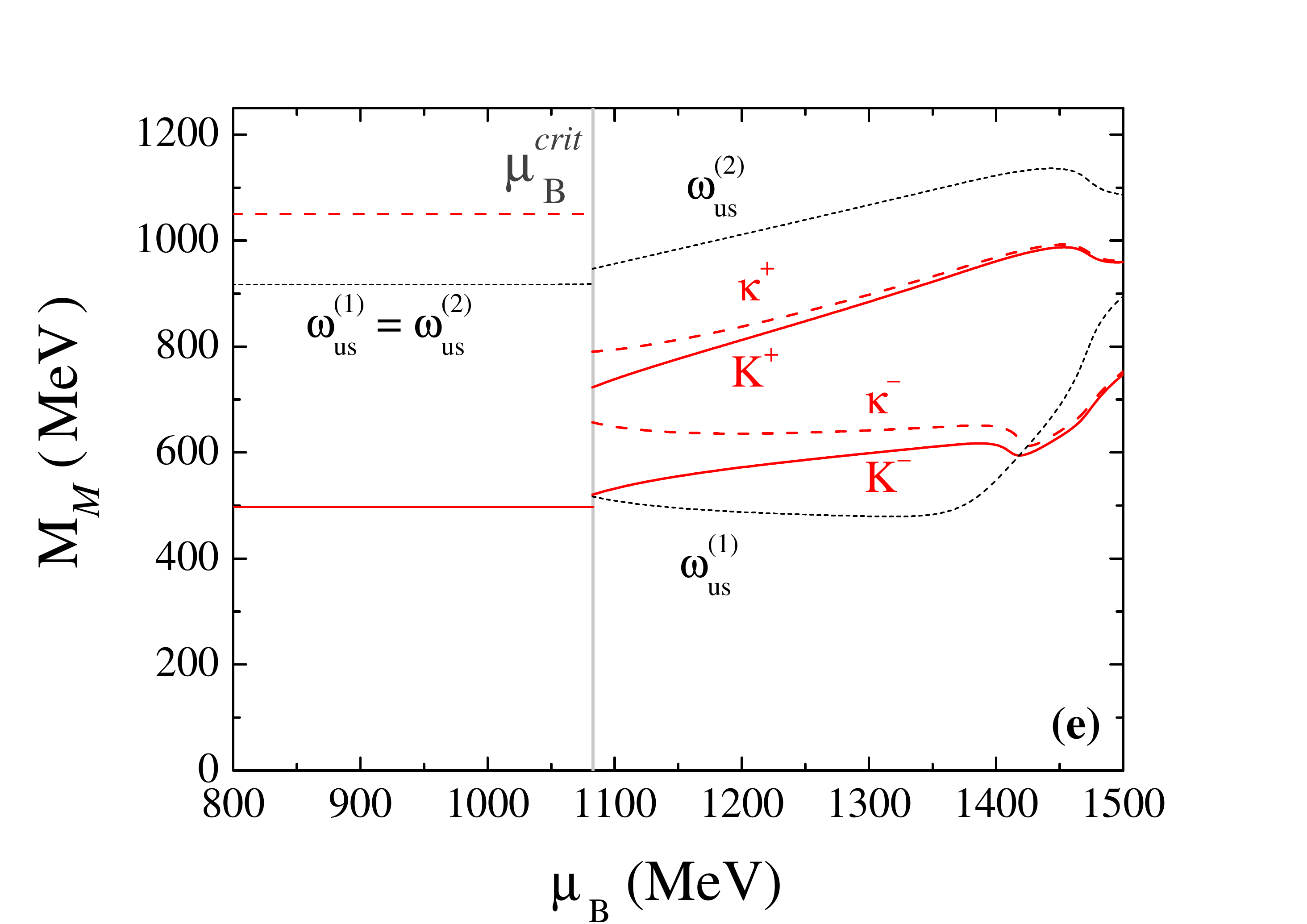}
\end{subfigure}%
\begin{subfigure}{.5\textwidth}
\centering
\vspace{-0.25cm}
\includegraphics[width=1.1\linewidth]{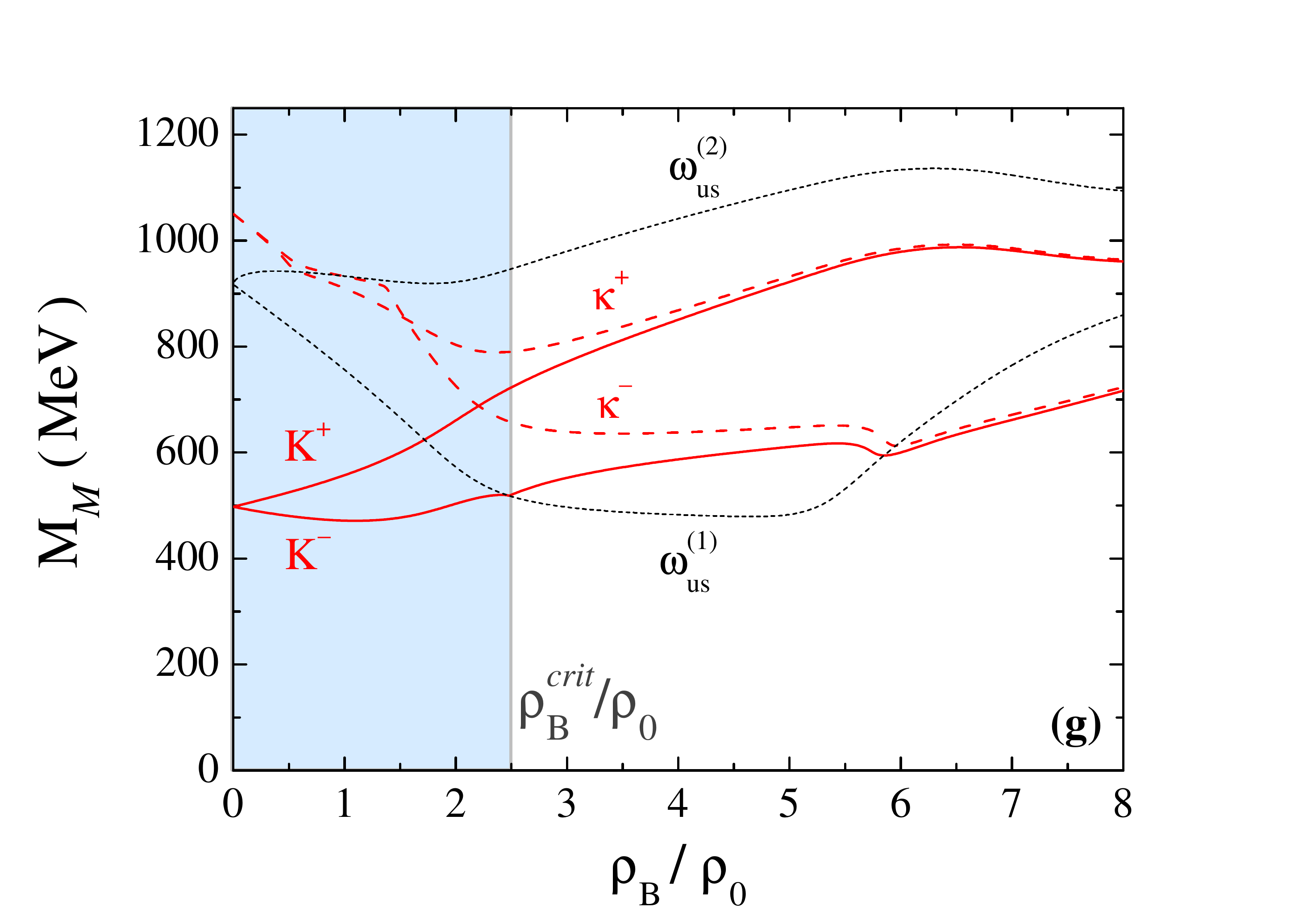}
\end{subfigure}
\vspace{-0.25cm}
\caption{
Masses of the neutral pseudoscalar and scalar mesons as a function of $\mu_B$
(panel (\textbf{a})) and as a function of $\rho_B/\rho_0$ (panel (\textbf{b})); 
the pseudoscalar and scalar mixing angles as a function of $\mu_B$
(panel (\textbf{c})) and as a function of $\rho_B/\rho_0$ (panel (\textbf{d}));
masses of the charged pseudoscalar and scalar mesons as a function of $\mu_B$
(panel (\textbf{e})) and as a function of $\rho_B/\rho_0$ (panel (\textbf{g})). 
All results are at $T=0$ (see text for~details).
}
\label{fig:Mesons_Dens}
\end{figure}

In Figure~\ref{fig:Mesons_Dens}, the meson masses are plotted as functions of
the baryonic chemical potential (panel~(a)) and of the density (panel (b)).
The chiral partners [$\pi,\sigma$] (blue curves) are always bound states.
Before~continuing the discussion, some words about the continuum
$\omega^{\prime}$ at zero temperature are needed.
From~the dispersion relations for the pion and kaons, at $T=0$, two kinds of
discrete bound states are allowed (looking to the limits of the regions of
poles in the integrals $I^{ij}_2$, Equation~(\ref{Iij2})):
particle-antiparticle modes of the Dirac sea, that are already present in the
vacuum and are related with the spontaneous breaking of chiral symmetry; and,
when the breaking of the flavor symmetry is considered, particle-hole
excitations of the Fermi sea, that only manifest themselves in the medium
(for a detailed description, see~\cite{deSousa:1997nu}). On the other hand,
the Fermi and Dirac sea continua are defined by the unperturbed solution.
In the isospin limit, the Dirac continuum starts at
$\omega^{\prime}=\omega_{u(d)}^{\prime}=2\mu_{u(d)}$ and at
$\omega_s^{\prime}=2\mu_{s}$ (at finite temperature, we have
$\omega_{u(d)} = 2M_{u(d)}$, and $\omega_{s} = 2M_{s}$).

Returning to meson masses, for all range of densities (chemical potentials), the
pion is a light quark system contrary to the $\sigma$ meson which has a strange
component at $\rho_{B}=0$ but never becomes a purely non-strange state because
$\theta_{S}$ never reaches the ideal mixing angle $35.264^{\circ}$
(Figure~\ref{fig:Mesons_Dens}, panels~(c) and (d)).
As $\rho_B$ ($\mu_B$) increases these mesons become degenerate
($\rho^{eff}_{B}\gtrsim2.9\rho_{0}$, $\mu_B^{eff}\gtrsim1126$ MeV, effective
restoration of chiral symmetry in the light sector, see blue curves of panels
(a) and (b)).

At the approximate same density, and $\mu_B$, the chiral partners [$\eta,a_{0}$]
(green curves in panels (a) and (b) of Figure~\ref{fig:Mesons_Dens}) also become
degenerate.
On one hand, the $\eta$-meson is permanently a bound state;
on the other hand, $a_{0}$ starts to be a resonance, because its mass is above
the $\bar{q}q$ continuum, and converts into a bound state for
$\rho_{B}\gtrsim0.6\rho_{0}$, and $\mu_B\gtrsim1121$ MeV.

However, as the baryonic chemical potential (panel (a)) or the density
(panel (b)) increases, the $a_{0}$ mass splits from the $\eta$ mass and goes
to degenerate with the $\eta^{\prime}$ mass. To clarify this behavior we look
to the behavior of the mixing angle $\theta_{P}$ in panels (d) and (e) of Figure
\ref{fig:Mesons_Dens}: it is noticed that the angle $\theta_{P}$, which starts
at $-5.8^{\circ}$, changes sign at $\mu_B\simeq1167$ MeV
($\rho_{B}\simeq 3.2\rho_{0}$) becoming positive and increasing rapidly, which
can be interpreted as a signal of a change of identity between both $\eta$ and
$\eta'$ mesons.
As~it was seen in Figure~\ref{fig:Quarks_dens}, the strange quark mass
decreases more rapidly when strange quarks appear in the system ($\mu_q>M_s$).
A consequence of this behavior is the changing in the percentage of strange,
$(q\bar q)_s=s\bar s$, and non-strange,
$(q\bar q)_{\rm ns}=\frac{1}{2}({u}\bar u+{d} \bar d)$, quark content in $\eta$
and $\eta'$ mesons: the $\eta^{\prime}$ has a bigger strangeness component than
the $\eta$ at low density, and the opposite takes place at high density
\cite{Costa:2002gk}.
Then, the $\eta^{\prime}$ mass will degenerate with the mass of the
$a_{0}$-meson that is permanently a non-strange state. As a final point, the
$f_{0}$-resonance is a strange state for all densities that denotes a tendency
to be degenerated with other mesons only at very high values of $\mu_B$
($\rho_B$).

A very interesting scenario happens for kaons and their chiral partners:
$K^{\pm}$ and $\kappa^{\pm}$ (Figure~\ref{fig:Quarks_dens}, panels (e) and (f)).
A first finding concerns to the separation between charge multiplets of kaons
and the respective $\kappa$-mesons. Also the the mass degeneracy of
[$K^+$, $\kappa^+$] and [$K^-$, $\kappa^-$] occur at different values of
density (chemical potential) which are higher then for [$\pi,\sigma$]. Indeed,
taking the difference ($M_{K^\pm}-M_{\kappa^\pm}$), now at $5\%$ of its vacuum
value, for [$K^+$, $\kappa^+$] the degeneracy of their masses takes place at
$\mu_B=1187$ MeV ($\rho_B=3.38\rho_0$) while for [$K^-$, $\kappa^-$] the
degeneracy takes place at $\mu_B=1419$ MeV ($\rho_B=5.86\rho_0$).

Another particularity is the fact that the $K^+$ is always a bound state, never
reaching the continuum, while the $K^-$ is bounded only until it gets in the
continuum of $s\bar{u}$ excitations of the Dirac sea at $\mu_B = 1083$ MeV
($\rho_B=2.5\rho_0$) and becomes thereafter a  $s\bar{u}$ resonance:
$M_{K^-}>\omega_{us}^{(1)}$
(with $\omega_{us}^{(1)}=(M_s^2+\lambda_u^2)^{1/2}+\mu_u$, being the chemical
potential $\mu_u=(M_u^2+\lambda_s^2)^{1/2}$, and the Fermi momentum of the
$i$-quark $\lambda_i = (\pi^2\rho_i)^{1/3}$).
However, when $\mu_B > 1419$ MeV ($\rho_B>5.85\rho_0$) the $K^-$ becomes once
again a bound state (this behavior was already found in
\cite{Costa:2001uz,Costa:2003uu}).
Still as far as the $K^-$ is concerned, it~was shown in
\cite{Ruivo:1996np,deSousa:1997nu,Costa:2001uz,Costa:2005cz} that there are
low bound states with quantum numbers of $K^{-}$ that appear below the inferior
limit of the Fermi sea continuum of particle-hole excitations being the Fermi
see bounded by $\omega_{low}=(M_s^2+\lambda_u^2)^{1/2}-\mu_u$ and
$\omega_{up}=(M_s^2+\lambda_s^2)^{1/2}-(M_u^2+\lambda_s^2)^{1/2}$. However, we
will not go further in this discussion.

Finally, with respect to the scalar mesons, $\kappa^{\pm}$, it is verified that
both mesons start to be unbounded, with the continuum now defined as
$\omega_{us}^{(2)}=\mu_s+(M_u^2+\lambda_s^2)^{1/2}$ (at $T=\rho_B=0$
$\omega_{us}^{(1)}=\omega_{us}^{(2)}$) but, as $\mu_B$ ($\rho_B$) increases,
they will turn into bound states and become degenerated with the respective
partners.

\subsection{Mesons Properties in Different Regions of the Phase Diagram}
\vspace{-4pt}

\subsubsection{Meson Masses in the Crossover Region}

\begin{figure}[h!]
\centering
\begin{subfigure}{.5\textwidth}
\centering
\includegraphics[width=1.1\linewidth]{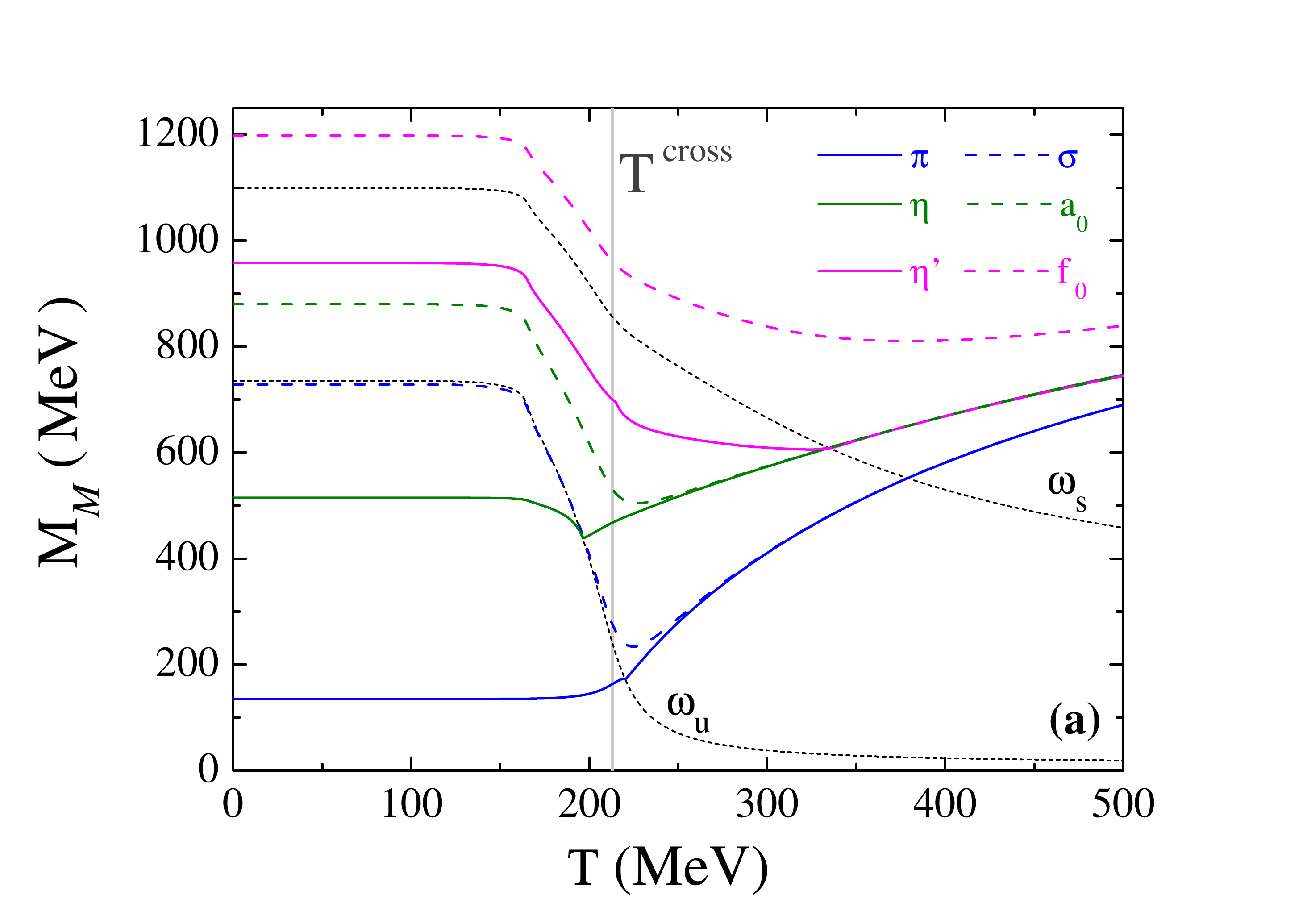}
\end{subfigure}%
\begin{subfigure}{.5\textwidth}
\centering
\includegraphics[width=1.1\linewidth]{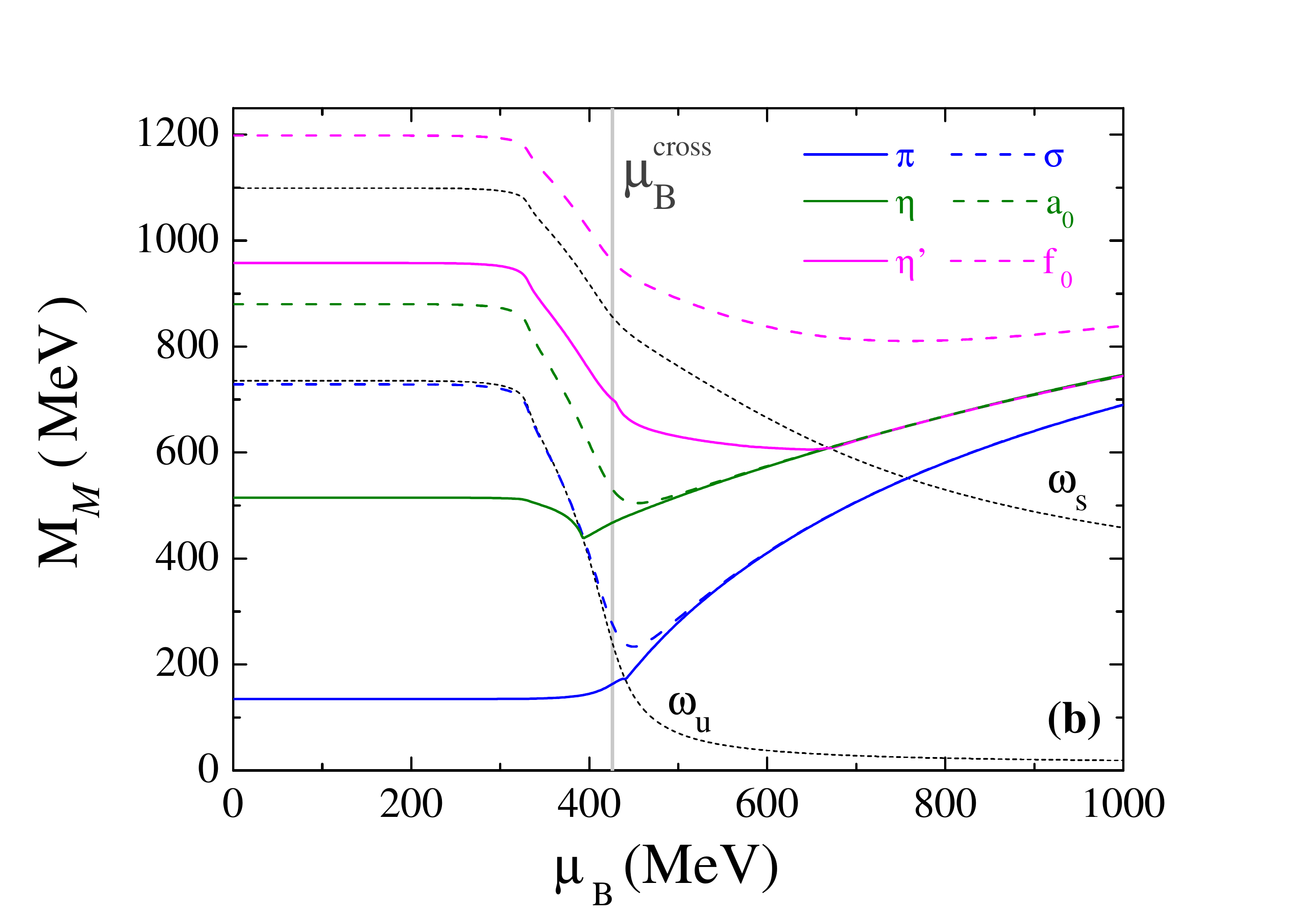}
\end{subfigure}
\begin{subfigure}{.5\textwidth}
\centering
\vspace{-0.5cm}
\includegraphics[width=1.1\linewidth]{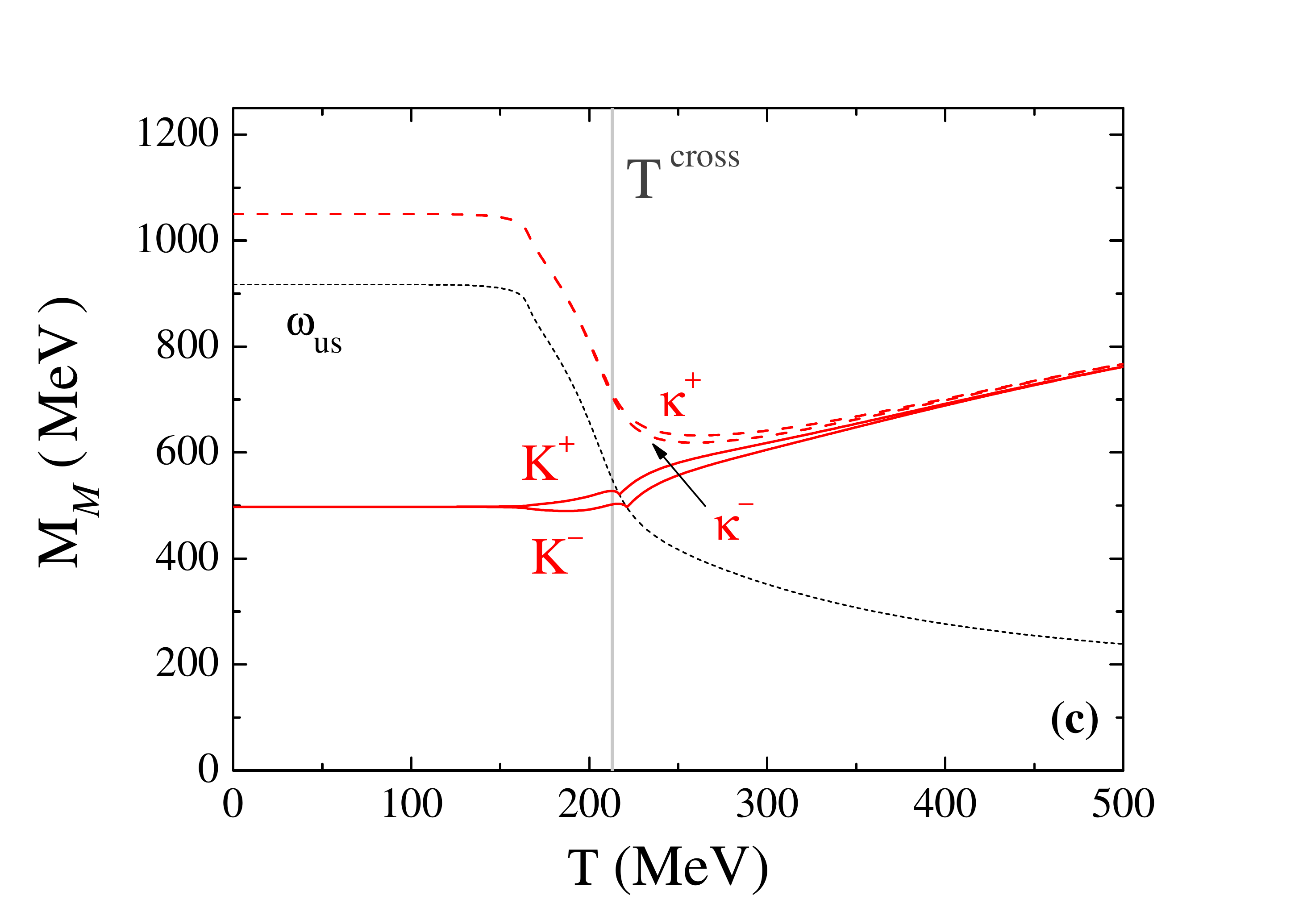}
\end{subfigure}%
\begin{subfigure}{.5\textwidth}
\centering
\vspace{-0.5cm}
\includegraphics[width=1.1\linewidth]{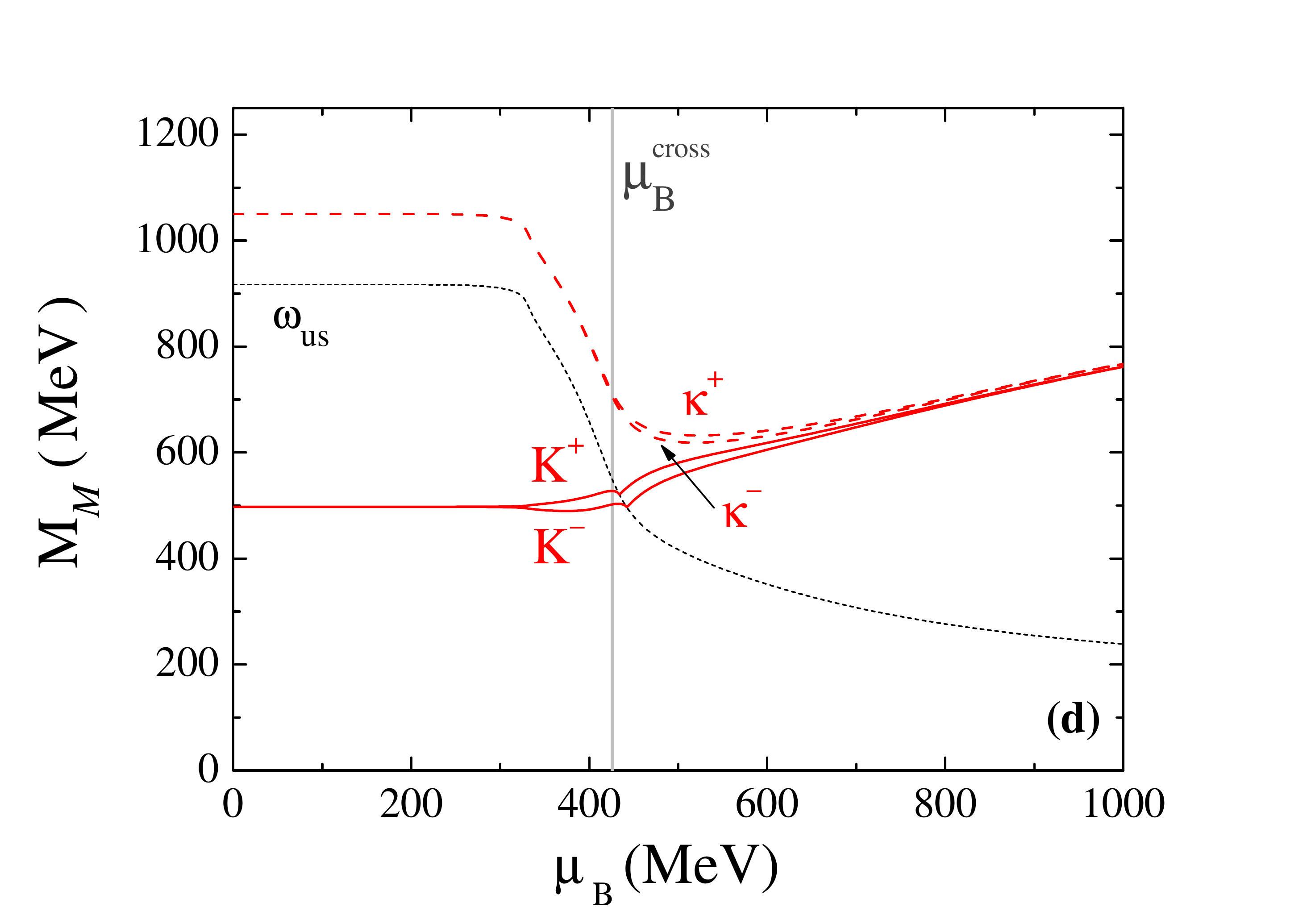}
\end{subfigure}
\vspace{-0.25cm}
\caption{Masses of the pseudoscalar and scalar mesons as function of the
temperature (panels (\textbf{a},\textbf{c})) and as function of $\mu_B$ (panels (\textbf{b},\textbf{d}))
along the path $T=0.5\mu_B$, the crossover region. \mbox{$T^{\text{cross}}=213$~MeV}
and $\mu_B^{\text{cross}}=426$ MeV are, respectively, the temperature and the
baryonic chemical potential for the~crossover.}
\label{fig:Mesons_T05muB}
\end{figure}

We continue the analysis of the behavior of mesons masses in symmetric matter,
by following a path in the ($T-\mu_B$)-plane passing through the crossover
region. As indicated in Figure~\ref{fig:PD}, we choose the path $T=0.5\mu_B$.
Chiral symmetry is, again, effectively restored only in the light quark sector.
In~Figure~\ref{fig:Mesons_T05muB} the meson masses are plotted as functions of
temperature and baryonic chemical potential, in~the left panels and right
panels, respectively.

The chiral partners [$\pi$,$\sigma$] and the $\eta$ meson are bound states up to
the respective Mott dissociation for a certain temperature and chemical
potential(density) as can be seen in panels (a) and (b)
Figure~\ref{fig:Mesons_T05muB}.
The pion still survives as a bound state after the crossover transition in a
small range of temperatures and chemical potentials.

Differently from the zero temperature case, and like the case at finite
temperature and zero density, the $a_0$ meson is always a resonance since its
mass always lies above the $\omega_u=2M_u$ continuum (see~Figure~\ref{fig:Mesons_T05muB} panel (a) and (b)).  As in the zero temperature case,
the degeneracy between each set of chiral partners, [$\pi$,$\sigma$] and
[$\eta$,$a_0$], happens at almost the same temperature and chemical potential.

The $\eta'$ meson is always a resonate state and it will reach the
$\omega_s=2M_s$ continuum, a temperature and a chemical potential where it can
decay also in a $\bar{s}s$ pair. Immediately, it degenerates with the $\eta$ and
$a_0$ mesons. Its chiral partner, $f_0$, is always above the $\omega_s=2M_s$
continuum.

Concerning the kaons and their chiral partners (Figure~\ref{fig:Mesons_T05muB},
panels (c) and (d)), the previously observed charge splitting at $T=0$ MeV is
also present (contrasting with the case $\mu_B=0$ MeV where no splitting
occurs).
However, its effect is much less visible then in the zero temperature case and
it happens, for the $K^\pm$, before the crossover while for $\kappa^\pm$ it
happens after the critical temperature and chemical potential.
Both $K^+$ and $K^-$ are bound states until they reach the $\omega_{us}=M_u+M_s$
continuum while their chiral partners, are always resonances.
The continuum is reached slightly above the critical temperature and baryonic
chemical potential of the crossover transition.

The degeneracy mechanism for these mesons in this path is different from
previously studied zero temperature case. Since the charge splitting in this
scenario is much less severe, we first observe degeneracy between each charge
multiplet $K^\pm$ and $\kappa^\pm$, happening at approximately the same
temperature and chemical potentials. Following this behavior the chiral partners
[$K$,$\kappa$] become degenerate at higher values of temperature and
baryonic chemical potential.

At high enough temperatures and chemical potentials (outside the range of
applicability of the model) it is expected that all mesons become degenerate.

\subsubsection{Mesons through the CEP}\label{MesonsCEP}

\begin{figure}[h!]
\centering
\begin{subfigure}{.5\textwidth}
\centering
\includegraphics[width=1.1\linewidth]{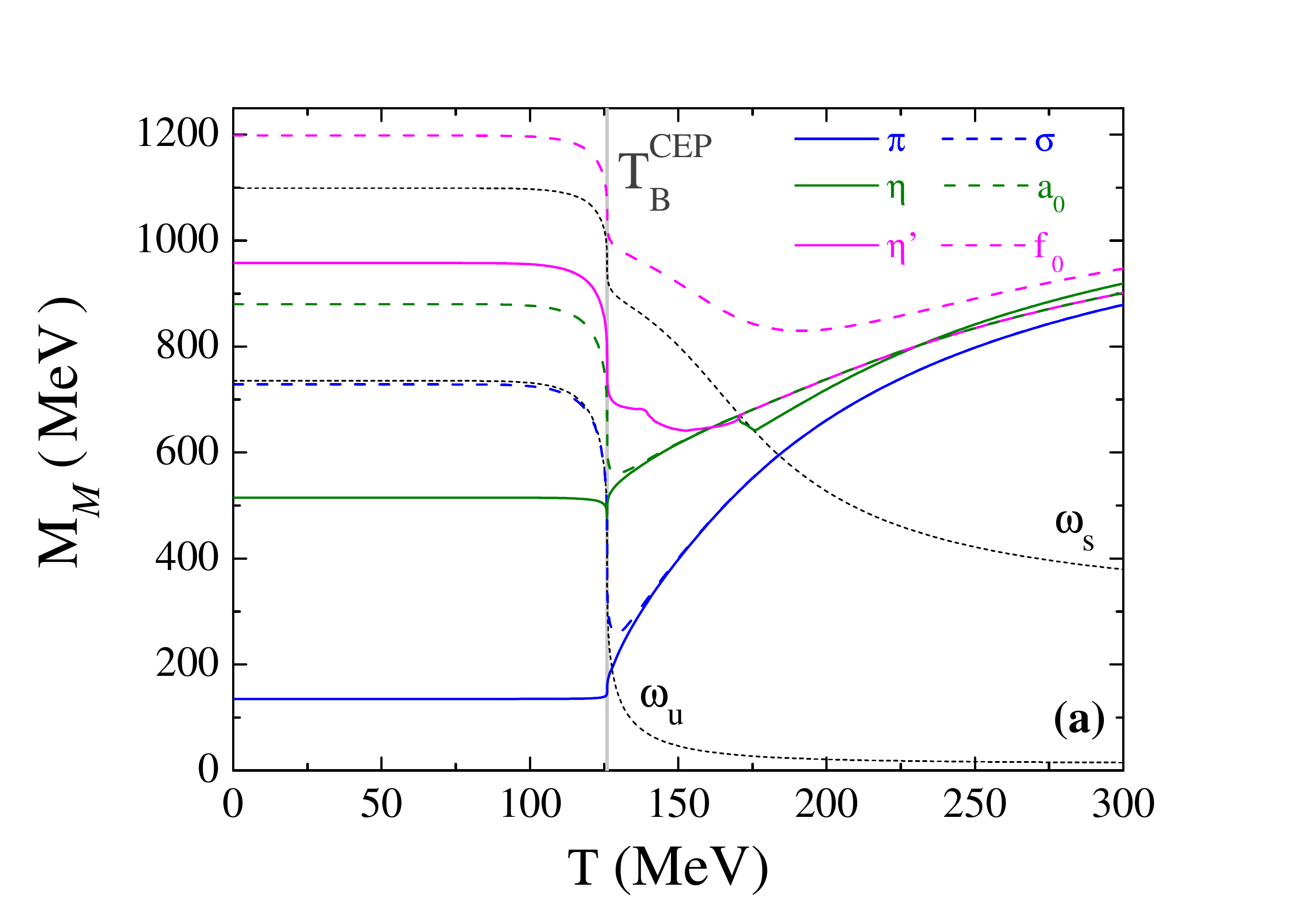}
\end{subfigure}%
\begin{subfigure}{.5\textwidth}
\centering
\includegraphics[width=1.1\linewidth]{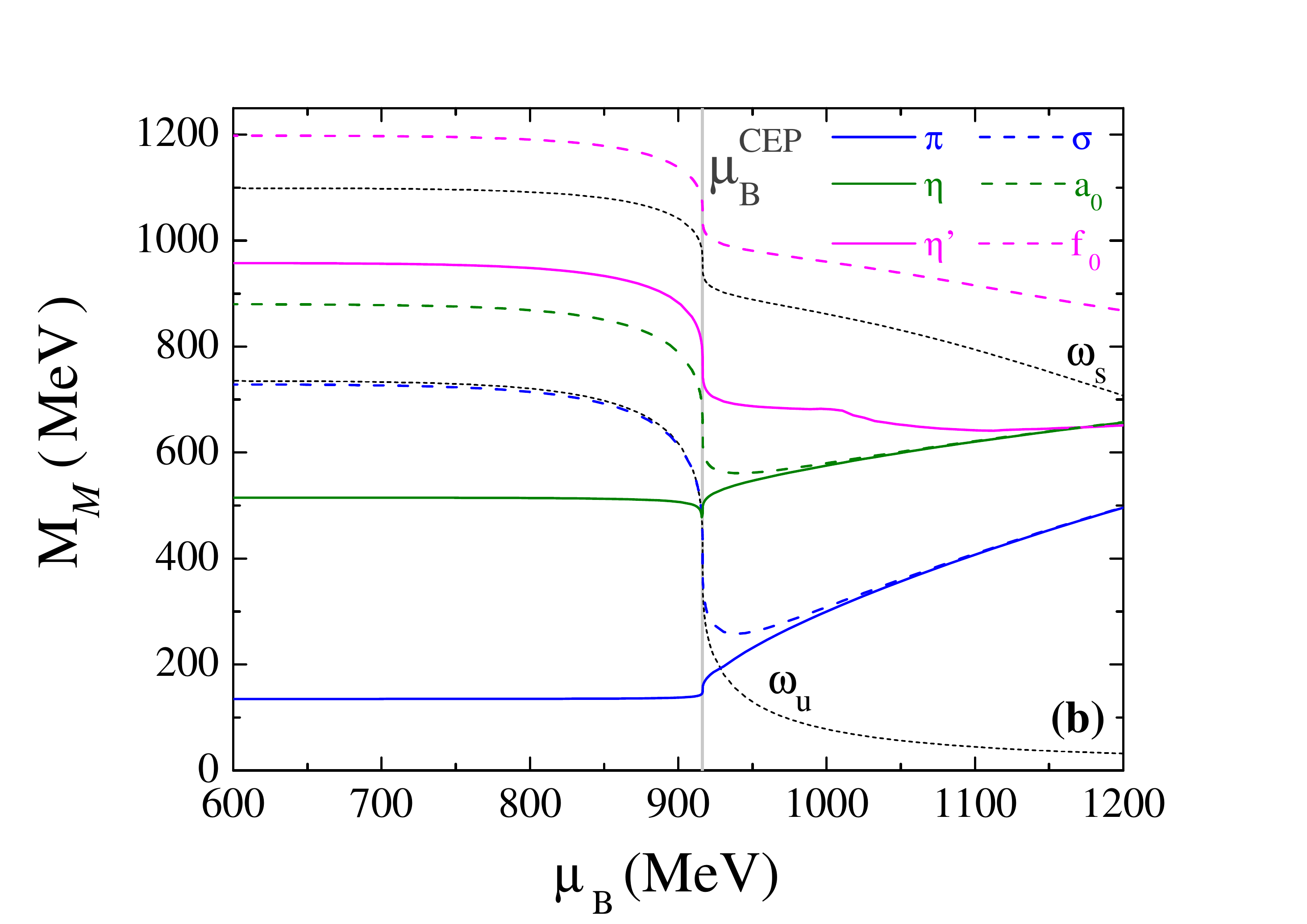}
\end{subfigure}
\begin{subfigure}{.5\textwidth}
\centering
\vspace{-0.5cm}
\includegraphics[width=1.1\linewidth]{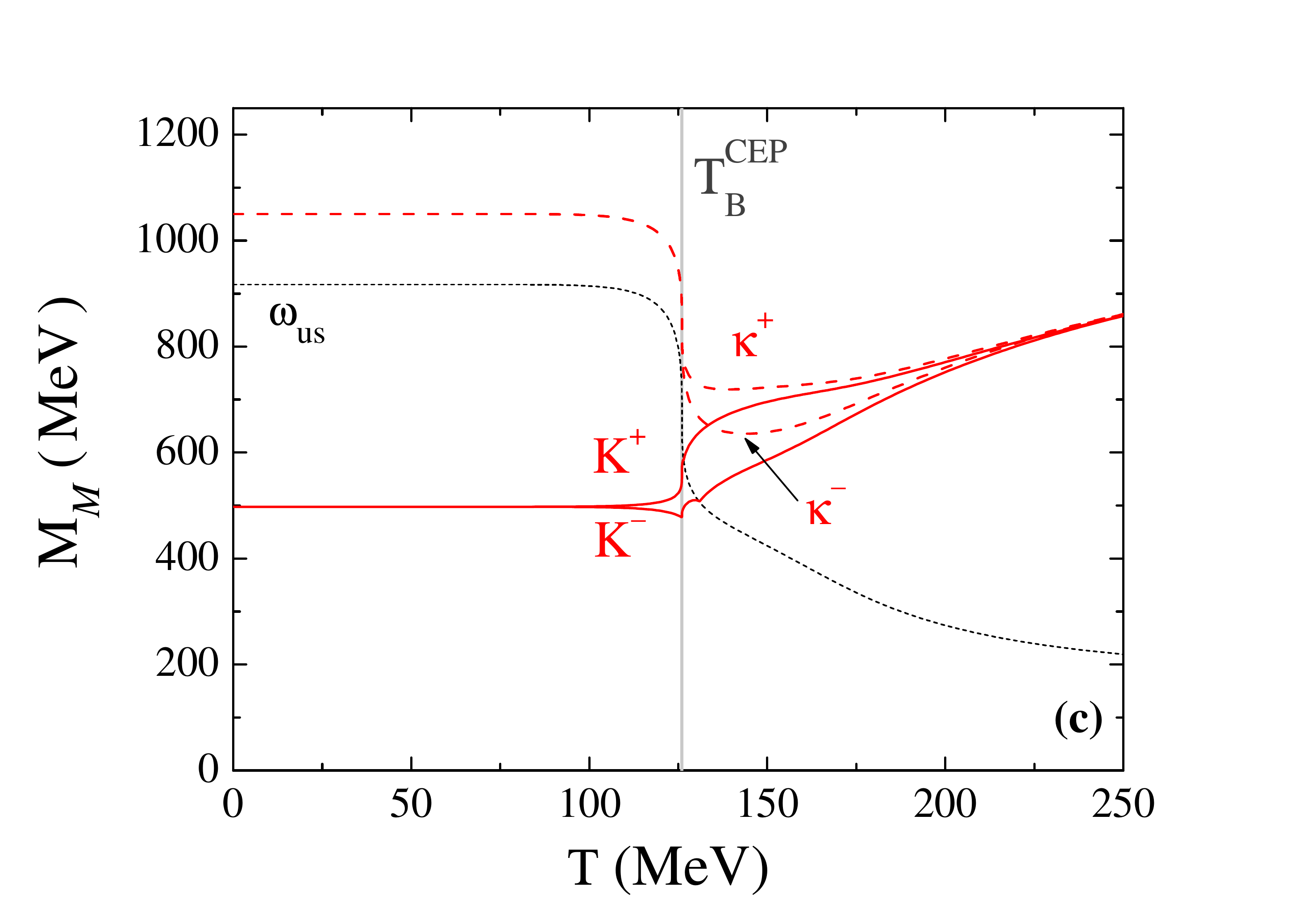}
\end{subfigure}%
\begin{subfigure}{.5\textwidth}
\centering
\vspace{-0.5cm}
\includegraphics[width=1.1\linewidth]{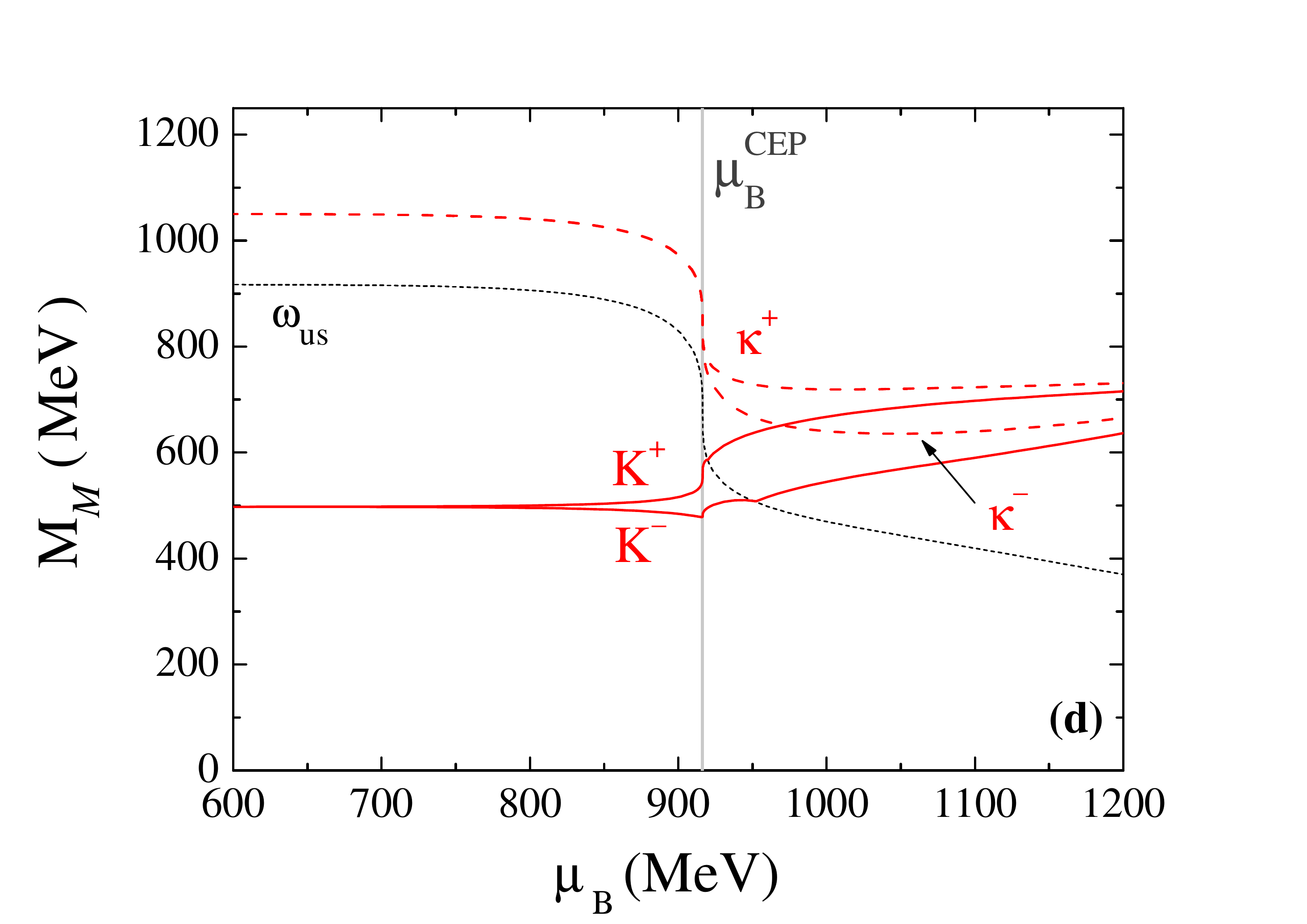}
\end{subfigure}
\vspace{-0.25cm}
\caption{Masses of the pseudoscalar and scalar mesons as function of the
temperature (panels (\textbf{a},\textbf{c})) and as function of $\mu_B$ (panels (\textbf{b},\textbf{d}))
along the path that crosses the CEP.}
\label{fig:Mesons_CEP}
\end{figure}

In this subsection we explore the meson behavior along a path that goes through
the CEP. This~path can be parameterized by $T=0.14~\mu_B$, as seen in Figure
\ref{fig:PD}. The~meson masses are plotted as functions of temperature and chemical potential,
in the left panels and right panels of Figure~\ref{fig:Mesons_CEP} respectively.


We first discuss the behavior of the [$\pi$,$\sigma$], [$\eta$,$a_0$] and
[$\eta'$,$f_0$] meson masses presented in panels (a) and (b) of Figure
\ref{fig:Mesons_CEP}.
The main and most important observation from Figure~\ref{fig:Mesons_CEP} is the
drastic continuous decrease that almost meson masses (except the pion and $K^+$)
displays in the temperature and chemical potential of the CEP.
In fact, this feature may be used as a signal for the CEP.
For example, processes such as $\sigma\rightarrow\pi\pi$ and
$\sigma\rightarrow\gamma\gamma$ productions are important tools for the search
of the CEP~\cite{Anticic:2009pe,Hatsuda:1998vb}. If~the $\sigma$-meson has an
abnormally small mass around the CEP, this means that peculiar experimental
signatures are expected to be observed through its spectral changes:
the dipion decay can be suppressed near the CEP due to the fast reduction of
the $\sigma$-meson mass.
Also the maximum in the $K^+/\pi^+$ ratio (``the horn'' effect)
\cite{Cleymans:2004hj}, that was discussed as a possible signal of the onset of
deconfinement, that also appears near CEP (being considered as a critical region
signal for the CEP~\cite{Friesen:2018ojv}), can be related with the increase of
the $K^+$ mass seen in Figure~\ref{fig:Mesons_CEP} (panels (c) and (d)) relative
to the decreasing $K^-$ mass.

The $\sigma$ and $\eta$ mesons are bound states up to the CEP, where both mesons
reach the $\omega_u$ continuum. The pion survives as a bound state after the CEP
for a very limited range of temperatures and chemical potentials, reaching the
continuum after the CEP.
After becoming resonances, the chiral partners [$\pi$,$\sigma$] degenerate.
The $\eta'$-meson mass is bigger then the $\omega_u$ continuum, meaning it
is always unbound. The~[$\eta$,$a_0$] also become degenerate up to the
$\omega_s$ continuum where the $\eta$ (green curve) and $\eta'$ (magenta curve)
switch roles (this is particularly seen in panel (a) of Figure
\ref{fig:Mesons_CEP} with the mesons as functions of the temperature), as we
will discuss in the following.
The $f_0$ meson is a resonance and its mass is always above the $\omega_s$
continuum.

The $\eta'$ mass reaches the $\omega_s$ continuum at
$\sim1.4T^\mathrm{CEP}$ and $\sim1.4\mu_B^\mathrm{CEP}$. Then, the $\eta$ and
$\eta'$ switch nature since the $\eta'$ immediately become degenerate with
the $a_0$ meson while the $\eta$ meson separates from the $a_0$ mass and
asymptotically degenerates with the $f_0$ meson mass.

As in the crossover scenario, there is charge splitting before and after the
CEP for the $K^\pm$ and $\kappa^\pm$, respectively (see panels (c) and (d) of
Figure~\ref{fig:Mesons_CEP}). The splitting for kaons become even more
pronounced just at the CEP. The $K^+$ is a bound state up to the CEP where it
reaches the $\omega_{us}$ continuum while the $K^-$ is a bound state that
survives the CEP, for a limited range of temperature and chemical potential.
After becoming resonances, the both kaons will degenerate with the respective
chiral partners of the same charge, $\kappa^\pm$, being these scalar mesons
always resonances. The charge splitting between mesons disappears as temperature
and chemical potential increases (faster for the temperature).

\subsubsection{Mesons through the First-Order Transition}

\begin{figure}[t!]
\centering
\begin{subfigure}{.5\textwidth}
\centering
\includegraphics[width=1.1\linewidth]{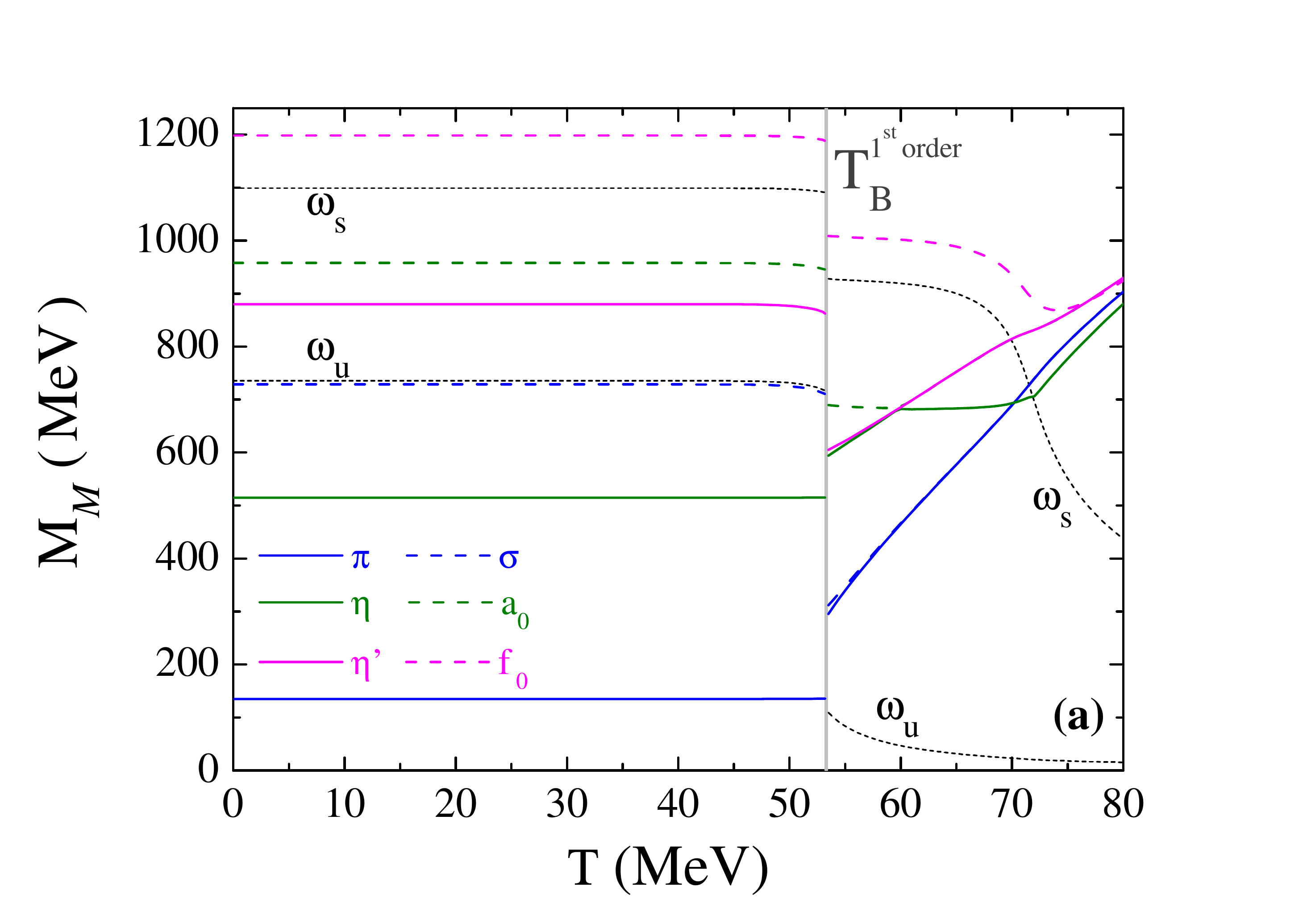}
\end{subfigure}%
\begin{subfigure}{.5\textwidth}
\centering
\includegraphics[width=1.1\linewidth]{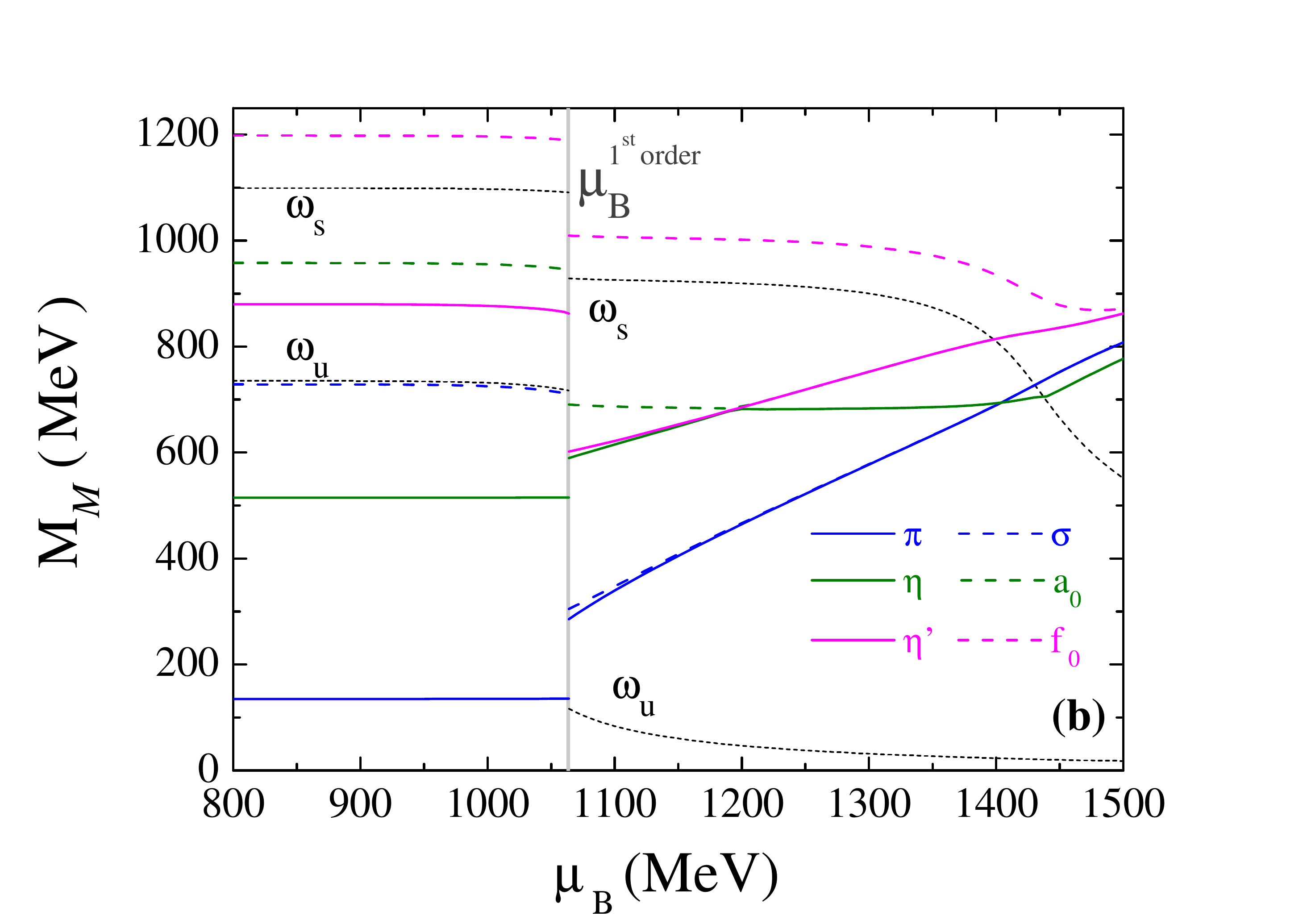}
\end{subfigure}
\begin{subfigure}{.5\textwidth}
\centering
\vspace{-0.5cm}
\includegraphics[width=1.1\linewidth]{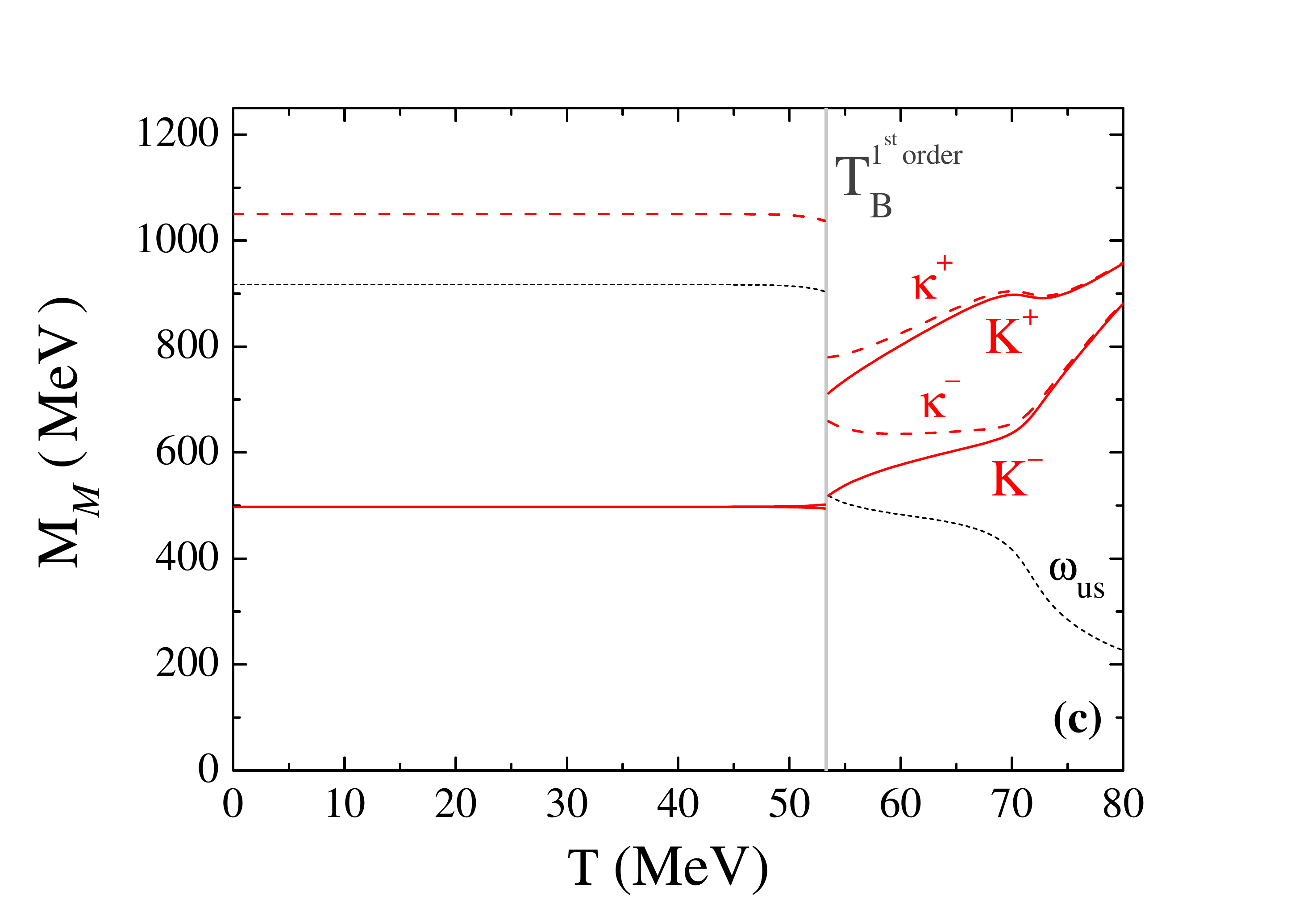}
\end{subfigure}%
\begin{subfigure}{.5\textwidth}
\centering
\vspace{-0.5cm}
\includegraphics[width=1.1\linewidth]{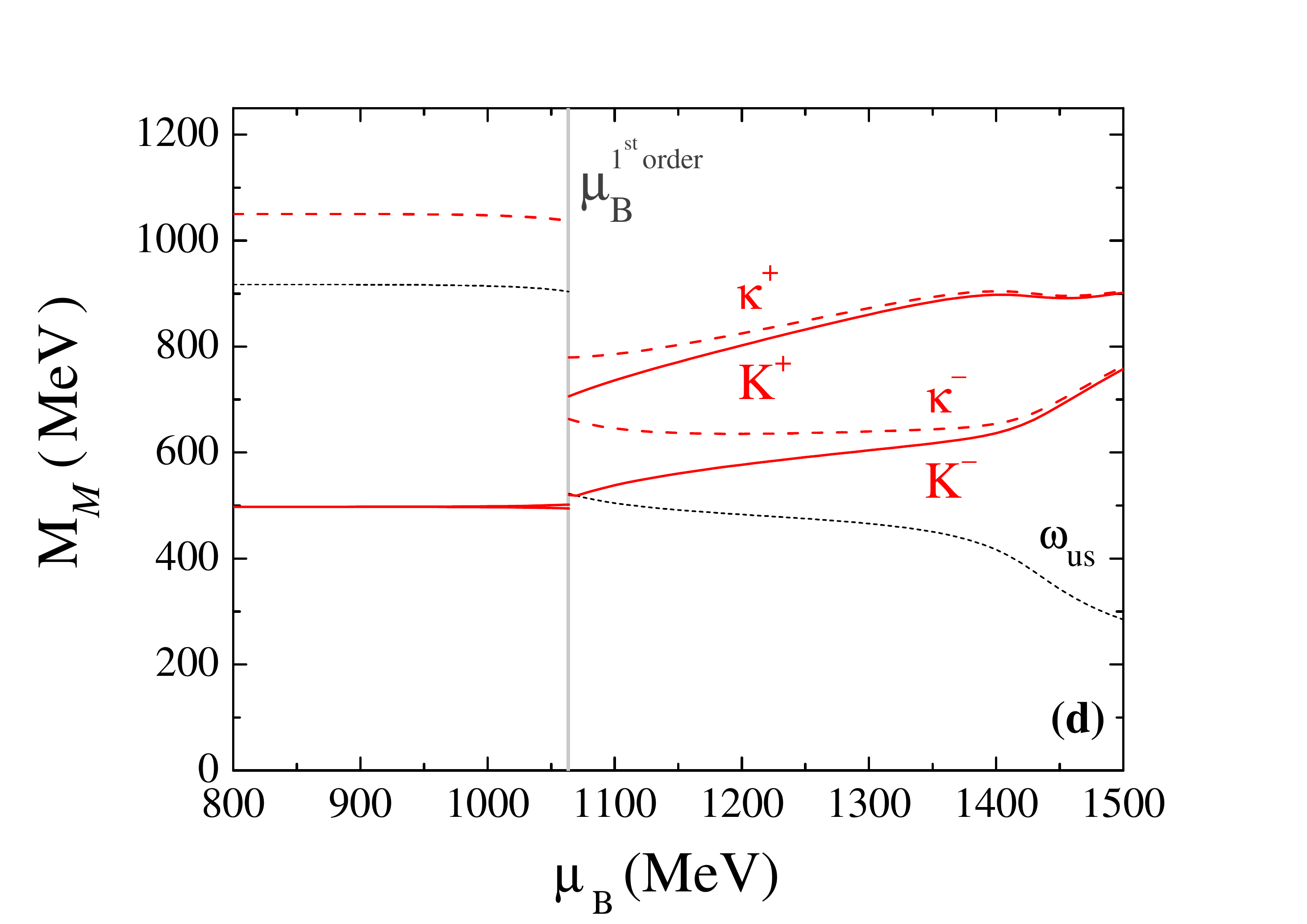}
\end{subfigure}
\vspace{-0.25cm}
\caption{Masses of the pseudoscalar and scalar mesons as function of the
temperature (panels (\textbf{a},\textbf{c})) and as function of $\mu_B$ 
(panels (\textbf{b},\textbf{d}))
along the the path $T=0.05\mu_B$, crossing the first-order region.}
\label{fig:Mesons_1st}
\end{figure}

We now analyze a path that goes through the first-order phase transition,
parameterized by $T=0.05~\mu_B$ (see Figure~\ref{fig:PD}). In the right and left
panels of Figure~\ref{fig:Mesons_1st}, are displayed the mesons masses as
functions of the temperature and chemical potential for this scenario.


The general behavior of the mesons in this case is very similar to the one
encountered in the zero temperature case since, in both cases, there is a
first-order phase transition. However, there are some~specifications.

The $\pi$, $\sigma$ and $\eta$ mesons are bound states up to the phase
transition. Immediately after this point they, discontinuously, become
resonances. In fact, the discontinuity in the meson masses is a consequence of
the nature of the first-order phase transition where the thermodynamical
potential has two degenerate minima and there is a discontinuous jump from one
stable minimum to the other. The~$a_0$ meson is always a resonance.

The $\eta'$ meson mass is always above $\omega_u$ up to the point where it
reaches the $\omega_s$ continuum at higher values of temperature and chemical
potential.
The $f_0$ meson is also always unbound with its mass above $\omega_s$.

After the first-order phase transition the pairs $\pi$ and $\sigma$ and $\eta$
and $\eta'$ degenerate almost immediately. When the $a_0$ mass (dashed green
curve) reaches the mass of the degenerate $\eta$ and $\eta'$ pair, the
$\eta$-meson (solid green curve) decouples from the $\eta'$.
After this point, the $\eta'$ and $a_0$ degenerate and evolve to degenerate
with the $f_0$ meson at high energies while, the $\eta$-meson, remains separate
from the [$\eta'$,$a_0$] pair. When the $\eta$ mass reaches the $\omega_s$
continuum its mass increases to asymptotically degenerate with the other mesons.

Regarding the kaons and their chiral partners, right before the first-order
phase transition the charge multiplets of the $K$ start to split (see Figure
\ref{fig:Mesons_1st} panels (c) and (d)). Before the phase transition only the
$K$ mesons are bound while the chiral partners, $\kappa$, are resonances. At
the first-order phase transition, as expected, there is a discontinuous split
of each charge multiplet of both $K$ and $\kappa$ and every meson is now a
resonance, since they all lie above the $\omega_{us}$ continuum. As temperature
and chemical potential increases, the $K^+$ and $\kappa^+$ mesons start to
degenerate, followed by the respective negative pair. For~larger energies the
charge multiplets of the chiral pair also show tendency to degenerate.

\subsubsection{Mesons along the Isentropic Trajectory That Passes over the CEP}

\begin{figure}[h!]
\centering
\begin{subfigure}{.5\textwidth}
\centering
\includegraphics[width=1.1\linewidth]{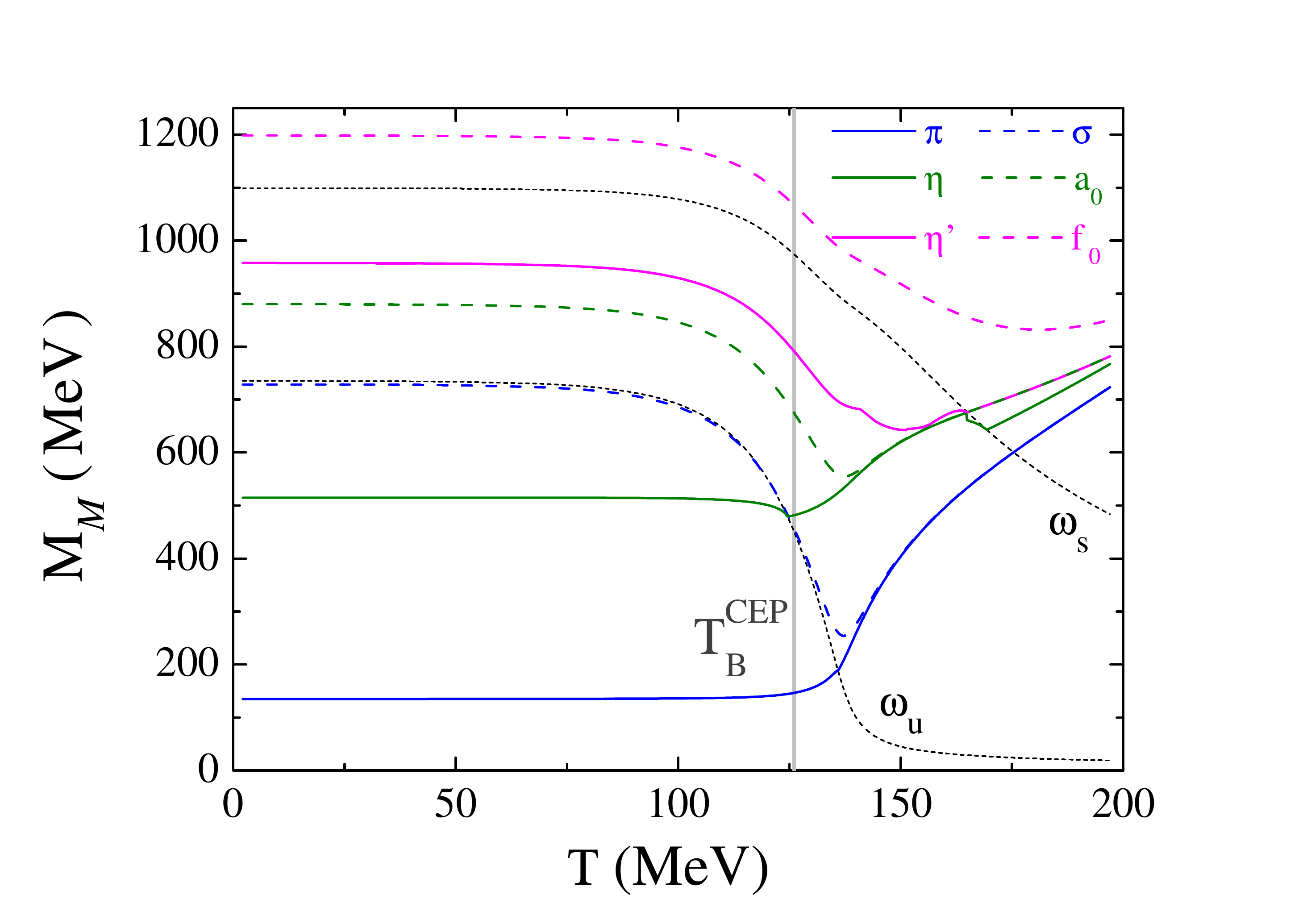}
\end{subfigure}%
\begin{subfigure}{.5\textwidth}
\centering
\includegraphics[width=1.1\linewidth]{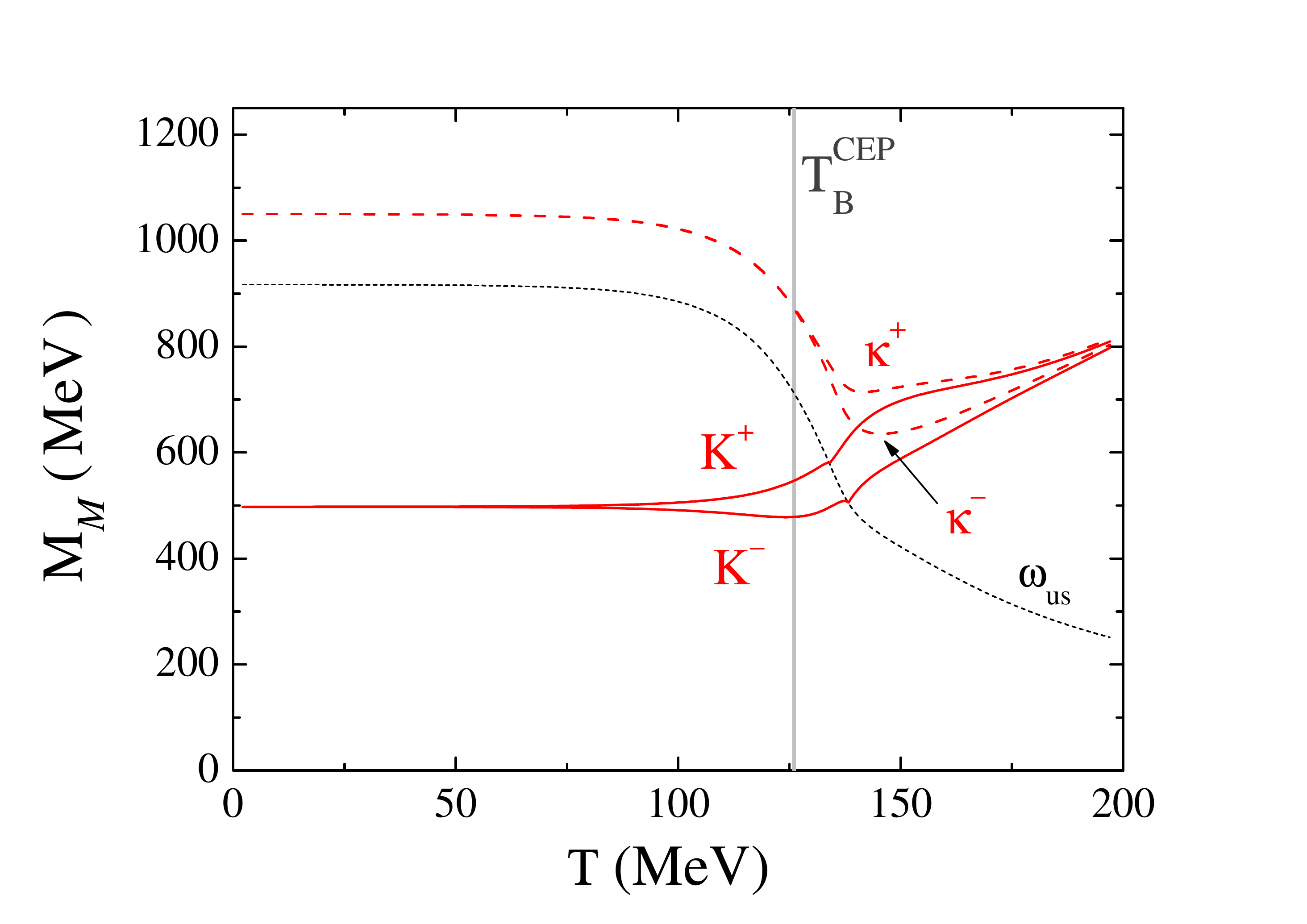}
\end{subfigure}
\vspace{-.25cm}
\caption{Masses of the pseudoscalar and scalar mesons as function of the
temperature (\textbf{left panel}) and as function of $\mu_B$ (\textbf{right panel})
along the the path $T=0.05\mu_B$, crossing the first-order region.}
\label{fig:Mesons_Isent}
\end{figure}

Previously, the isentropic trajectories relevance was presented, namely the
fact that the expansion of the QGP in HIC is accepted to be a hydrodynamic
expansion of an ideal fluid and it approximately follows trajectories of
constant entropy. This argument motivates the presentation of the behavior of
mesons masses as functions of the temperature along the nearest isentropic
trajectory of the CEP, $s/\rho_B\approx5.6$, in Figure~\ref{fig:Mesons_Isent}.
There is a substantial difference for the scenario studied in Section
\ref{MesonsCEP} (the~path $T=0.14\mu_B$): the abrupt decrease of the meson
masses is not seen. Indeed, in this case, along the isentropic trajectory both,
the temperature and the baryonic chemical potential, are varying in such a way
that $s/\rho_B$ is held constant.
So, the approximation to the CEP is not direct, being the behavior of
the mesons masses more smooth in the neighborhood of the CEP.
This may be an indication of different critical behavior near the CEP, depending
on the direction the CEP is approached in the $(T,\mu_B)$ plane.
Indeed in~\cite{Costa:2008gr}, the critical exponents in the PNJL model were
studied and it was found that their values slightly changed depending on the
direction.
Nevertheless, from the right panel of Figure~\ref{fig:Mesons_Isent}, we see that
the separation between $K^+$ and $K^-$ is the same ($\sim70$ MeV) like in
Section~\ref{MesonsCEP}, as it should.
\vspace{-8pt}

\subsection{Effective Restoration of Chiral Symmetry and Mott Dissociation of
$\pi$ and $\sigma$ along the Phase Diagram}

\begin{figure}[h!]
\centering
\includegraphics[width=.55\linewidth]{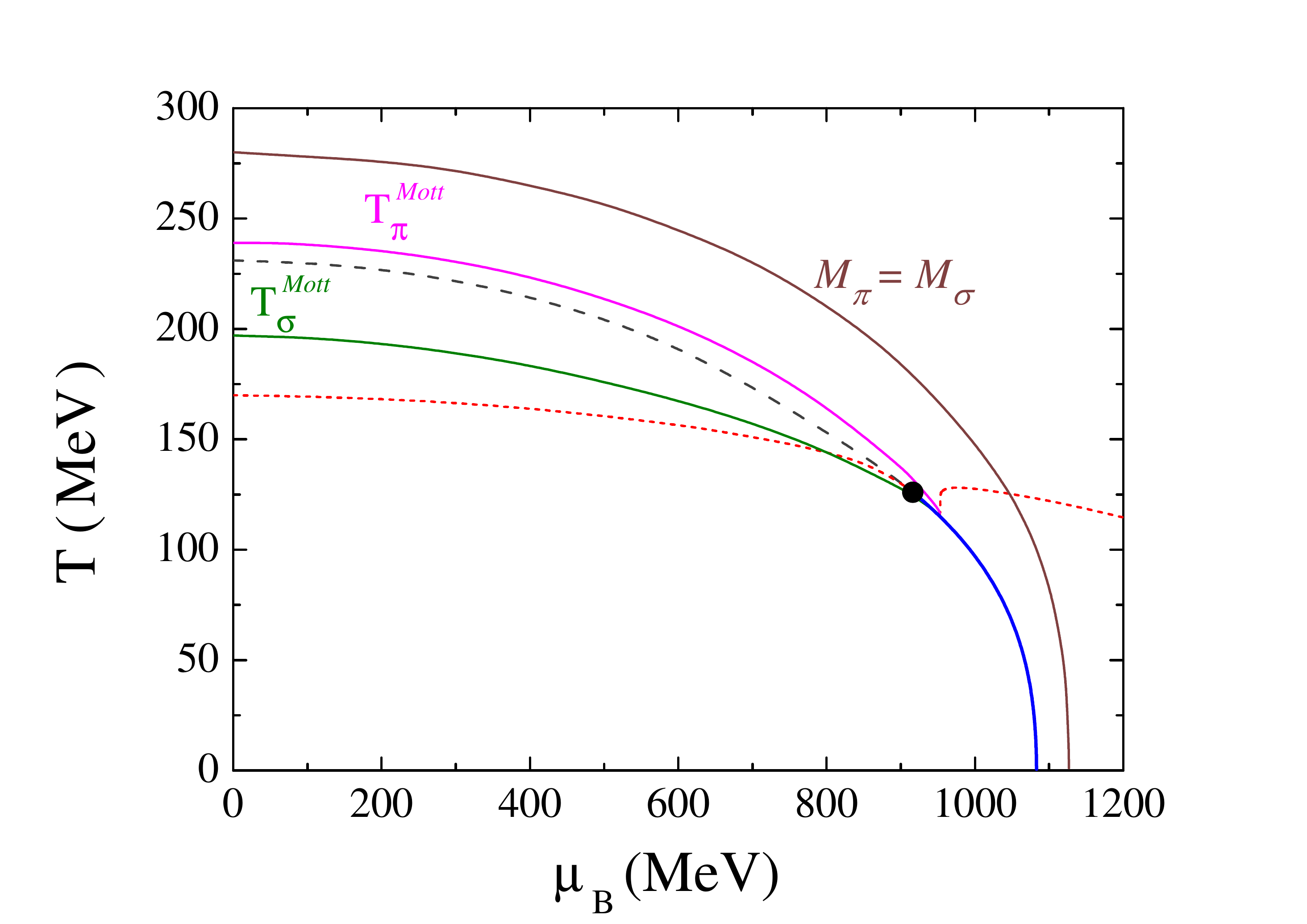}
\vspace{-.25cm}
\caption{Line of effective restoration of chiral symmetry (brown curve) and the
Mott dissociation line for $\pi$ (magenta) and $\sigma$ (green)
along the ($T-\mu_B$)-plane.}
\label{fig:DF_Tmott}
\end{figure}

Since the pion and the sigma mesons dissociate in a $\bar{q}q$ pair, in Figure
\ref{fig:DF_Tmott} we represent the respective ``Mott lines''. In the region
$\mu_B<\mu_B^{CEP}$, the Mott line for $\sigma$-meson occurs below chiral
crossover, while~the correspondent Mott line for $\pi$ occurs slightly above.
It is interesting to note that above $\mu_B\sim\mu_B^{CEP}$ both Mott lines
occurs inside the first-order region.

When we use the degeneracy of the $\pi$ and $\sigma$-meson masses as criterion
to define the point of the effective restoration of chiral symmetry, we
guarantee that all quantities that violate the chiral symmetry are already
sufficiently small.
We then can draw a line for the effective restoration of chiral symmetry (brown
line in Figure~\ref{fig:DF_Tmott}). For example, when finally
$M_{\pi} = M_{\sigma}$, the constituent masses of the $u$ and $d$ quarks are
already sufficiently close to the respective currents masses values (also the
light quark condensates already have a very low values).

\section{Conclusions}\label{sec:conclusions}

In this work, we have explored the phase diagram of strongly interacting matter
and discussed the scalar-pseudoscalar meson spectrum, with related properties,
in connection with the restoration of chiral symmetry and deconfinement for
different regions of the phase diagram. We have used the (2~+~1) PNJL model with
explicit breaking of the U$_A$(1) anomaly, which includes flavor mixing, and the
coupling of quarks to the Polyakov loop.
The PNJL model has proven to be very useful in understanding the underlying
concepts of the phase diagram of strongly interacting matter, namely~the chiral
symmetry breaking and restoration mechanisms. It is also an important tool to
study the dynamics of pseudoscalar and scalar mesons with regard to the
effective restoration of chiral symmetry.

Some of the presented results have shown a great relevance:

\begin{enumerate} [leftmargin=2.3em,labelsep=3.2mm]
\item[(i)] the survivability of some meson modes, especially the pion, as a
bound state after the transition to the QGP (this tendency to a slightly longer
survival as bound state is also shown by the behavior of meson-quark coupling
constants for $\pi$, $\sigma$ and $\eta$ mesons);

\item[(ii)] the change of identity between $\eta$ and $\eta'$ at finite density
for scenarios at lower temperatures;

\item[(iii)] the meson masses change abruptly when choosing a path that passes
through the CEP (this can be very important for the signatures of the CEP);

\item[(iv)] in relation to kaons, with the exception of the limiting cases for
$T=0$ and $\mu_B=0$, a kaon charge splitting before critical temperature/baryonic
chemical potential occurs. At CEP and first-order cases, kaons first degenerate
with the respective chiral partners and only then with charge multiplet,
contrary to the crossover scenario where charge multiplets degenerate first.
At~the~CEP, there is a accentuated splitting for kaons, with $K^+$ sharply
increasing, a splitting that is still pronounced just after the CEP;

\item[(v)] above certain critical values of temperature and chemical potentials
($T^\chi_{eff}$, ${\mu_B}_{eff}^\chi$) the masses of the chiral partners
[$\pi$,$\sigma$] will degenerate, meaning that chiral symmetry is effectively
restored. All~quantities that violate chiral symmetry are guaranteed to be
already sufficiently small .
\end{enumerate}

New extensions of the model (e.g., with explicit chiral symmetry breaking
interactions that lead to a very good reproduction of the overall
characteristics of the LQCD data~\cite{Moreira:2018xsp}, or with spin-0 and
spin-1 U(3)$\times$U(3) symmetric four-quark interactions to deal with
vector–scalar and axial-vector–pseudoscalar transitions~\cite{Morais:2017uvn})
or going beyond the mean-field approximation, by modifying the gap equations
due to mesonic fluctuations in the scalar and pseudoscalar channels, can lead
the (P)NJL type models to new insights of low energy QCD.


\acknowledgments{
The Authors would like to thank Jo\~{a}o Moreira for helpful discussions.
This work was supported by ``Funda\c{c}\~{a}o para a Ci\^{e}ncia e Tecnologia", Portugal, 
under Grant PD/BD/128234/2016 (R.P.).}

\appendix
\section{}
\label{append}

The polarization operator for the meson channel $M$ given in Equation~
(\ref{eq:polarization}) in the static limit, $\vec{q}=\vec{0}$ can be calculated
to yield:
\begin{equation}\label{PI_ij}
\Pi_{ij} (q_0) = 4 \left((I_1^i + I_1^j)	-   [q_0^2-(M_i-M_j)^2]\,\,I_2^{ij}(q_0)\right).
\end{equation}
where $M_i$ is the effective mass of the quark $i$. At  finite temperature and
with two chemical potentials, $\mu_i$ and $\mu_j$, the integral $I_1^i$ can be
written as:
\begin{align}
I_1^i(T,\mu_i) = - \frac{N_c}{4\pi^2} \int \frac{{ p}^2 d{ p}}{E_i} \left(
\nu_i - \bar{\nu}_i
\right).
\label{Ii1}
\end{align}
The integral $I_2^{ij}$ is given by:
\begin{align}
I_2^{ij} (q_0,T,\mu_i,\mu_j) =
& - N_c \int \frac{d^3{  p}}{(2\pi)^3}
\Biggl[
\frac{1}{2E_i}
\frac{ \nu_i }{(E_i+q_0- (\mu_i-\mu_j))^2-E_j^2}
-  \frac{1}{2E_i}  \frac{ \bar{\nu}_i }{(E_i-q_0+ (\mu_i-\mu_j))^2-E_j^2}
\nonumber \\
& +  \frac{1}{2E_j}\frac{ \nu_j }{(E_j-q_0+ (\mu_i-\mu_j))^2-E_i^2}
- \frac{1}{2E_j}\frac{ \bar{\nu}_j }{(E_j+q_0- (\mu_i-\mu_j))^2-E_i^2}
\Biggr].
\label{Iij2}
\end{align}

%
Here, $E_i$ is the quasiparticle energy, $E_{i}=\sqrt{\mathbf{p}^{2}+M_{i}^{2}}$,
$\mu_i$ is the chemical potential for the quark $i$. The distribution functions
are given by Equations (\ref{eq:gap.PHI.PNJL}) and (\ref{eq:gap.PHIbar.PNJL}).

\section{}
\label{append2}

In the $T \to 0$ limit, the grand canoncial potential of the SU$_f$(3) PNJL and
the SU$_f$(3) NJL model are identical. We can see this by defining the zero
temperature limit of Equation~(\ref{eq:pot.termo.PNJL}) as $\Omega^{0}$ and
writing:
\begin{myequation}
\begin{array}{lll}
\Omega^{0} & =
\lim_{T \to 0} \Omega (T)
\vspace{6pt}\\
& =  \lim_{T \to 0} \mathcal{U}\left(\Phi,\bar{\Phi};T\right)
+ g_{_{S}} \sum_{i=u,d,s} \left\langle\bar{q}_{i}q_{i}\right\rangle^2
+ 4 g_{_{D}}\left\langle\bar{q}_{u}q_{u}\right\rangle
\left\langle\bar{q}_{d}q_{d}\right\rangle\left\langle\bar{q}_{s}q_{s}\right\rangle
-  2N_c \sum_{i=u,d,s} \int \frac{d^3p}{(2\pi)^3} E_i
\vspace{6pt}
\\
& \quad \quad \quad - 2 \sum_{i=u,d,s} \int \frac{d^3p}{(2\pi)^3} \left[
\lim_{T \to 0}\mathcal{F}\left(\vec{p},T,\mu_i \right) +
\lim_{T \to 0}\mathcal{F}^*\left(\vec{p},T,\mu_i \right)  \right] .
\numberthis
\end{array}
\end{myequation}
The~Polyakov loop potential, $\mathcal{U}\left(\Phi,\bar{\Phi};T\right)$, as
defined in Equation~(\ref{eq:Polyakov.loop.potential}), vanishes in this limit.
The~limits of the thermal functions (\ref{eq:Fthermal}) and (\ref{eq:F*thermal})
are given by:
\begin{align}
\lim_{T \to 0} \mathcal{F}\left(T,{\mu}_i\right) & = 3\left( {\mu}_i - E_i \right)\theta \left(E_i-\mu_i \right) ,
\\
\lim_{T \to 0} \mathcal{F}^*\left(T,\mu_i\right) & = 0 .
\end{align}
where $\theta(E_i-{\mu}_i)$ is the Heaviside step function. One can define the
Fermi momentum of the quark of flavor $i$ as:
\begin{align}
\lambda_{F_i} = \sqrt{\mu_i^2 - M_i^2 }.
\end{align}
The grand canonical potential can then be written as:
\begin{align*}
\Omega^{0} & =
g_{_{S}} \sum_{i=u,d,s} \left\langle\bar{q}_{i}q_{i}\right\rangle^2
+ 4 g_{_{D}}\left\langle\bar{q}_{u}q_{u}\right\rangle
\left\langle\bar{q}_{d}q_{d}\right\rangle\left\langle\bar{q}_{s}q_{s}\right\rangle
-  \frac{N_c}{\pi^2} \sum_{i=u,d,s} \int^{\Lambda}_{\lambda_{F_i}} dp  p^2 E_i
-
\sum_{i=u,d,s} {\mu}_i \frac{\lambda_{F_i}^3}{\pi^2}  .
\numberthis
\end{align*}
The $i-$quark density is:
\begin{align}
\rho_i = \frac{\lambda_{F_i}^3}{\pi^2} .
\label{quark.density.NJL}
\end{align}
Applying the stationary conditions yields the respective $i-$quark condensate,
\begin{align}
\left\langle\bar{q}_{i}q_{i}\right\rangle=-\frac{N_c}{\pi^2} \int^{\Lambda}_{\lambda_{F_i}} dp  p^2 \frac{M_i}{E_i} .
\end{align}
These set of equations, which define the SU$_f$(3) PNJL in this limit, are
identical to those defining the SU$_f$(3) NJL at zero temperature.

\bibliography{biblio}        
\bibliographystyle{apsrev}

\end{document}